\documentclass[3p,twocolumn,9pt]{elsarticle}
\usepackage[utf8]{inputenc}
\usepackage[english]{babel}
\usepackage{graphicx}
\usepackage{subfig}
\usepackage{subfiles} 
\usepackage{blindtext}
\usepackage{pagecolor}
\usepackage{color}
\usepackage{soul}
\usepackage{lineno}
\usepackage{algorithm}
\usepackage{algorithmicx}
\usepackage{algpseudocode}
\usepackage[colorinlistoftodos,prependcaption,textsize=tiny]{todonotes}
\usepackage{amsfonts}
\usepackage{amsmath}
\usepackage{amssymb}
\usepackage{amsthm}
\usepackage[T1]{fontenc}
\usepackage[title]{appendix}
\usepackage{xr}
\definecolor{lightyellow}{RGB}{255, 255, 240}
\pagecolor{lightyellow}

\newtheorem{corollary}{Corollary}[section]

\newtheorem{proposition}{Proposition}[section]

\newtheorem{definition}{Definition}[section]

\newtheorem{remark}{Remark}[section]

\newtheorem{theorem}{Theorem}[section]

\usepackage{listings}
\usepackage{color}
 
\definecolor{codegreen}{rgb}{0,0.6,0}
\definecolor{codegray}{rgb}{0.5,0.5,0.5}
\definecolor{codepurple}{rgb}{0.58,0,0.82}
\definecolor{backcolour}{rgb}{0.95,0.95,0.92}
 
\lstdefinestyle{mystyle}{
    backgroundcolor=\color{backcolour},   
    commentstyle=\color{codegreen},
    keywordstyle=\color{magenta},
    numberstyle=\tiny\color{codegray},
    stringstyle=\color{codepurple},
    basicstyle=\footnotesize,
    breakatwhitespace=false,         
    breaklines=true,                 
    captionpos=b,                    
    keepspaces=true,                 
    numbers=left,                    
    numbersep=5pt,                  
    showspaces=false,                
    showstringspaces=false,
    showtabs=false,                  
    tabsize=2
}
 
\algnewcommand\algorithmicassumption{\textbf{Assumptions:}}
\algnewcommand\ASSUMPTIONS{\item[\algorithmicassumption]}

\algnewcommand\algorithmicinput{\textbf{Inputs:}}
\algnewcommand\INPUTS{\item[\algorithmicinput]}

\algnewcommand\algorithmicoutput{\textbf{Outputs:}}
\algnewcommand\OUTPUTS{\item[\algorithmicoutput]}

\algnewcommand\algorithmicproc{\textbf{Procedure:}}
\algnewcommand\PROCEDURE{\item[\algorithmicproc]}

\begin{document}
\begin{frontmatter}
\title{\MakeUppercase{\textbf{What Makes An Asset Useful?}}}

\author[rvt]{Dr. Yves-Laurent Kom Samo\corref{cor1}\fnref{fn1}}
\author[rvt]{Dr. Dieter Hendricks\fnref{fn2}}
\fntext[fn1]{Founder \&  CEO, Pit.AI Technologies Inc.}
\fntext[fn2]{Senior  Researcher, Pit.AI Technologies Inc.}
\address[rvt]{Pit.AI Technologies Inc., 1735 N. 1st St., Suite 305, San Jose, CA 95112, United States}
\cortext[cor1]{Corresponding author: ml@pit.ai}
	
\begin{abstract}
Given a new candidate asset represented as a time series of returns, how should a quantitative investment manager be thinking about assessing its usefulness? This is a key qualitative question inherent to the investment process which we aim to make precise. We argue that the usefulness of an asset can only be determined relative to a reference universe of assets and/or benchmarks the investment manager already has access to, or would like to diversify away from, for instance standard risk factors, common trading styles and other assets. 

We identify four features that the time series of returns of an asset should exhibit for the asset to be useful to an investment manager, two primary and two secondary. As primary criteria, we propose that the new asset should provide sufficient \emph{incremental diversification} to the reference universe of assets/benchmarks, and its returns time series should be \emph{sufficiently predictable.} As secondary criteria, we propose that the new asset should \emph{mitigate tail risk}, and the new asset should be \emph{suitable for passive investment} (e.g. buy-and-hold or short-and-hold). We discuss how to quantify incremental diversification, returns predictability, impact on tail risk, and suitability for passive investment, and for each criteria, we provide a scalable algorithmic test of usefulness.
\end{abstract}
\end{frontmatter}


\section{\MakeUppercase{Introduction}}

The idea of diversification is ancient. The related well-known phrase `don't put all your eggs in one basket' can be traced back to the classical novel \emph{Don Quixote} by \emph{Miguel de Cervantes Saavedra} as early as 1605 \cite{de1842don}, where it is phrased as ``\emph{It is the part of a wise man to keep him­self to­day for to­mor­row, and not ven­ture all his eggs in one bas­ket.}''  The idea of diversification itself is much older, and can be found for instance in the book of Ecclesiastes ---\emph{``But divide your investments among many places, for you do not know what risks might lie ahead''}, Ecclesiastes 11:2---, which is thought to have been written around 935 B.C. This notion was formalized in the context of portfolio construction by Harry Markowitz in the celebrated Modern Portfolio Theory (also referred to as Mean-Variance Analysis) introduced in his seminal paper \cite{markowitz1952portfolio}, where it is illustrated that an investor can reduce portfolio risk simply by holding combinations of assets that have the same expected return and are not perfectly (positively) correlated.

The main limitation of the existing literature in analyzing the value in investing in multiple assets is that two fundamentally different questions are often entangled: i) how much incremental value can an investment manager derive from trading a given additional asset,  and ii) how should an investment manager go about extracting the potential incremental value inherent to trading a specific additional asset? 

In the literature, the former question is seldom investigated, whereas attempts to address the latter abound \cite{grinold2000active, ilmanen2011expected}. It is typically assumed that the universe of assets to trade is given, and the emphasis is placed on optimizing capital allocation across these assets. In the Mean-Variance framework and related approaches, this typically requires estimating expected returns ---or equivalently forecasting returns---, and forecasting risk by estimating the cross-covariance matrix of returns, so as to find a portfolio that maximizes a utility function trading off expected returns and risk \cite{grinold2000active}. Naturally, if one is able to figure out how to extract value out of trading a given additional asset, then it means that the additional asset does indeed add intrinsic value to the existing pool. However, attempting to determine optimal allocation within a large pool of related assets can pose serious numerical challenges, such as ill-conditioning of the covariance matrix, capable of impairing the optimization process. In such situations, and more generally, one would be better off first determining \emph{how much incremental value} can a new asset add to an existing pool, and then determining \emph{how to extract incremental value}, if any. Intuitively, one should include an asset in one's trading universe only if it is expected to add sufficient value. Our aim in this paper is to propose a rigorous approach to quantifying the value an asset intrinsically adds to a reference pool, without making any restrictive assumption on how one would go about extracting such value.

Interestingly, judging by the proliferation of fund-of-funds, practitioners have long grasped the importance of mitigating risk concentration across risk factors such as fund managers, asset classes, strategy styles, geographic locations etc., on top of or as constraints to the optimization process.  However, their approaches to mitigating risk concentration are often intuitively grounded, but may lack scientific rigor. The approach we take in this paper does not treat these risk factors differently \emph{a priori}; instead we use the following generic definition of an asset, which has exchange-traded assets (e.g. stocks, bonds, commodities, currencies, futures, etc.) as well as synthetic assets (e.g. any fund irrespective of its manager, mandate or strategy style, any trading strategy) as special cases, and we propose a data-driven approach for constituting a pool of assets that mitigates risk concentration, among other features.

\begin{definition}
We denote an \textbf{asset} as any investment resulting in a periodic stream of returns, realized or marked-to-market.
\end{definition}

Throughout the rest of this paper we identify an asset by its time series of returns, and we consider that two assets having identical time series of returns are identical for all investment purposes. In particular, our approach to quantifying the usefulness of an asset primarily relies on its time series of returns. Moreover, we assume all returns time series have the same sample frequency. For multi-frequency series, one can normalize frequencies by marking lower frequency returns series to market up to the highest frequency, or aggregating higher frequency returns up to the timescale of the lowest frequency, the former being preferred as the latter would result in data loss. To avoid any confusion, we use the expression `pool of assets' to denote a universe or set of assets, and we reserve the expressions `portfolio of assets' and `fund' to any specific allocation of capital across assets in a pool. Additionally, we denote `static portfolio' any portfolio of assets whose target capital allocation does not change over time.

The rest of the paper is structured as follows. In Section  \ref{section:motivation} we provide an intuitive answer to what makes an asset useful. We argue that the usefulness of an asset can only be considered relative to a reference pool of assets, and we argue that for an asset to be useful, it needs to sufficiently add to the diversification of the reference pool of assets, its returns time series should be sufficiently predictable, it needs to mitigate tail events, and it needs to be suitable for passive investment. In Section \ref{section:diversification} (resp. Section \ref{section:predictability}) we provide a mathematical framework for quantifying incremental diversification (resp. predictability of returns), and we propose a scalable implementation thereof. In Section \ref{section:tails} we propose a formalism for quantifying the impact of a new asset on tail events, and in Section \ref{section:passive} we discuss quantifying the suitability of an asset for passive investment. We conclude in Section \ref{section:conclusion}.

\section{\MakeUppercase{What Exactly Makes An Asset Useful?}}
\label{section:motivation}
\documentclass[../main.tex]{subfiles}
So what exactly makes an asset useful to an investment manager? Intuitively, the answer ought to depend on what assets the investment manager already has access to, and what risk factors and benchmarks, if any, he/she wants to avoid exposure to. Indeed, no matter how high an asset's returns are, if the asset's time series of returns can be mimicked or replicated using assets the investment manager already has access to, and/or factors or benchmarks the investment manager would like to avoid exposure to, then it is fair to conclude that the new asset presents little incremental usefulness to the investment manager. Thus, the usefulness of an asset to an investment manager can be thought of as the incremental usefulness that the asset adds to a reference pool of assets and/or factors/benchmarks the investment manager trades and/or would like to avoid exposure to. 

Prior to any technical discussion, let us review four intuitive features we would expect an asset to exhibit to consider it incrementally useful relative to a reference pool of assets. Each of the four features corresponds to a motivation an investment manager might have for widening the universe of assets he/she trades.

\subsection{\textbf{\textsc{Incremental Diversification}}}
Perhaps the most fundamental reason why an investment manager might want to consider broadening the pool of assets he/she trades is diversification. To make matters precise, we provide the following definition for diversification.

\begin{definition}
Throughout this paper we denote \textbf{diversification} as the act through which one aims at reducing the level of risk of a portfolio, for the same level of expected return, by adding one or more assets.
\end{definition}

We note however that not all new assets have the same potential for reducing the risk of a portfolio; some provide more diversification potential than others. Intuitively, adding shares of a U.S. bank to a universe of U.S. financial stocks might not present the same diversification benefits as adding a commodity futures contract to the same pool. The former can be regarded as mostly driven by the same market dynamics as U.S. financial stocks, whereas the latter appears fairly unrelated to financial stocks on the surface. We note that, as much as the actual reduction of risk incurred by trading a new asset depends on the specific allocation of capital across all assets (including the new one), in some cases, the new asset will not provide any risk reduction (for a given level of expected return) irrespective of the allocation. An important criteria of usefulness of an asset is therefore the extent to which it \emph{can} provide a reduction of risk (for a given level of expected return), which is a function of the reference pool of assets, and is \emph{independent} of asset allocation.

\subsection{\textbf{\textsc{Predictability of Returns}}} 
Central to most popular portfolio optimization approaches, is the need to estimate expected values of asset returns, which practitioners typically do by forecasting future asset returns \cite{grinold2000active, ilmanen2011expected}. However, if a time series of returns is pure noise, any attempt to forecast it would be vain, and most portfolio optimization processes would fail to make use of the new asset.\footnote{More precisely, the best forecast for future returns of the new asset would be the average of all past returns, which would be constant and, mostly likely, very close to $0$. Hence, rational portfolio allocation procedures wouldn't allocate capital to the new asset as it will be thought to have too low a return per unit of risk.} Therefore, for a new asset to be useful, its time series of returns needs to be sufficiently predictable.

\subsection{\textbf{\textsc{Reduced Tail Risk}}}
Another reason that can motivate investment managers to add a new asset to their trading universe is to mitigate the likelihood that their portfolio can undergo a significant idiosyncratic move, thereby possibly causing their investors to panic and withdraw assets. Unlike the incremental diversification requirement previously discussed, this requirement focuses on the tail (or extreme) events. 

The rationale for widening the trading universe as a way of mitigating tail events is that, doing so could reduce the proportion of total exposure that is concentrated in a single asset, thereby reducing the sensitivity of the overall portfolio to idiosyncratic shocks. However, trading more assets doesn't always result in fewer tail events. A new asset that has frequent and large idiosyncratic moves might adversely affect the tail behavior of a reference pool of assets, especially if it is positively correlated with assets in the pool. However, if the new asset has lighter tails than, or is negatively correlated with assets in the reference pool, its idiosyncratic moves might be lighter or coincide and cancel out shocks in other assets in the reference pool so that, overall, including the new asset in the trading universe would reduce tail events. The usefulness of an asset here is related to its potential to reduce tail risk inherent to the reference pool of assets, rather than tail risk of a specific portfolio. Once more, this criteria is not specific to the investment manager's capital allocation procedure.

\subsection{\textbf{\textsc{Suitability for Passive Investment}}}
Perhaps the most wide-spread expectation one can have of an asset is that it appreciates over time. We slightly relax this requirement, and instead require of an asset that it be suitable for passive investment in order to be considered useful. In other words, it should be possible to find strategies that do not change target holdings in the asset too often ---for instance buy-and-hold and short-and-hold strategies--- and that perform well in the long-run.

\subsection{\textbf{\textsc{Relative Importance of Usefulness Criteria}}}
Incremental diversification and predictability of returns are primary criteria of incremental usefulness in that, if either one is not met, no investment manager will find the new asset incrementally useful. A new asset whose returns can be perfectly replicated using existing assets intuitively wouldn't add any utility to the reference pool of assets. The alert reader might be thinking of Exchange Traded Funds (ETFs) as a counter-example. We note however that investors who prefer buying/selling ETFs over transacting in the underlying assets directly, do so for multiple reasons (e.g. lower execution costs, more favorable taxation, etc.), all of which eventually translate to a difference in net returns between the ETF and the tracked portfolio of underlying assets. Similarly, if the new asset has returns time series that is pure noise, active investment managers would have no hope of forecasting returns, and passive investment managers would have no reason to believe that the new asset would either appreciate over time or depreciate over time --both being equally likely. 

The other two criteria on the other hand are secondary in that a new asset could be useful to some investment managers, even if those criteria are not met. For instance, a new asset that exhibits strongly predictable returns can be exploited by an active investment manager, even if it isn't suitable for passive investment. Additionally, in case the new asset undergoes more or sharper idiosyncratic moves than assets in the reference pool, but otherwise has highly predictable returns and significantly diversifies the reference pool, an active investment manager might still want to consider adding it to the reference pool, and hedging the incremental tail risk with derivative products. Indeed, high predictability of returns and high incremental diversification provide a guarantee that, although buy-and-hold and short-and-hold strategies on the new asset do not perform particularly well, it is possible to find a (more active) trading strategy on the new asset that both performs well and yields a returns stream that is decorrelated with existing assets.

\section{\MakeUppercase{Quantifying Incremental Diversification}}
\label{section:diversification}

In order to motivate our approach to quantifying incremental diversification, let us first make precise scenarios in which we would intuitively conclude an asset incrementally diversifies a reference pool, and cases where we would consider the new asset to be redundant.

\subsection{\textbf{\textsc{Intuition}}}
The guiding principle to determining whether a new asset adds incremental diversification to a reference pool of assets ought to be that, \emph{if it is easy to replicate the stream of returns of the new asset using assets and other factors in the reference pool, then the new asset doesn't add diversification to the reference pool.} Thus, assets whose time series of returns are impossible to replicate using assets and factors in the reference pool should have the highest incremental diversification. Similarly, assets whose returns are easy to replicate using assets and factors in the reference pool should have the lowest incremental diversification. 

To affine our intuition, let us consider some concrete examples that will help derive stylized features that a suitable quantitative measure of incremental diversification should exhibit.

 We start by considering a fund $\pi$ whose (one-period) returns we assume are independent across time, and drawn from the same random variable $r_\pi$ with mean $\mu$ and standard deviation $\sigma$. We consider another asset A, whose (one-period) returns we assume are also independent across time, and drawn from a random variable $r_A$ whose mean $\mu$ and standard deviation $\sigma$ are the same as that of $r_\pi$, and we denote $\rho$ the correlation between $r_\pi$ and $r_A$. Furthermore, we denote $\pi^\prime$ the portfolio consisting of investing a fraction $0 \leq \omega \leq 1$ of our wealth in fund $\pi$ and the rest in buying asset A. The (one-period) returns of the new portfolio are easily found to be independent draws across time from the random variable $$r_{\pi^\prime} = \omega r_\pi + (1-\omega) r_A,$$ 
whose mean and standard deviation read $$\mu_{\pi^\prime} := \mathbb{E}(r_{\pi^\prime}) = \mu$$ and 
\begin{align}
\label{form:red_std}
\sigma_{\pi^\prime} :&= \sqrt{ \mathbb{V}\text{ar}\left( r_{\pi^\prime}\right)} \nonumber \\ &= \sigma \sqrt{1- 2\omega (1-\omega)(1-\rho)} \\
& \leq \sigma \nonumber,
\end{align}
where the last inequality stems from the fact that $0 \leq \omega \leq 1$ and $\rho \leq 1$. It is worth noting that the inequality is always strict, unless either $2\omega (1-\omega)=0$, which corresponds to only investing in $\pi$ or A, or $\rho=1$, that is when $\pi$ and A are perfectly correlated.  This leads us to the most basic, yet fundamental, observation about diversification. \emph{Expected return and variance being equal, adding a new asset to a portfolio always results in a higher return per unit of risk,\footnote{Here we use the standard deviation of returns as a measure of risk. However, our aim is not to equate the notion of risk to the standard deviation of returns, which might fail to capture subtle tail behaviors in non-Gaussian distributions.} unless the new asset is perfectly correlated with the existing portfolio, in which case expected mean and standard deviation of returns remain unchanged.}

We also note that, for a fixed $\omega$, the lower the correlation between $\pi$ and A, the lower the standard deviation of the new portfolio, and consequently the more `diversification value' we get out of adding A to the portfolio. In this toy example, portfolio $\pi$ is regarded as a single asset, that is, we do not have control over its allocation. The generalization of our previous comment to the multi-assets case is that, the lower the correlation between the new asset A and its best replicating portfolio, the more `diversification value' we would expect A to provide. Let us further formalize this expectation.

\begin{definition}\label{def:brp}Let P be a reference pool of assets with returns $\pmb{x}$, and let A be an asset not in P, with return $r_A$. We denote the portfolio that best replicates A using P as the portfolio of assets in P whose allocation, which we denote $\omega^*$, satisfies
\begin{align}
\label{prob:brp}
 \omega^* :&= \underset{\omega}{\text{argmin}} ~~ \mathbb{V}\text{ar}\left( r_A - r_A^\prime \right) \nonumber \\
 &= \underset{\omega}{\text{argmin}} ~~ \mathbb{V}\text{ar}\left( r_A - \omega^T \pmb{x} \right) 
\end{align}
where $r_f$ denotes the risk-free interest rate, and $$r_A^\prime := \omega^T \pmb{x}  + \left( 1- 1^T\omega \right)r_f$$ denotes the return of the portfolio with allocation $\omega$ across assets in P and whose excess cash (resp. net leverage) earns (resp. is funded by borrowing at) the risk-free interest rate.\footnote{We do not assume that the replicating portfolio is fully-invested (i.e. $1^T\omega = 1$), but instead we allow borrowing and lending at the (deterministic) risk-free rate $r_f$, in which case $1^T\omega$ can go above or below $1$ depending on whether we borrowed to fund our positions in assets in the pool, or lent the part of our wealth that isn't invested in assets in the pool.}
\end{definition}

Optimization Problem (\ref{prob:brp}) is quadratic and is easily found to have solution
\begin{align}
\omega^* = \mathbb{C}\text{ov}\left(\pmb{x}, \pmb{x} \right)^{-1} \mathbb{C}\text{ov}\left(\pmb{x}, r_A \right) .
\end{align}
Denoting $r_A^*$ the return of the best replicating portfolio, it follows that 
\begin{align}
\label{eq:cov_brp}
\mathbb{C}\text{ov}\left(r_A, r_A^* \right) & = \mathbb{V}\text{ar}\left( r_A^* \right)  \\
& = \mathbb{C}\text{ov}\left(r_A, \pmb{x} \right) \mathbb{C}\text{ov}\left(\pmb{x}, \pmb{x} \right)^{-1} \mathbb{C}\text{ov}\left(\pmb{x}, r_A \right). \nonumber
\end{align}
Additionally, the residual return of the replication ---which we also refer to as tracking error---, namely $r_A-r_A^*$, has variance that is found to be
\begin{align}
\mathbb{V}\text{ar}\left( r_A-r_A^* \right) = \mathbb{V}\text{ar}\left( r_A \right) - \mathbb{V}\text{ar}\left(r_A^* \right).
\end{align}
Intuitively, a low variance of residual returns of replication indicates that we are able to replicate the stream of returns of the new asset using assets in the reference pool fairly well, which in turn implies, according to our foregoing guiding principle, a low potential for incremental diversification. In other words, a suitable quantitative measure of incremental diversification should never be high when the correlation between the return of the new asset and that of its best replicating portfolio is high, or when the variance of the return of the best replicating portfolio is high relative to that of the replicated asset.\\

\textbf{Stylized Fact 1:} \emph{A good quantitative measure of the incremental diversification an asset A adds to a reference pool of assets should never be high when the correlation between the new asset A and the portfolio of assets in the reference pool that best replicates A is high.}\\

What about the reverse? What scenarios can we intuitively conclude wouldn't result in incremental diversification? The first that comes to mind from our previous discussion is when the correlation between the new asset and its best replicating portfolio of assets in the reference pool is 1, in particular when the returns of the new asset A can be written down as a linear combination of returns of assets in the reference pool, and consequently can be perfectly and easily replicated using assets in the reference pool.\\

 \textbf{Stylized Fact 2:} \emph{A good quantitative measure of the incremental diversification an asset A adds to a reference pool of assets should be the lowest when returns of the new asset A can be obtained as a linear combination of returns of assets in the existing pool.}\\

\begin{remark}
The alert reader might be wondering what to make of trading strategies such as pairs trading, that are only possible when some assets exhibit strong relationships. We note that these strategies are only profitable if the spread between the two assets forming the pair deviates enough from its equilibrium regime, certainly enough to cover transaction costs. The more the spread deviates from its equilibrium, the bigger the profit opportunity, and therefore the more useful both assets are. At the same time, the farther the spread deviates from its equilibrium, the harder it is to exactly replicate the returns time series of one asset using that of the other, and therefore the more incremental diversification one asset adds to the other. Thus, successful pairs trading strategies are consistent with our discussion so far. The subtlety here is that, to be useful, such strategies as pairs trading rely on both a valid long-term equilibrium model for the relationship between asset returns, and a potential for a large short-term deviation from the equilibrium model.
\end{remark}

The second stylized fact deals with assets whose returns are a linear combination of returns of assets in the reference pool, and consequently can easily be emulated using a portfolio of assets in the reference pool, the allocation of which doesn't change over time and can be obtained using linear regression.  Additionally, we consider undesirable any trading strategy whose return over a time period is fully determined by current and past returns of assets in the reference pool, and current and past values of other factors in the reference pool.\\

\textbf{Stylized Fact 3:} \emph{Let} $\{\pmb{x}_t\} := \{(x^1_t, \dots, x^n_t)\}$ \emph{be the time series of asset returns and factor values of a reference pool of $n$ assets and factors, and} $\{y_t\}$ \emph{the times series of returns of a new asset A. A good quantitative measure of the incremental diversification asset A adds to the reference pool should be the lowest when returns of the new asset A can be obtained as a function of present and past values of returns and factors in the pool, that is} $y_t = f(\pmb{x}_t, \dots, \pmb{x}_{t-m})$, \emph{for some function} $f$ \emph{and memory} $m \geq 0$.\\

It is worth stressing that the time series characteristic of the reference pool, namely $\{\pmb{x}_t\}$, is discrete-time, and each unit of discrete-time corresponds to the same wall-clock time. Thus, Stylized Fact 3 aims at discarding new trading strategies or assets that exploit information present in existing assets or factors at the same timescale as the reference time series, or equivalently only considering novel trading strategies or assets that either exploit information about the reference pool at a higher resolution than the sampling period of $\{\pmb{x}_t\}$, or that are driven by signals exogenous to the reference pool's characteristic time series $\{\pmb{x}_t\}$.

Stylized Fact 3 is however \emph{not} to say that, if two fund managers trade the same universe of assets, their funds do not provide diversification to the universe of assets they trade. Diversification doesn't arise solely as a result of \emph{what} a fund manager trades, but also, and perhaps more importantly, as a result of \emph{how} he/she trades. Stylized Fact 3 simply implies that, if a fund manager solely trades based on current and past daily returns on a certain universe of assets, and current and past daily factor values, then his daily returns shouldn't be regarded as any different from the reference universe of assets and factors. However, a different fund manager, trading the exact same universe of assets, but using more granular data or alternative data to drive his trading decision, will produce a fund that, as an asset, diversifies the universe of assets he/she trades. Crucially, because two fund managers trade the same universe of exchange-traded assets (e.g. stocks, bonds, futures etc.) does not mean that one cannot diversify the other! This is such an important distinction that we believe it warrants a fourth Stylized Fact. \\

\textbf{Stylized Fact 4:} \emph{A good quantitative measure of incremental diversification should allow for manager diversification. That is, two funds with identical constituents but different time-varying allocations driven by different (random) signals, should be able to diversify each other, despite their identical constituents.}\\

Asset returns can be scaled up and down through leverage. Thus, the scale of a time series of returns should intuitively bear no relevance on whether the corresponding asset incrementally diversifies a reference pool, or more generally is incrementally useful. This observation gives rise to our fifth and last Stylized Fact.\\

\textbf{Stylized Fact 5:} \emph{A good quantitative measure of incremental diversification should be scale-invariant. Equivalently,\footnote{To be precise, when stationarity holds, so that it makes sense to talk about standard deviation of processes rather than samples.} a good quantitative measure of incremental diversification should neither depend on the standard deviations of returns time series of assets in the reference pool, nor should it depend on the standard deviation of returns of the new asset.}\\

%
%
%
%
\subsection{\textbf{\textsc{Differential Mutual Information Timescale as a Measure of Incremental Diversification}}}
To motivate our measure of incremental diversification, we start with the simple case of two assets $\pi$ and A. \\

\noindent \textbf{Case 1: Two assets, i.i.d. Gaussians}

\noindent Returns of $\pi$ (resp. A) are assumed to be independent draws across time from the same distribution $r_\pi$ (resp. $r_A$), with mean $\mu$ and variance $\sigma^2$.  We assume that $(r_\pi, r_A)$ is jointly Gaussian and the correlation between $r_\pi$ and $r_A$ is $\rho$. As previously discussed the lower the correlation $\rho$, the higher the return per unit of risk we can obtain by combining $\pi$ and A. Moreover, we also discussed encouraging strategies that are driven by sources of information that are exogenous to the reference pool of assets, or at the very least information that are endogenous to the reference pool of assets but aren't fully captured at the resolution of the returns series. Either way, this implies that, if A incrementally diversifies $\pi$, then knowing $r_\pi$ shouldn't drastically reduce our uncertainty about $r_A$. Given that Gaussian random variables are fully determined by their first two moments, the uncertainty remaining in $r_A$ after knowing $r_\pi$ is well characterized by the conditional variance $\mathbb{V}\text{ar}\left(r_A \vert r_\pi \right)$, which in this case is easily shown to be 
\begin{align}
\label{form:01}
\mathbb{V}\text{ar}\left(r_A \vert r_\pi \right) = \sigma^2 \left( 1- \rho^2 \right).
\end{align}

When $\rho \geq 0$, $\mathbb{V}\text{ar}\left(r_{\pi^\prime} \right)$ (Equation (\ref{form:red_std})) increases with $\rho$, whereas $\mathbb{V}\text{ar}\left(r_A \vert r_\pi \right)$ decreases with $\rho$, and the two requirements for diversification, namely high expected return per unit of risk and unrelated returns, are consistent. However, when $\rho < 0$, $\mathbb{V}\text{ar}\left(r_{\pi^\prime} \right)$ still increases as a function of $\rho$ but $\mathbb{V}\text{ar}\left(r_A \vert r_\pi \right)$ now also increases with $\rho$. In other words,  when $\rho < 0$, the potential to increase the return per unit of risk by combining $\pi$ and A increases as the correlation decreases, but $\pi$ and A share more underlying driving factors. It might seem as though $\mathbb{V}\text{ar}\left(r_{\pi^\prime} \right)$ should be preferred over $\mathbb{V}\text{ar}\left(r_A \vert r_\pi \right)$ to measure diversification in such a case. However, the previous discussion on $\mathbb{V}\text{ar}\left(r_{\pi^\prime} \right)$ only holds when both $r_\pi$ and $r_A$ have the same expected return and variance. When either the variances or the expectations differ, we can no longer draw simple conclusions as to whether the return per unit of risk can be increased solely based on $\mathbb{V}\text{ar}\left(r_{\pi^\prime} \right)$. On the other hand the conditional variance, which in the general case reads 
\begin{align}
\label{form:02}
\mathbb{V}\text{ar}\left(r_A \vert r_\pi \right) = \mathbb{V}\text{ar}\left(r_A\right) \left( 1- \rho^2 \right),
\end{align} still provides valuable insight on shared information between A and $\pi$, namely that knowing $r_\pi$ never increases the uncertainty about $r_A$; the uncertainty is preserved when the two assets are decorrelated ($\rho = 0$), and decreases otherwise. A natural measure of the diversification A adds to $\pi$ is therefore
\begin{align}
\label{form:1}
D(A; \pi) = \mathbb{V}\text{ar}\left(r_A \vert r_\pi \right).
\end{align}
\begin{remark}
Expected returns being equal, Equation (\ref{form:1}) as a measure of incremental diversification penalizes equally new assets with $\rho \approx -1$ and new assets with $\rho \approx 1$, which could be perceived as a limitation, as the former can be used to construct portfolios with much higher return per unit of risk than the latter. This should however not pose a problem in practice, as two assets that have the same expected return are unlikely to have a correlation close to -1; this would be an arbitrage opportunity.\\
\end{remark}
\noindent \textbf{Case 2: n assets, i.i.d. Gaussians}

\noindent This intuitive measure of incremental diversification easily extends to the multi-assets case. If we consider a pool of $n$ assets and factors P, with corresponding returns and factor values drawn independently (across time) from a Gaussian random vector $\pmb{x}$ that is also assumed to be jointly Gaussian with $r_A$, Equation (\ref{form:1}) can be extended to quantify the incremental diversification A adds to the pool P as follows:
\begin{align}
\label{form:2}
D(A; P) = \mathbb{V}\text{ar}\left(r_A \vert \pmb{x} \right).
\end{align}
It immediately follows from Gaussian identities that
\begin{align}
\label{form:3}
D(A; P) = \frac{\text{det}\left(\mathbb{C}\text{ov}\left([\pmb{x}, r_A], [\pmb{x}, r_A]\right) \right)}{\text{det}\left( \mathbb{C}\text{ov} \left(\pmb{x}, \pmb{x}\right) \right)},
\end{align}
where we assume that $\pmb{x}$ is non-degenerate,\footnote{When $\pmb{x}$ is degenerate, it can be replaced by its largest non-degenerate subset without loss of generality.} from which we recover Equation (\ref{form:02}) in the two assets special case.\\

\noindent \textbf{Case 3: Beyond Gaussianity}

\noindent In the non-Gaussian case, the conditional variance $\mathbb{V}\text{ar}\left(r_A \vert \pmb{x} \right)$ might very well be a function of $\pmb{x}$, so that a more suitable candidate to quantify incremental diversification is obtained by taking the expectation with respect to $\pmb{x}$:
\begin{align}
\label{form:4}
D(A; P) = \mathbb{E}_{\pmb{x}} \left( \mathbb{V}\text{ar}\left(r_A \vert \pmb{x} \right) \right).
\end{align}

Expected conditional variance as measure of incremental diversification only captures the first two moments of the joint-distribution. This is sufficient for Gaussian distributions as they are fully determined by their first two moments. However, non-Gaussian distributions typical exhibit tail behaviors that are not captured by the first two moments; such tail behaviors play a role in our intuitive understanding of risk, and should therefore be embedded in our measure of incremental diversification. Another way to look at this is that, although knowing $\pmb{x}$ might not reduce the variance of $r_A$, if it does affect higher moments of $r_A$, then A should be regarded as more related to the reference pool than if $r_A$ and $\pmb{x}$ were independent. Our measure of incremental diversification should therefore be capable of differentiating statistical independence from decorrelation,\footnote{The former implying the latter.} which as per Proposition \ref{prop:1} conditional variance cannot.
\begin{proposition}
\label{prop:1}
Let $x$ and $y$ be two squared-integrable random variables.  Then
 \begin{align}
 \mathbb{E}_y\left(\mathbb{V}\text{ar} \left(x \vert y \right) \right) \leq \mathbb{V}\text{ar} \left(x\right),
 \end{align}
and the inequality is an equality if and only if $$\mathbb{E}(y \vert x) = \mathbb{E}(y) \text{ a.s.},$$
or equivalently, if and only if $\mathbb{C}\text{ov}\left(y, f(x)\right)=0$ for any $f$.
\end{proposition}
\begin{proof}Hint: This follows from the law of total variance and Hilbert's projection theorem.
\end{proof}

The canonical measure of the amount of information in a random variable with probability measure $\mathbb{P}$ and admitting pmf or pdf $p(x)$, is the notion of entropy (expressed in bits) defined as 
\begin{align}
h(x) = \mathbb{E}_\mathbb{P}\left[-\log_2 p(x)\right].
\end{align}
Unless stated otherwise, throughout the rest of this paper we assume $\mathbb{P}$ admits a pdf. When we need both cases, we will use the expression \emph{continuous entropy} or \emph{differential entropy} to emphasize that $\mathbb{P}$ admits a pdf, and \emph{discrete entropy} or \emph{Shanon entropy} when $\mathbb{P}$ admits a pmf, in which case we will use the notation $H$ instead of $h$. 

A related notion is that of conditional entropy, which can be defined as
\begin{align}
h(y \vert x) = h(x, y) - h(x),
\end{align}
when $h(x, y)$ and $h(x)$ exist, and that measures the amount of information contained in random variable $y$ that is not already contained in random variable $x$. In the multi-assets Gaussian case, this measure of incremental diversification reads
\begin{align}
h(r_A \vert \pmb{x}) &= \frac{1}{2} \log_2 \frac{\text{det}\left( 2\pi e \mathbb{C}\text{ov}([\pmb{x}, r_A], [\pmb{x}, r_A]) \right)}{\text{det} \left(2\pi e  \mathbb{C}\text{ov}(\pmb{x}, \pmb{x}) \right)} \nonumber \\
&= \frac{1}{2} \log_2 \mathbb{V}\text{ar}\left(r_A \vert \pmb{x} \right) + \frac{1}{2} \log_2 2\pi e.
\end{align}
In other words, in the Gaussian case, conditional entropy and conditional variance are equivalent measures of incremental diversification as one is fully determined by the other and is an increasing function of the other. In general however, conditional entropy is a more general measure of incremental diversification in the following sense (Theorem 8.6.1 in \cite{cover}).
\begin{proposition}
\label{prop:2}
Let $x$ and $y$ be two random variables having finite entropies $h(x)$ and $h(y)$.  Then
 \begin{align}
h \left(y \vert x \right) \leq h \left(y \right),
 \end{align}
and the inequality is an equality if and only if $x$ and $y$ are independent.
\end{proposition}
Unlike expected conditional variance that cannot differentiate decorrelation from independence, conditional entropy, as a measure of incremental diversification, is informative about the full distribution tails, and is maximized (for a given entropy $h \left(r_A \right)$) when $r_A$ is independent from $\pmb{x}$, which we recall implies, but is not equivalent to, $\mathbb{C}\text{ov}\left(r_A, f(x)\right)=0$ for any $f$.\\

\noindent \textbf{Case 4: Beyond temporal independence}

\noindent Conditional entropy as a measure of incremental diversification satisfies both Stylized Facts 1 and 2. To see why, we note that, in the Gaussian case, $\mathbb{V}\text{ar}\left(r_A \vert \pmb{x} \right)$ is also the variance of the residual return of the best replicating portfolio, and using Equation (\ref{eq:cov_brp}) we obtain
\begin{align}
h(r_A \vert \pmb{x}) &=  \frac{1}{2} \log_2 \left[1- \mathbb{C}\text{orr}\left(r_A, r_A^* \right)  \sqrt{\frac{\mathbb{V}\text{ar}\left( r_A^* \right)}{\mathbb{V}\text{ar}\left( r_A \right)}} \right] \nonumber \\  &+h\left( r_A \right)  
\end{align}
which confirms that $h(r_A \vert \pmb{x})$ decreases with the correlation between A and its best replicating portfolio (Stylized Fact 1), and is lowest when $r_A$ is a linear combination of $\pmb{x}$ ---i.e. where $r_A^*=r_A$ (Stylized Fact 2). In the general case, using the fact that Gaussian random variables are maximum-entropy among distributions having the same covariance matrix to upper bound $h(r_A, \pmb{x})$, it follows that 
\begin{align}
\label{eq:entropy_bound}
h(r_A \vert \pmb{x}) & \leq  \frac{1}{2} \log_2 \left[1- \mathbb{C}\text{orr}\left(r_A, r_A^* \right)  \sqrt{\frac{\mathbb{V}\text{ar}\left( r_A^* \right)}{\mathbb{V}\text{ar}\left( r_A \right)}} \right] \nonumber \\ &+  h\left( r_A \right)+ h\left(\pmb{\hat{x}} \right) -h\left(\pmb{x} \right)
\end{align}
where $\pmb{\hat{x}}$ is a Gaussian distribution with equal covariance matrix to that of $\pmb{x}$. Hence, $h(r_A \vert \pmb{x})$ can be made arbitrarily small by increasing $\mathbb{C}\text{ov}\left(r_A, r_A^* \right)$ or equivalently by jointly increasing $\mathbb{C}\text{orr}\left(r_A, r_A^* \right)$ and $\sqrt{\frac{\mathbb{V}\text{ar}\left( r_A^* \right)}{\mathbb{V}\text{ar}\left( r_A \right)}}$, which is consistent with Stylized Facts 1 and 2.

However, conditional entropy as a measure of incremental diversification does not satisfy Stylized Fact 3, as we have been ignoring the temporal aspect of our time series because of our i.i.d. assumption (across time). To see why, we consider $y_t = f(\pmb{x}_{t-i}), ~ i > 0$, and note that, under our memoryless assumption on the reference pool characteristic time series  $\pmb{x}_t$, $y_t$ is independent from $\pmb{x}_t$ and consequently has the highest conditional entropy for a given $h \left(y_t \right)$. The main issue here is that, as a measure of incremental diversification, conditional entropy does not capture similarities across time. Independence of returns corresponding to the same time period, $\pmb{x}_t$ and $y_t$, should not be the ideal diversification scenario, \emph{independence of the underlying stochastic processes $\{\pmb{x}_t\}$ and $\{y_t\}$}  should be.

The notion of entropy of random variables is extended to discrete-time stochastic processes by the notion of \emph{entropy rate} which is defined as 
\begin{align}
h(\{\pmb{x_t} \}) = \lim_{T \to \infty} \frac{1}{T} h(\pmb{x}_1, ..., \pmb{x}_T),
\end{align} when the limit exists. The notion of conditional entropy is then extended to define conditional entropy rate as 
\begin{align}
h\left(\{ y_t \} \vert \{\pmb{x_t} \}\right) = h\left(\{ y_t, \pmb{x_t}\}\right) - h\left(\{\pmb{x_t} \}\right)
\end{align}
when $h\left(\{ y_t, \pmb{x_t}\} \right)$  and $h\left(\{\pmb{x_t} \} \right)$ exist.

Similarly to the random variable case, the conditional entropy rate measures the amount of information per unit of time contained in stochastic process $\{ y_t \}$ that is not already reflected in $\{\pmb{x_t} \}$. Moreover, the conditional entropy rate fully captures dependencies between time series across time as stated in the following proposition, which follows from Proposition \ref{prop:2}.

\begin{proposition}
 \label{prop:cond_entr_rate}
Let $\{\pmb{x_t}\}$ and $\{y_t\}$ be two discrete-time stationary stochastic processes having finite entropy rates $h\left(\{\pmb{x_t} \}\right)$ and $h\left(\{y_t \}\right)$. Then
 \begin{align}
 \label{eq:cer}
h \left(\{y_t\} \vert \{\pmb{x_t}\} \right) \leq h \left(\{y_t\} \right).
 \end{align}

If we further assume that $\{\pmb{x_t}\}$ and $\{y_t\}$ have bounded memory in the sense that there exists $\alpha,  \beta > 0$ and $m \geq 1$ such that
$$\forall k, ~~ \alpha k I(m) \leq I(km) \leq \beta k I(m),$$
where $$I(n) = h \left(y_1, \dots, y_n\right) - h \left(y_1, \dots, y_n \vert \pmb{x_1}, \dots, \pmb{x_n} \right),$$ then the inequality in Equation (\ref{eq:cer}) is an equality if and only if $\{\pmb{x_t}\}$ and $\{y_t\}$ are independent. 
\end{proposition}
\begin{proof}
The inequality in Equation (\ref{eq:cer}) is a direct consequence of Proposition \ref{prop:2}. Moreover if $\{\pmb{x_t}\}$ and $\{y_t\}$ are independent, then it is easy to see that $h \left(\{y_t\} \vert \{\pmb{x_t}\} \right) = h \left(\{y_t\} \right)$. 

To prove the reverse, we note that 
\begin{align}
h \left(\{y_t\} \right)-h \left(\{y_t\} \vert \{\pmb{x_t}\} \right) &= \underset{k \to +\infty}{\lim} ~~ \frac{I(km)}{km} \nonumber \\
&\geq \alpha \frac{I(m)}{m} \nonumber \\
& \geq 0. \nonumber
\end{align}
Hence, if $h \left(\{y_t\} \vert \{\pmb{x_t}\} \right) = h \left(\{y_t\} \right)$ then $I(m)=0$. As $I(km) \leq \beta k I(m)$ for every $k$, it follows that $I(km)=0$ for every $k$, which implies that $(y_1, \dots, y_{n})$ and $(\pmb{x_1}, \dots, \pmb{x_{n}})$ are independent for every $n$ or, equivalently, $\{\pmb{x_t}\}$ and $\{y_t\}$ are independent.
\end{proof}

\begin{remark}
By definition, $h \left(\{y_t\} \vert \{\pmb{x_t}\} \right) = h \left(\{y_t\} \right)$ if and only if the mutual information between $\left(y_1, \dots, y_n\right)$ and $\left(\pmb{x_1}, \dots, \pmb{x_n}\right)$ grows too slowly with $n$, specifically in $o(n)$. Such slow growth can only be attributed to excessive cross-sectional and/or temporal coupling as $n$ increases. Moreover, it is easy to see that when $\{y_t, \pmb{x_t}\}$ has no memory, then $h \left(\{y_t\} \vert \{\pmb{x_t}\} \right) = h \left(\{y_t\} \right)$ if and only if $\{\pmb{x_t}\}$ and $\{y_t\}$ are independent. Hence, when $h \left(\{y_t\} \vert \{\pmb{x_t}\} \right) = h \left(\{y_t\} \right)$ and $\{\pmb{x_t}\}$ and $\{y_t\}$ are not independent, the slow growth in the mutual information between $\left(y_1, \dots, y_n\right)$ and $\left(\pmb{x_1}, \dots, \pmb{x_n}\right)$ can only be attributed to excessive temporal coupling/memory. By placing limitations on the memory of $\{y_t, \pmb{x_t}\}$, we are able to ensure that the conditional entropy rate $h\left(\{y_t\} \vert \{\pmb{x_t}\} \right)$ is maximized only in the event of independence between input processes $\{\pmb{x_t}\}$ and $\{y_t\}$. We stress that assuming that financial time series do not have excessive memory is consistent with empirical evidence, so that, in what follows, we might omit the bounded memory condition as a requirement for Proposition \ref{prop:cond_entr_rate} to hold.
\end{remark}
The following corollary is a direct consequence of Proposition \ref{prop:cond_entr_rate}.
\begin{corollary}
\label{corr:best_case}
Let $\{\pmb{x_t}\}$ and $\{y_t\}$ be two discrete-time stationary stochastic processes such that $$h \left(\{y_t\} \vert \{\pmb{x_t}\} \right) = h \left(\{y_t\} \right),$$ then for any function $f$, time $t$, memory $m \geq 0$, and lag $p \geq 0$. The random variables $y_{t+p}$ and $f\left(\pmb{x}_{t}, \dots, \pmb{x}_{t-m}\right)$ are independent.
\end{corollary}
Another perspective on Corollary \ref{corr:best_case} is that, using \emph{conditional entropy rate} as measure of incremental diversification, the best case scenario corresponds to assets whose current and future returns are independent from (and therefore cannot be predicted using) past returns of assets and values of factors in the reference pool, irrespective of how far back we look, in which case it would indeed be impossible to replicate the stream of returns of the new asset using the reference pool of assets and factors (at the same resolution as the shared sampling frequency).

The flip side of the foregoing observation is provided in Proposition \ref{corr:worst_case}.
\begin{proposition}
\label{corr:worst_case}
Let $\{\pmb{x_t}\}$ and $\{y_t\}$ be two discrete-time stochastic processes that admit finite entropy rates. If there exist a function $f$, and $m > 0$ such that \begin{equation}\label{eq:determin}\forall  t>m, ~ y_t = f\left(\pmb{x}_t, \dots, \pmb{x}_{t-m}\right),\end{equation} then $$h \left(\{y_t\} \vert \{\pmb{x_t}\} \right) = -\infty.$$
\end{proposition}
\begin{proof} This proposition is a direct consequence of $$h\left(y_{m+1}, ..., y_{m+n} \vert \pmb{x}_1, ..., \pmb{x}_{m+n}= * \right) = -\infty,$$ which follows from Equation (\ref{eq:determin}).\end{proof}
Proposition \ref{corr:worst_case} shows that conditional entropy rate, as a measure of incremental diversification an asset A adds to a reference pool of assets and factors P, satisfies both Stylized Facts 2 and 3. 

Let's study the consistency of conditional entropy rate with Stylized Fact 1. Entropy rates do not always exist in general, nor are there generic analytic formulae to compute them when they exist. For stationary stochastic processes however, entropy rates are guaranteed to exist. 

The notion of best replicating portfolio in the mean-squared sense (Definition \ref{def:brp}) is easily extended to the non-i.i.d. case as the portfolio that has dynamic allocation that is solution to the Optimization Problem
\begin{align}
\label{prob:brp2}
 \omega^*_t := \underset{\omega}{\text{argmin}} ~~ \mathbb{V}\text{ar}\left( y_t - \omega^T \pmb{x}_t \right),
\end{align}
whose solution is found to read
\begin{align}
\label{eq:alloc_is_const}
\omega^*_t = \mathbb{C}\text{ov}\left(\pmb{x}_t, \pmb{x}_t \right)^{-1} \mathbb{C}\text{ov}\left(\pmb{x}_t, y_t \right),
\end{align}
and the covariance between the return of the new asset $y_t$ and the return of the best replicating portfolio $y_t^*$ reads
\begin{align}
\label{eq:cov_brp2}
\mathbb{C}\text{ov}\left(y_t, y_t^* \right) & = \mathbb{V}\text{ar}\left( y_t^* \right)  \\
& = \mathbb{C}\text{ov}\left(y_t, \pmb{x}_t \right) \mathbb{C}\text{ov}\left(\pmb{x}_t, \pmb{x}_t \right)^{-1} \mathbb{C}\text{ov}\left(\pmb{x}_t, y_t \right). \nonumber
\end{align}
\begin{remark}\label{rmk:stat_implies_static}
When $\{y_t, \pmb{x}_t\}$ is jointly stationary, it is easy to see from Equations (\ref{eq:alloc_is_const}) and (\ref{eq:cov_brp2}) that the best replicating portfolio is in fact a static portfolio (i.e. its target allocation is constant over time), $\{y^*_t\}$ is stationary,\footnote{We recall that $y^*_t:=  \pmb{x}_t^T\omega^*_t  + \left( 1- 1^T\omega^*_t \right)r_f$.} and the correlation between the new asset and its best replicating portfolio is constant over time. 
\end{remark}
Using the following property of stationary processes $$h\left(\{\pmb{x}_t, y_t\}\right) \leq h\left(\pmb{x}_t, y_t\right),$$ and using the maximum-entropy property of Gaussian random variables to upper bound $h\left(\pmb{x}_t, y_t\right)$, we obtain a generalization of Equation (\ref{eq:entropy_bound}) to the non-i.i.d. case:
\begin{align}
\label{eq:entropy_bound2}
h\left(\{y_t\} \vert \{\pmb{x}_t \} \right) & \leq  \frac{1}{2} \log_2 \left[1- \mathbb{C}\text{orr}\left(y_t, y_t^* \right)  \sqrt{\frac{\mathbb{V}\text{ar}\left( y_t^* \right)}{\mathbb{V}\text{ar}\left( y_t \right)}} \right] \nonumber \\ &+  h\left( y_t \right)+ h\left(\pmb{\hat{x}} \right) -h\left(\pmb{x}_t \right),
\end{align}
where $\pmb{\hat{x}}$ a Gaussian with the same covariance matrix as $\pmb{x}_t $. This confirms that, in the stationary case, using conditional differential entropy rate as measure of incremental diversification is consistent with Stylized Fact 1 in that the correlation between an asset and its best replicating portfolio acts as a cap on the amount of incremental diversification the new asset A provides to the reference pool.

Finally, conditional differential entropy rate also satisfies Stylized Fact 4. To see why, we consider two funds $\pi$ and $\pi^\prime$ whose constituents are the same and have returns $\pmb{x}_t$. Let's denote $$\omega^\pi_t\left(s_t^\pi\right) \text{ and } \omega^{\pi^\prime}_t\left(s_t^{\pi^\prime}\right)$$ the funds' respective allocations, each driven by a different time series of signals $\{s_t^\pi\}$ or $\{s_t^{\pi^\prime}\}$. The funds' respective time series of returns read
$$\{r_t^\pi\} := \{\pmb{x}_t^T\omega^\pi_t\} \text{  and  } \{r_t^{\pi^\prime}\} := \{\pmb{x}_t^T\omega^{\pi^\prime}_t\}.$$ It is clear that the amount of diversification $\pi^\prime$ adds to $\pi$, $h\left(\{ r_t^{\pi^\prime} \} \vert \{ r_t^\pi\}  \right)$ depends on the joint law of $\{s_t^{\pi^\prime}, s_t^\pi\}$, and is certainly \emph{not} always $-\infty$. Thus, the two funds can diversify each other; the lower the stochastic similarity between their signal processes, the more they can diversify each other. 

\begin{remark}
We stress that, to ensure manager diversification, conditional differential entropy rate does not require knowing what underlying assets or asset classes the fund manager is trading, what his/her trading thesis is, or what types of data (alternative or otherwise) drive his/her trading decisions. Our approach is solely based on the returns of his/her fund.
\end{remark}

Conditional differential entropy rate, however, is scale-sensitive, and consequently does not satisfy Stylized Fact 5.\footnote{In fact, no candidate measure of incremental diversification considered thus far, including expected conditional variance, satisfies Stylized Fact 5.} Indeed, for any scalar $\alpha$ and vector $\pmb{\beta} \in \mathbb{R}^n$, $$h\left(\{\alpha y_t\} \vert  \{\pmb{\beta} \odot \pmb{x}_t\}\right) = h\left(\{y_t\} \vert  \{ \pmb{x}_t\}\right)  + \log_2 \vert \alpha \vert,$$ where $\odot$ denotes the Hadamard product. Another limitation of its use as incremental diversification is that it can be negative. Both drawbacks are related to the difference between differential and Shanon/discrete entropies. Strictly speaking, unlike their discrete counterparts that quantify information in absolute terms, differential entropy and differential entropy rate only quantify information in relative terms. For instance, $h\left(\{y_t\} \vert  \{ \pmb{x}_t\} \right)$ is a relative measure of the amount of information per unit of time contained in $\{y_t\}$ that is not already contained in $\{ \pmb{x}_t\}$, but the \emph{differential mutual information rate}
\begin{align}
\label{eq:mutual_info}
I\left(\{y_t\};\{ \pmb{x}_t\}\right) :&= h\left(\{y_t\}\right)  - h\left(\{y_t\} \vert  \{ \pmb{x}_t\}\right)
\end{align}
is an absolute measure of the amount of information per unit of time contained in $\{y_t\}$ that is also contained in $\{\pmb{x}_t\}$. It is always non-negative and invariant by any smooth change of variable (a.k.a. diffeomorphism), including linear rescaling. A large differential mutual information rate corresponds to higher similarity between $\{y_t\}$ and $\{\pmb{x}_t\}$. Whence, the \emph{differential mutual information timescale}, defined as the inverse of the differential mutual information rate, $$\frac{1}{I\left(\{y_t\};\{ \pmb{x}_t\}\right)},$$ is a candidate measure of incremental diversification.

\begin{definition}
\label{def:id}
Let P be a reference pool of assets and factors, whose time series of returns and factor values we denote $\{\pmb{x}_t\}$. Let A be an asset not in P, whose time series of returns we denote $\{y_t\}$. Let us further assume that entropy rates of $\{y_t\}$ and $\{\pmb{x}_t\}$ exist, and are possibly infinite. We define measure of incremental diversification the asset A adds to the reference pool P the differential mutual information timescale between $\{y_t\}$ and $\{\pmb{x}_t\}$, namely
\begin{align}
\mathbb{ID}\left(A; P\right) := \frac{1}{I\left(\{y_t\};\{ \pmb{x}_t\}\right)},
\end{align}
where we use the convention $1/0^+ = +\infty$ and $1/{+\infty} = 0$.
\end{definition}
$\mathbb{ID}\left(A; P\right)$ represents the amount of time required to see $1$bit of mutual/shared information between returns of the new asset A and returns of the reference pool P. Intuitively, it is always non-negative, takes its lowest value $0$ when A is fully determined by P (i.e. knowing returns and factor values of P is sufficient to know returns of A), and takes its highest value $+\infty$ when values of returns of A can never be inferred from P, no matter how long a history of returns and factor values of P we have. Moreover, differential mutual information timescale is invariant by rescaling of asset returns through leverage and any other smooth change of representation. This is formalized in the following proposition.

\begin{theorem}
\label{th:fund}
For any reference pool P and new asset A satisfying the conditions of Definition \ref{def:id}, 

(a) $\mathbb{ID}\left(A; P\right) \geq 0,$

(b) When  $\vert h\left(\{y_t\}\right) \vert  < \infty,$ $\mathbb{ID}\left(A; P\right) = 0$ if and only if $h\left(\{y_t\} \vert  \{ \pmb{x}_t\}\right) = -\infty.$

(c) $\mathbb{ID}\left(A; P\right) = +\infty$ if and only if $\{y_t\}$ and $\{\pmb{x}_t\}$  are independent.

(d) For any continuously differentiable bijections $f: \mathbb{R} \to \mathbb{R}$ and $g: \mathbb{R}^n \to \mathbb{R}^n$, $$\mathbb{ID}\left(A; P\right) = \frac{1}{I\left(\{f(y_t)\};\{ g(\pmb{x}_t)\}\right)}.$$
\end{theorem}
\begin{proof} Let us generically denote $p(u)$ the probability density function of random variable $u$. We note that, $$I\left(\{y_t\};\{ \pmb{x}_t\}\right) = \underset{T \to +\infty}{\lim} \frac{1}{T} D_{KL}\left( \mathcal{J}_T \vert \vert \mathcal{I}_T \right)$$ where $$ \mathcal{I}_T = p(y_1, \dots, y_T)p(\pmb{x}_1, \dots, \pmb{x}_T)$$ and $$\mathcal{J}_T = p(y_1, \pmb{x}_1, \dots, y_T, \pmb{x}_T),$$ and  $D_{KL}$ denotes the Kullback-Leibler divergence \cite{cover}. (a) Follows from the non-negativity of KL-divergence (also from Proposition \ref{prop:cond_entr_rate}), and (d) follows from the invariance of the KL-divergence by smooth bijections. As for (b), $\mathbb{ID}\left(A; P\right) = 0$ when $I\left(\{y_t\};\{ \pmb{x}_t\}\right) = +\infty$, which given $\vert h\left(\{y_t\}\right) \vert  < \infty$, is equivalent to $h\left(\{y_t\} \vert  \{ \pmb{x}_t\}\right) = -\infty.$ With regard to (c), $\mathbb{ID}\left(A; P\right) = + \infty$ if and only if  $I\left(\{y_t\};\{ \pmb{x}_t\}\right)$ goes to $0$ (which always happens from above), or equivalently $h(\{y_t\}) = h\left(\{y_t\} \vert  \{ \pmb{x}_t\}\right)$, and we use Proposition \ref{prop:cond_entr_rate} to conclude.
\end{proof}

\begin{proposition} The measure of incremental diversification $(A, P) \to \mathbb{ID}\left(A; P\right)$ satisfies Stylized Facts 1-5.
\end{proposition}
\begin{proof} As previously discussed $(A, P) \to h\left(\{y_t\} \vert  \{ \pmb{x}_t\}\right)$ satisfies Stylized Fact 1 (see Equation (\ref{eq:entropy_bound2})). Moreover, for every finite $h\left(\{y_t\} \right)$, the function $$h\left(\{y_t\} \vert  \{ \pmb{x}_t\}\right) \to  \mathbb{ID}\left(A; P\right) = \frac{1}{h\left(\{y_t\} \right)-h\left(\{y_t\} \vert  \{ \pmb{x}_t\}\right)}$$ is a strictly increasing function, whence $\mathbb{ID}\left(A; P\right)$ also satisfies Stylized Fact 1. This also implies that $\mathbb{ID}\left(A; P\right)$ is lowest if and only if $h\left(\{y_t\} \vert  \{ \pmb{x}_t\}\right)$ is lowest. Hence, the fact that conditional entropy rate satisfies Stylized Facts 2 and 3 extends to $\mathbb{ID}\left(A; P\right)$. In a similar reasoning, the fact that conditional entropy rate allows for manager diversification (Stylized Fact 4) extends to differential mutual information timescale. Finally, consistency with Stylized Fact 5 is a direct consequence of Theorem \ref{th:fund}-(d).
\end{proof}
%
%
\subsection{\textbf{\textsc{Differential Entropy Rates From Discrete Entropy Rates}}}
\label{sct:discrete_cont_entropy}
In the previous Section, we discussed two major differences between differential (continuous) and Shanon (discrete) entropies, namely that, unlike discrete entropy, differential entropy is neither non-negative nor invariant by rescaling. Another oddity of the differential (continuous) entropy of a continuous random variable is that it is not obtained as the limit of the discrete entropy of a discretization thereof, as the discretization mesh size/volume goes to zero, as is typical of discrete-to-continuous transitions. Nonetheless, there is a link between the differential entropy of a random variable and the discrete entropy of a discretized version thereof, in the limit when the discretization error is arbitrarily small. This link, which we recall below, forms the foundation of the model-free estimation of incremental diversification that we develop in Section \ref{sct:mofree}.
\begin{theorem}
\label{theo:discrete_cont_entropy}
Let $\pmb{z}$ be a  random variable taking values in $\mathbb{R}^n$ and that admits differential entropy $h(\pmb{z})$, and let $\pmb{z}^m$ be the random vector taking values in $\mathbb{Z}^n$ and satisfying $$\pmb{z}^m[i] = k, ~~ m \in \mathbb{R}, ~~ k \in \mathbb{Z}$$ if and only if $$\frac{k}{2^m} \leq \pmb{z}[i]  < \frac{k+1}{2^m},$$
and that admits discrete entropy $H(\pmb{z}^m)$.
Denoting $p$ the probability density function of $\pmb{z}$, if the following properties are met,

(a) $p$ is continuous and bounded,

(b) $\int | p(\pmb{z}) \log p(\pmb{z}) | d\pmb{z} < \infty$,

(c) $H(\pmb{z}^0) < \infty$, \\
then, 
\begin{align}
h(\pmb{z}) = \underset{m \to +\infty}{\lim} H(\pmb{z}^m) - mn.
\end{align}
\end{theorem}
\begin{proof}
This is the multivariate extension of Theorem 1.3.1 in \cite{ihara}. The proof is almost identical, except that the intervals of size $2^{-m}$ become hypercubes with volume $2^{-mn}$. A similar result is provided by Theorem 8.3.1 of \cite{cover}, where condition (a) is replaced by Riemann integrability, condition (b) is replaced by $h(\pmb{z}) < \infty$, and $\Delta = 2^{-mn}$.
\end{proof}

\begin{corollary}
\label{cor:approx_diff_ent_disc}
Let $\{y_t\}$ be a real-valued discrete-time stochastic process, and let $\{\pmb{x}_t\}$ be an $\mathbb{R}^n$-valued discrete-time stochastic process such that every marginal of the joint process $\{y_t, \pmb{x}_t\}$ admits a probability density function that satisfies conditions (a), (b) and (c) above. Let $\{y_t^m, \pmb{x}_t^m\}$ be the discretized process as per Theorem \ref{theo:discrete_cont_entropy}. Then
\begin{align}
\label{eq:approx_diff_ent_disc}
h\left(\{y_t\}  \right) = \underset{m \to +\infty}{\lim} H\left(\{ y_t^m \}\right) -m,
\end{align}
\begin{align}
\label{eq:approx_diff_ent_disc_multi}
h\left(\{\pmb{x}_t\}  \right) = \underset{m \to +\infty}{\lim} H\left(\{ \pmb{x}_t^m \}\right) -mn,
\end{align}
\begin{align}
\label{eq:approx_cond_diff_ent_disc}
h\left(\{y_t\} | \{\pmb{x}_t \} \right) &= \underset{m  \to +\infty}{\lim}  H\left(\{ y_t^m \} \vert \{ \pmb{x}_t^m \} \right) -m,
\end{align}
and
\begin{align}
I\left(\{y_t\}  | \{\pmb{x}_t \}  \right) = \underset{m \to +\infty}{\lim} H\left(\{ y_t^m \}\right) - H\left(\{ y_t^m \} \vert \{ \pmb{x}_t^m \} \right).
\end{align}
\end{corollary}
In other words, if we can estimate discrete entropy rates, we can estimate differential entropy rates and differential mutual information rates, and consequently incremental diversification. Interestingly, unlike the differential entropy rate, the differential mutual information rate is indeed obtained as the limit of the discrete/Shanon mutual information rate between the discretization of $\{y_t\}$ and that of $\{\pmb{x}_t \}$, namely $H\left(\{ y_t^m \}\right) - H\left(\{ y_t^m \} \vert \{ \pmb{x}_t^m \} \right)$, as the discretization error $\Delta = 2^{-mn}$ goes to $0$.

\subsection{\textbf{\textsc{Estimating Incremental Diversification}}}
Differential mutual information timescale does not always exist. When the process $\{y_t, \pmb{x}_t\}$ is stationary, the differential mutual information timescale $1/I\left(\{y_t\} \vert \{\pmb{x}_t\} \right)$ is guaranteed to exist. Thus, throughout the rest of this paper, we assume that $\{y_t, \pmb{x}_t\}$ is jointly (strongly) stationary and (strongly) ergodic \cite{hamilton1994time}. We note however that these assumptions are not restrictive as they are impossible to invalidate experimentally \emph{with a finite sample}.\footnote{Most, if not all, statistical tests of stationarity make additional assumptions, such as the fact that the process is an AR(p), which the popular unit-root tests (e.g. Dickey-Fuller, Phillips-Perron etc.) rely on. These tests make the implicit assumption that the sample to be tested spans over a time range that is longer than the memory of the underlying process. Consequently, when one of these stationarity tests typically fails, one of multiple assumptions could be to blame: the memory of the underlying process could be much longer than the sample size or said differently the sample could be too short to characterize the underlying process, or the diffusion model (e.g. AR(p)) could be ill-suited to the underlying process, or the process could be non-stationary. The only way to conclude non-stationarity from a unit-root test failure, is to treat the other hypotheses as axioms, in which case the test is no longer a stationarity test (i.e. one that has non-stationarity as sole null hypothesis), but one where the null hypothesis is really the combination of non-stationarity, a specific diffusion model, and assuming the memory of the underlying process isn't longer than the range of the sample tested. The skeptic reader might find it useful to simulate a mean-zero Gaussian process with Squared Exponential covariance function $\gamma(u, v) = \exp\left(-(u-v)^2/10 \right)$ on a fine grid on $[0, 1]$, and notice that draws will most likely fail any stationarity test, irrespective of mesh size of the grid, and therefore irrespective of the sample size, even though they were simulated from a stationary process. The skeptic reader would also note that, if the same test is run using samples simulated on $[0, 100]$, they will likely pass traditional stationarity tests.} 

In order to estimate differential mutual information timescale $$\frac{1}{I\left(\{y_t\} \vert \{\pmb{x}_t\} \right)}= \frac{1}{h(\{y_t\}) + h(\{\pmb{x}_t\}) -h(\{y_t, \pmb{x}_t\})},$$ it is sufficient to be able to estimate the entropy rate of any vector-valued discrete-time stationary ergodic stochastic process $h(\{\pmb{z}_t\})$ from a single sample path $(\pmb{z}_1, \dots, \pmb{z}_T)$; this is what we focus our discussion on.

Considering that differential mutual information rate is invariant by rescaling, we assume coordinate processes of $\{\pmb{z}_t\}$ all have the same variance $\frac{2}{\pi e}$, and we normalize sample path $(\pmb{z}_1, \dots, \pmb{z}_T)$ accordingly, if needed. We make this specific choice of variance to ease estimation interpretation and debugging. Indeed, under this constraint, $h(\{\pmb{z}_t\}) \leq n$ and the equality holds if and only if i) $\{\pmb{z}_t\}$ is memoryless, that is its samples are independent, ii) its coordinate processes are independent, and iii) it is a Gaussian process. An estimated entropy rate higher than $n$ is an indication of an implementation bug. An estimated entropy rate strictly lower than $n$ is an indication of either temporal dependency (memory), cross-dependency, or non-Gaussianity.

In the following, we discuss and compare three estimation approaches, namely model-free estimation in Section \ref{sct:mofree}, nonparametric estimation in Section \ref{sct:mobased} and maximum-entropy estimation in Section \ref{sct:maxent}. The model-free approach places no assumption on the diffusion of $\{\pmb{z}_t\}$. The nonparametric approach does not assume a parametric model for the diffusion of $\{\pmb{z}_t\}$, but instead assumes it is a Gaussian process. Finally the maximum-entropy approach adopts the modeling principle of the same name, which stipulates that, among all models that are consistent with empirical evidence, one should always choose the one that is the most uncertain/ignorant about everything but what has been observed.

\subsubsection{\textsc{Model-Free Estimation}}
\label{sct:mofree}
We recall from Section \ref{sct:discrete_cont_entropy} that differential entropy rates can be estimated by first discretizing the input process, then estimating the discrete entropy rate of the discretized process, and finally adjusting for discretization precision. We also note that, when the input process is strongly ergodic and stationary, so is its discretization.

The notion of complexity of a sequence of characters emitted by a stochastic source is tightly coupled with the discrete entropy rate of the emitting source. Of particular interest is the link between the \emph{Lempel-Ziv complexity} introduced in \cite{lz76c}, and for which we provide a Python implementation in Listing 1, and the discrete entropy rate of a stationary ergodic process \cite{lz78}, which we recall below.

\begin{theorem}
\label{th:lzcomp_sh}
Let $\{\pmb{a}_t\}$ be a discrete-time stationary ergodic stochastic process taking values in a countable set $\mathcal{A}$, and that has discrete entropy rate $H\left(\{\pmb{a}_t\}\right)$. If we denote $c(T)$ the Lempel-Ziv complexity (as per Listing 1) of a sample path of length $T$ of this process, then 
\begin{equation}
H\left(\{\pmb{a}_t\}\right) = \underset{T \to \infty}{\lim} \frac{c(T) \log_2 T}{T} \text{ a.s.}
\end{equation}
\end{theorem}

\begin{corollary}
\label{cor:lzcomp_sh}
Let $\{\pmb{a}_t\}$ be a discrete-time stationary ergodic stochastic process taking values in a countable set $\mathcal{A}$, and such that  $0 < H\left(\pmb{a}_t\right) < \infty$. Let us consider a path $(\hat{\pmb{a}}_1, \dots, \hat{\pmb{a}}_T)$ with Lempel-Ziv complexity $c(T)$. Let $\hat{\pmb{a}}_t^i$ with $1 \leq t \leq T$ and $1 \leq i \leq k$ be $kT$ independent draws from $\{\hat{\pmb{a}}_1, \dots, \hat{\pmb{a}}_T\}$ sampled uniformly at random with replacement, and let us denote $c_i(T)$ the Lempel-Ziv complexity of the sequence $(\hat{\pmb{a}}_1^i, \dots, \hat{\pmb{a}}_T^i)$. Then for every $k>1$,

\begin{equation}
\frac{H\left(\{\pmb{a}_t\}\right)}{H\left(\pmb{a}_t\right)} = \underset{T \to \infty}{\lim} \frac{c(T)}{\frac{1}{k}\sum_{i=1}^k c_i(T)} \text{ a.s.}
\end{equation}
\end{corollary}
\begin{proof}Hint: For every $i$, $H\left(\{\pmb{a}_t^i\}\right) = H\left(\pmb{a}_t^i\right) = H\left(\pmb{a}_t\right) = \underset{T \to \infty}{\lim} \frac{c_i(T) \log_2 T}{T} \text{ a.s.}$
\end{proof}

In summary, $\frac{c(T) \log_2 T}{T}$ is a consistent estimator of discrete entropy rate. However, in practice, we find the convergence of Theorem \ref{th:lzcomp_sh} to be slower than that of Corollary \ref{cor:lzcomp_sh}. Whence, we choose instead to estimate the discrete entropy rate given a sequence of characters as 
\begin{align}\label{eq:mofest} H\left(\{\pmb{a}_t\}\right) \approx \hat{H}\left(\pmb{a}_t\right) \frac{c(T)}{\frac{1}{k}\sum_{i=1}^k c_i(T)},\end{align}
where $$\hat{H}\left(\pmb{a}_t\right) = - \sum_i p_i \log_2 p_i$$ and terms $p_i$ represent frequencies of occurrence of distinct symbols in $(\hat{\pmb{a}}_1, \dots, \hat{\pmb{a}}_T)$. The estimate of Equation (\ref{eq:mofest}) is a consistent estimate of discrete entropy rate for every $k$, but larger $k$ can help reduce estimation variance. Finally, using the results of Section \ref{sct:discrete_cont_entropy}, we may estimate the differential entropy rate of any vector-valued stationary ergodic process, and consequently we may estimate incremental diversification. 

We stress that this approach does not require placing any assumption on the diffusion of $\{y_t, \pmb{x}_t\}$ other than ergodicity and stationarity and, in that sense, is model-free. Algorithm \ref{alg:direct_comp} provides a summary. \\

\noindent \textbf{Choice of Discretization Precision $m$}: Corollaries \ref{cor:approx_diff_ent_disc} and \ref{cor:lzcomp_sh} guarantee convergence of Algorithm \ref{alg:direct_comp} to the true entropy rate as both $m$ and $T$ go to infinity. However, for a given sample size $T$, as will be the case in practice, estimation error can vary greatly with $m$. Too small an $m$ and the estimation error in Corollary \ref{cor:approx_diff_ent_disc} will be large. Too large an $m$ and the discretized sample will have distinct characters irrespective of the source, its Lempel-Ziv complexity will be the sample size $T$, and our estimate for the entropy rate, which will be close to $ \log_2 T$, will overshoot (See Figure (\ref{fig:lz_efficiency_m})). In practice, we find that choosing $m$ such that $2^{-m}$ is between $\frac{1}{5}\sqrt{\frac{2}{\pi e}}$ and $\frac{1}{2}\sqrt{\frac{2}{\pi e}}$ works well for a range of sample sizes. \\

\noindent \textbf{Data Efficiency}: Care should be taken before applying this approach when the dimensionality $n$ of the input process is large. Indeed, if we denote $a$ the number of distinct characters we expect to commonly see in the discretization of a coordinate process of input $\mathbb{R}^n$-valued process $\{\pmb{z}_t\}$, then the discretization of $\{\pmb{z}_t\}$ can require up to $a^n$ distinct characters. We note that, despite $a  \ll T$, $a^n$ is bound to exceed $T$ even for a moderate $n$ (e.g. $n \approx 30$) and consequently, all characters in $(\hat{\pmb{z}}_1, \dots, \hat{\pmb{z}}_T)$ will be distinct, and both $c(T)$ and $c_i(T)$ will be equal to $T$, irrespective of the diffusion of the underlying source. This issue is well illustrated in the experiment of Figure (\ref{fig:lz_efficiency_n}). We generated $10$ independent draws from a standard Gaussian white noise, each with sample size $T=100$. Figure (\ref{fig:lz_efficiency_n}) illustrates the evolution of the ratio between the Lempel-Ziv complexity and sample size as a function of the dimensionality $n$. It can be seen that for $n=1$ the Lempel-Ziv complexity is less than half the sample size, but for $n=3$  it increases to about $95\%$ of sample size. From $n=5$ onwards, all characters are distinct and $c(T)= T$. In general, for large $n$, the sample size $T$ required to achieve a satisfactory estimation accuracy can be unusually large. As a rule of thumb, an estimated entropy rate close to $\log_2(T)-mn$ is an indication that the sample size $T$ is too small.\\

\noindent \textbf{Scalability}: Algorithm \ref{alg:direct_comp} scales linearly with both the sample size $T$ and the dimensionality of the input process $n$. However, as previously discussed, the number of samples $T$ needed for accurate estimation itself depends on $n$. For a fixed estimation accuracy, the number of samples $T$ required, and consequently time complexity, will typically grow exponentially in $n$.\\

\begin{figure*}
  \centering
  \subfloat[Lempel-Ziv complexity of a univariate standard Gaussian white noise as a function of discretization precision $m$, proportionally to the sample size ($T=100$), and averaged over $1000$ random draws.]{\label{fig:lz_efficiency_m}\includegraphics[width=0.45\textwidth]{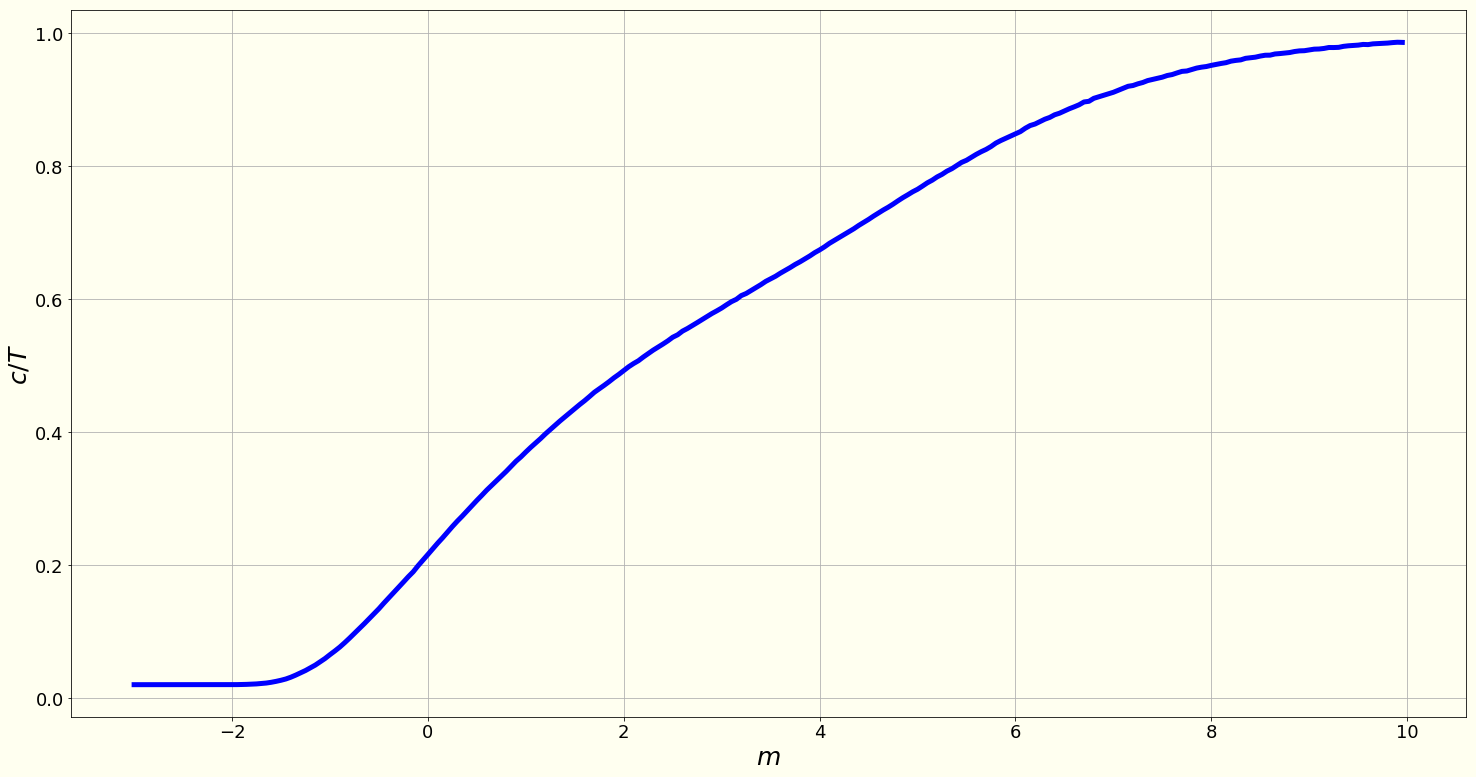}}
  \hfill
  \subfloat[Lempel-Ziv complexity of multivariate standard Gaussian white noises as a function of the number of series $n$, for a discretization precision $m=1.58$, and proportionally to the sample size ($T=100$).]{\label{fig:lz_efficiency_n}\includegraphics[width=0.45\textwidth]{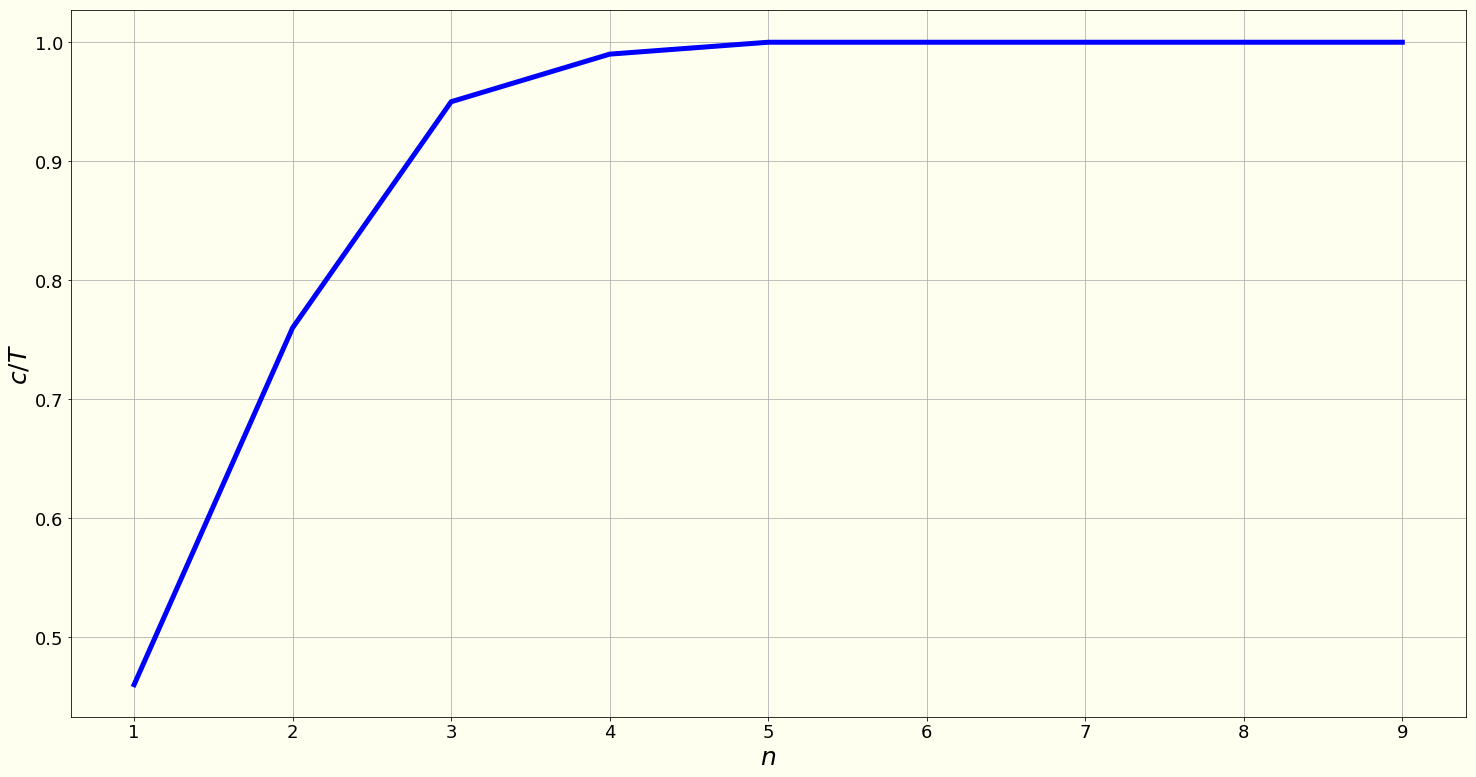}}
  \caption{Lempel-Ziv complexity of discretized time series as a function of input dimensionality and discretization precision.}
\end{figure*}

\subsubsection{\textsc{Nonparametric Estimation}}
\label{sct:mobased}
Analytic formulae to compute entropy rates are not always available. When $\{\pmb{z}_t\}$ is a stationary Gaussian process however, its entropy rate exists and is available in closed-form. More precisely, if we denote $$\Gamma \left( h \right) := \mathbb{C}\text{ov} \left(\pmb{z}_t, \pmb{z}_{t+h} \right)$$ the autocovariance function of $\{\pmb{z}_t\}$, and if we assume that 
$$\sum_{h=-\infty}^{+\infty} || \Gamma\left( h \right) ||  < +\infty,$$
where $||.||$ denotes any matrix norm, then the matrix-valued spectral density function
\begin{align}
\label{eq:spectal_density}
g(\omega) := \frac{1}{2\pi} \sum_{h=-\infty}^{+\infty} \Gamma \left( h \right) e^{-ih\omega}
\end{align}
is well-defined, forms a Fourier pair with the autocovariance function,
\begin{align}
\label{eq:fourier_autocov}
\Gamma \left( h \right)  = \int_0^{2\pi} g(\omega) e^{-ih\omega} d\omega
\end{align}
and the entropy rate reads:
\begin{align}
\label{eq:fourier_entropy_rate}
h\left( \{ \pmb{z}_t \} \right) = \frac{1}{4\pi} \int_0^{2\pi} \log_2 \det \left[4 \pi^2 e g(\omega) \right] d\omega.
\end{align}
Thus, entropy rates can be estimated in the stationary Gaussian case by first estimating the spectral density function, and  then using Equation (\ref{eq:fourier_entropy_rate}), where the integral can be approximated numerically. 

A na\"ive estimator of the spectral density function is obtained as the piecewise constant extension to $[0, 2\pi]$ of the periodogram, defined as 
\begin{align}
\hat{g}(\omega_k) = \frac{1}{2\pi} d_k d_k^*
\end{align}
with $$d_k = \frac{1}{\sqrt{T}} \sum_{t=1}^T \pmb{z}_t e^{-i (t-1) \omega_k},$$
where $d_k^*$ denotes the transpose of the complex conjugate of $d_k$, and for $\omega_k = \frac{2\pi k}{T}, ~ k = 0, \dots, \lfloor \frac{T}{2}\rfloor$. The periodogram is not a consistent estimator of the spectral density function. It is typically  improved and made consistent thanks to smoothing. Reviewing spectral density estimation methods is beyond the scope of this paper. We refer the reader to \cite{priestley1981spectral, book:512425} and references therein for a more detailed discussion on smoothed periodograms.

\begin{remark}
The Gaussian assumption can be relaxed by assuming that, although $\{y_t, \pmb{x}_t\}$ might not be Gaussian, there exists a mapping $\phi$ such that the stochastic process $\{ \pmb{\varphi}_t \} := \{ \left(y_t, \phi\left( \pmb{x}_t \right) \right)\}$ is Gaussian, stationary and ergodic. Both the periodogram of $\{ \pmb{\varphi}_t \}$ and smoothed versions thereof can then be obtained in closed-form and only depend on the spectral density function of $\{y_t, \pmb{x}_t\}$ and the reproducing kernel induced by the feature mapping $\phi$, which enables us to consider possibly infinite dimensional feature spaces through the kernel trick (see \cite{bachandjordan} for more details). As for the choice of kernel, the Generalized Spectral Kernels of \cite{samo2015generalized} provide a family that is provably arbitrarily flexible. Equation (\ref{eq:fourier_entropy_rate}) can then be used to estimate $I\left(\{ y_t \} ;  \{ \phi\left( \pmb{x}_t \right) \} \right)$ which, although in general differs from $I\left(\{ y_t \} ; \{ \pmb{x}_t \} \right)$, can be used as a proxy for incremental diversification. As previously discussed, if the feature mapping $\phi$ is chosen to be a smooth bijection, then $I\left(\{ y_t \} ;  \{ \phi\left( \pmb{x}_t \right) \} \right)=I\left(\{ y_t \} ; \{ \pmb{x}_t \} \right)$.
\end{remark}

\noindent \textbf{Data Efficiency:} As in the model-free case, care should be taken before applying this approach when the dimensionality $n$ of the input process is too large, but for a different reason. The rationale here is that estimating the spectral density function typically scales poorly with dimensionality $n$, and can hardly cope with $T \ll n$. To estimate incremental diversification using this method as is, the operator might find it useful to first reduce the dimensionality of $\{\pmb{x}_t\}$ using one of the wide range of techniques available (e.g. PCA and kernel PCA \cite{scholkopf1998nonlinear}, GP-LVM \cite{lawrence2004gaussian}, autoencoders \cite{vincent2010stacked, kingma2013auto}, manifold learning \cite{roweis2000nonlinear, belkin2003laplacian, gorban2008principal} etc.), and then use the differential mutual information timescale between $\{y_t\}$ and the compressed version of $\{\pmb{x}_t\}$ as a proxy for incremental diversification. In Section \ref{sct:scaling_up} we propose an approximation to incremental diversification that is more data-efficient in that it does not require estimating large-dimensional spectral density functions. \\

\noindent \textbf{Scalability:} Time complexity scales cubically with dimensionality $n$ due to the need to evaluate $\det \left[ g(\omega) \right]$ at several frequencies to numerically compute the integral in Equation (\ref{eq:fourier_entropy_rate}), and linearly with $T$ because of the computation of the smoothed periodogram. The integral in Equation (\ref{eq:fourier_entropy_rate}) would be very costly to compute when $n$ is large, and Bayesian quadrature \cite{o1991bayes} can prove more efficient than traditional quadrature techniques, as it typically results in fewer function evaluations. Similarly, memory requirement scales quadratically with $n$ and linearly with $T$.  Overall, this approach cannot scale to very large $n$ as is. In Section \ref{sct:scaling_up} we propose an approximation of incremental diversification for which this nonparametric approach can be scaled to very large $n$. \\

\subsubsection{\textsc{Maximum-Entropy Estimation}}
\label{sct:maxent}
Our last estimation approach is based on the \emph{principle of maximum-entropy} pioneered by E. T. Jaynes in \cite{maxent, maxent2}. The maximum-entropy principle stipulates that, when faced with an estimation problem, among all models that are consistent with empirical evidence, one should always choose the one that is the most uncertain/ignorant about everything other than what has been observed. \\

Given a sample path $(\hat{\pmb{z}}_1, \dots, \hat{\pmb{z}}_T)$ of a vector-valued discrete-time stochastic process $\{\pmb{z}_t\}$, the sample autocovariance function defined as
\begin{align}
\hat{C}(h) = 
  \begin{cases}
    \frac{1}{T} \sum_{t=1+h}^T \left( \hat{\pmb{z}}_t - \bar{\pmb{z}} \right) \left(  \hat{\pmb{z}}_{t-h} - \bar{\pmb{z}}  \right)^T, & \text{if } h \geq 0\\ \\
    \hat{C}(-h)^T, & \text{if } h < 0 \nonumber
  \end{cases}
\end{align}
with $\bar{\pmb{z}} = \frac{1}{T} \sum_{t=1}^T  \hat{\pmb{z}}_t,$ provides reliable empirical evidence about the autocovariance of $\{\pmb{z}_t\}$, in that it is a consistent and asymptotically unbiased estimator thereof \cite{brockwell2013time}. As for measuring `uncertainty/ignorance' about everything else, the entropy rate happens to be the canonical measure for this purpose. Burg's maximum-entropy theorem \cite{ihara, choi1993multivariate}, which we recall below, provides an answer to the maximum-entropy optimization problem for discrete-time stationary processes under autocovariance constraints.

\begin{theorem}
\label{th:burg}
Let $\{\pmb{z}_t\}$ be a stationary $\mathbb{R}^n$-valued discrete-time stochastic process. Among all stationary processes whose (matrix-valued) autocovariance functions coincide with that of $\{\pmb{z}_t\}$ from lag $h=0$ to lag $h=p$, the mean-zero Gaussian Vector Autoregressive process of order $p$ (VAR(p)) has the highest entropy rate, and we have
$$h(\{\pmb{z}_t \}) = \frac{n}{2} \log_2 \left(2 \pi e\right) + \frac{1}{2} \log_2 \left[ \frac{\text{det} \left(\Sigma_p\right)}{\text{det} \left( \Sigma_{p-1} \right)} \right],$$
where $\Sigma_p$ is the block-matrix such that \begin{align}\label{eq:sample_autocov}\Sigma_p[i, j] :&= \mathbb{C}\text{ov}(\pmb{z}_{t+i},  \pmb{z}_{t+j})\\  :&= C(i-j), \nonumber\end{align}
with $0 \leq i, j \leq p.$
\end{theorem}
The maximum-entropy approach to estimating a differential entropy rate therefore consists of first computing the corresponding sample autocovariance function, and then choosing as $p$ the largest lag up to which we can reliably estimate autocovariance terms with finite sample size $T$.\footnote{Specifically, following the standard approach of \cite{schwert1989does}, we use $p= \left\lfloor 12 \left(\frac{T}{100}\right)^{\frac{1}{4}} \right\rfloor$.} This is summarized in Algorithm \ref{alg:maxent}.

\begin{remark} Theorem \ref{th:burg} is quite profound. It states that the most principled approach to modeling stochastic processes under an assumption as generic as known autocovariance terms,  follows a very simple and well-studied diffusion model whose entropy rate is available in closed-form. A subtle point to note however is that, under additional constraints such as higher sample moments (e.g. Negative skewness or excessive kurtosis), the mean-zero Gaussian VAR(p) is no longer maximum-entropy optimal, as higher moments of a Gaussian process are fully determined by its first two moments, and consequently a Gaussian VAR(p) does not have enough degrees of freedom to cope independently with second order and higher order constraints.
\end{remark}
\subsection{\textbf{\textsc{Scaling Up Incremental Diversification Estimation}}}
\label{sct:scaling_up}

\subsubsection{\textsc{The Source of Scalability Issues}}
As previously discussed, all three methods we proposed for estimating the differential entropy rate of an $\mathbb{R}^n$-valued stationary ergodic process $\{\pmb{z}_t\}$ scale poorly with dimensionality $n$. This should come as no surprise since, $h(\{\pmb{z}_t\})$ reflecting the total amount of information per unit of time in the process, should factor-in possible redundancies across coordinates, and therefore should somehow keep track of how each coordinate process of $\{\pmb{z}_t\}$ relates to all others. This is typically done through an $n \times n$ matrix, of which we either need to compute the determinant or the inverse, so as to get a sense of how coordinate processes depart from the i.i.d. case. This results in cubic time complexity and squared memory requirement, which is impractical for large $n$. In the nonparametric case, the $n \times n$ matrix is the value of the spectral density function at any frequency, and in the maximum-entropy case, the $n \times n$ matrix is the covariance matrix $\mathbb{C}\text{ov}(\pmb{z}_t, \pmb{z}_t)$, which is the upper-left corner block of $\Sigma_p$ and needs to be evaluated irrespective of the number of maximum-entropy autocovariance constraints $p$. 

The model-free approach does not directly suffer from this problem because the discretization step effectively turns the multivariate problem into a univariate one, at the cost of increasing the discrete entropy rate of the resulting discretized process. However, in the best case scenario, the impact of such entropy increase on computing resource requirements for a fixed estimation accuracy is in fact worse than the limitations of the nonparametric and maximum-entropy approaches. To see why, let's perform a back-of-the-envelop calculation to estimate how many samples one would need to reliably estimate the discrete entropy rate of a stationary stochastic process taking value in a finite alphabet, in the best case scenario. Let's denote $H$ the ground-truth discrete entropy rate. Ideally, each character in the alphabet should appear in our sample at least once. This happens for the smallest sample size when samples are uniformly drawn from the alphabet and independent across time. In this case, if we denote $\alpha$ the probability of occurrence of any symbol in the alphabet, then $\alpha = 2^{-H}$, and the smallest sample size we need to see all characters is $T= \frac{1}{\alpha} = 2^{H}$. In other words, as a rule of thumb, the number of samples required to have any hope of decently estimating the discrete entropy rate of a stationary process grows exponentially with the true entropy rate. In the model-free approach, if coordinate processes happen to be independent or loosely related, then the entropy rate of the discretized process will grow linearly with dimensionality $n$, and consequently the number of samples required to keep estimation accuracy constant will grow exponentially with $n$. This is worse than the maximum-entropy and nonparametric approaches since the time complexity of the model-free approach, which is linear in the sample size, grows exponentially with the number of assets for a fixed estimation accuracy.

The root cause of this lack of scalability is the absence of a structured model expressing how coordinate processes of $\{\pmb{z}_t \}$ relate to each other. This can for instance be done through a dimensionality reduction technique (e.g. PCA and kernel PCA \cite{scholkopf1998nonlinear}, GP-LVM \cite{lawrence2004gaussian}, autoencoders \cite{vincent2010stacked, kingma2013auto}, manifold learning \cite{roweis2000nonlinear, belkin2003laplacian, gorban2008principal} etc.). We do not follow this idea as it is very sensitive to the dimensionality reduction technique used, and most of them have scalability issues of their own. We choose instead to relax the implicit requirement that we should  understand how each asset relates to all the others.

\subsubsection{\textsc{Order-$q$ Incremental Diversification}}
Let $0 \leq q \leq n$ and $\pmb{\pi}_q$ be a partition of $\{1, \dots, n\}$ into subsets of size $k$, where $k=q$ for all but at most $1$ element in the partition. Let $\{y_t\}$ be a real-valued discrete-time process that is jointly ergodic and stationary with $\mathbb{R}^n$-valued process $\{\pmb{x}_t\}$. It follows from standard results on mutual information that for each element $\pmb{\pi}_q^i$ of the partition $$ I\left(\{y_t\} ; \{\pmb{x}_t\}) \geq I(\{y_t\} ; \left\{ \pmb{x}_t\left[\pmb{\pi}_q^i\right]\right\}\right),$$ where $\left\{\pmb{x}_t\left[\pmb{\pi}_q^i\right]\right\}$ is the vector-valued process whose coordinate processes are the ones of $\{\pmb{x}_t\}$ whose indices are in $\pmb{\pi}_q^i$. Consequently, denoting $\pmb{\Pi}_q$ the set of all possible partitions of $\{1, \dots, n\}$ into subsets of size $q$, it follows that 
\begin{align}
\label{eq:order_q_bound}
I(\{y_t\} ; \{\pmb{x}_t\}) \geq  I_q(\{y_t\} ; \{\pmb{x}_t\}),
\end{align}
where $$I_q(\{y_t\} ; \{\pmb{x}_t\}) = \underset{\pmb{\pi}_q \in \pmb{\Pi}_q }{\max} ~ \underset{\pmb{\pi}_q^i \in \pmb{\pi}_q }{\max} ~ I\left(\{y_t\} ; \{ \pmb{x}_t\left[\pmb{\pi}_q^i\right]\}\right).$$

When $\{y_t\}$  represents the time series of returns of a new asset, and $\{\pmb{x}_t\}$ those of assets in the reference pool, $1/I_q(\{y_t\} ; \{\pmb{x}_t\})$ reflects the least amount of incremental diversification the new asset adds to any subset of $q$ assets in the reference pool.
\begin{definition}
We denote order-$q$ incremental diversification a new asset A adds to a reference pool P, the least amount of incremental diversification A adds to a subset of size $q$ of P, namely
\begin{align}
\mathbb{ID}^q\left(A; P\right) := \underset{P_q  \in \mathcal{P}_q}{\min} ~ \mathbb{ID}\left(A; P_q \right),
\end{align}
where $\mathcal{P}_q$ is the set of all subsets of P of size $q$.
\end{definition}
It follows from Equation (\ref{eq:order_q_bound}) that, as we would expect, if a new asset adds no incremental diversification to any subset of $q$ assets in the reference pool, then it adds no incremental diversification to the reference pool. Moreover, it is easy to see that $I_q(\{y_t\} ; \{\pmb{x}_t\})$ is an increasing function of $q$, and that $I_n(\{y_t\} ; \{\pmb{x}_t\}) = I(\{y_t\} ; \{\pmb{x}_t\})$. For $q <n$, the difference $$I(\{y_t\} ; \{\pmb{x}_t\})-I_q(\{y_t\} ; \{\pmb{x}_t\})$$ reflects the amount of information about the new asset that can only be obtained from the reference pool by considering more than $q$ assets at a time. 

As a measure of incremental diversification, $\mathbb{ID}^q\left(A; P\right)$ satisfies Stylized Facts 1 and 2 under the sparsity constraint that the best replicating portfolio does not have more than $q$ non-zero allocations. $\mathbb{ID}^q\left(A; P\right)$ also satisfies Stylized Fact 3 providing that returns of the new asset do not depend on current and past returns of more than $q$ assets in the reference pool. As for Stylized 4, it is always met by $\mathbb{ID}^q\left(A; P\right)$ since, by the inequality of Equation (\ref{eq:order_q_bound}), given that $\mathbb{ID}\left(A; P\right)$ allows for manager diversification, so does $\mathbb{ID}^q\left(A; P\right)$ for any $q$. $\mathbb{ID}^q\left(A; P\right)$ is also trivially found to satisfy Stylized Fact 5.

We recall that the guiding principle we used to determine whether an asset adds incremental diversification to a reference pool is that, if it is easy to replicate returns of the new asset using those of the reference pool of assets and factors, then the new asset is not needed. In practice however, if the number of assets required to replicate the new asset is very large, it would not be far fetched to consider the new asset somewhat useful. Indeed, attempting to replicate the new asset with a large number of existing assets might result in excessive operating cost (e.g. transaction costs, borrowing rates for short-sells, tracking slippage due to rounding errors etc.). In that sense, although we introduced order-$q$ incremental diversification to scale-up inference, $q$ can be regarded as a sparsity factor chosen to reflect the largest number of assets the investment manager would consider practical to use to replicate a candidate new asset with assets and factors he/she already has access to, as an alternative to trading the new asset directly.\\

\noindent \textbf{Scalability}: Estimating $$\underset{\pmb{\pi}_q^i \in \pmb{\pi}_q }{\max} ~ I\left(\{y_t\} ; \left\{ \pmb{x}_t\left[\pmb{\pi}_q^i\right]\right\}\right)$$ using either the model-free approach, or the nonparametric approach, or the maximum-entropy approach scales linearly with the number of assets $n$. Rather than taking the $\max$ across all possible partitions of $\{1, \dots, n\}$ into subsets of size $q$, which would be intractable, we choose instead to randomly sample a smaller number of partitions, and take the $\max$ across sampled partitions. Once partitions have been sampled, evaluating $I\left(\{y_t\} ;  \left\{ \pmb{x}_t\left[\pmb{\pi}_q^i\right]\right\}\right)$ can be performed in parallel, and the double $\max$ can be calculated efficiently using map-reduce. This is summarized in Algorithm \ref{alg:order_q}.

\subsection{\textbf{\textsc{Extension to a Pool of New Assets}}}
Our method for quantifying incremental diversification can be extended to quantifying the amount of diversification a universe of new assets $\pmb{A} = (A_1, \dots, A_p)$, for which no asset is fully determined by the others, collectively adds to a reference pool of assets. If we denote $$\{\pmb{y}_t\} := \{(y_t^1, \dots, y_t^p)\}, ~ p > 1$$ the vector-valued time series of returns of assets in the new universe $\pmb{A}$, and $\{\pmb{x}_t\}$ the time series of returns and factor values of the existing reference pool of assets, then the amount of diversification the new universe of assets add to the existing one is \begin{align}\mathbb{ID}\left(\pmb{A}; P \right) :&= \frac{1}{I\left( \{\pmb{y}_t  \} ; \{ \pmb{x}_t\} \right)} \\ &= \frac{1}{h\left( \{\pmb{y}_t  \} \right) + h\left( \{\pmb{x}_t  \} \right)- h\left( \{\pmb{x}_t, \pmb{y}_t  \} \right)},\nonumber\end{align} and can be computed using previously established results.
%
%

\subsection{\textbf{\textsc{Illustration}}}
\label{sct:id_illustration}
In this section we empirically illustrate the pertinence of our measure of incremental diversification, as well as estimation methods previously discussed. We first provide a comparative analysis between model free, nonparametric and maximum-entropy approaches. Then we empirically illustrate that our finite-sample estimation approach of choice, namely maximum-entropy estimation, is consistent with all $5$ Stylized Facts. Finally, we apply our measure of incremental diversification to real financial data, first comparing pairwise incremental diversification and pairwise correlation, and then investigating information clustering across asset classes.

\subsubsection{\textsc{Model Comparison}}
We begin by comparing the three approaches we proposed for estimating differential entropy rates on synthetic data, starting with real-valued time series. \\

\noindent $\pmb{n=1}$, \textbf{Varying} $\pmb{T}$: In the interest of assessing how our three approaches perform in the presence of memory and leptokurticity, we consider an AR(1) time series with Student-t noise, namely 
\begin{align}
\label{eq:student_ar1}
y_t = \frac{1}{2} y_{t-1} + \xi_t,
\end{align}
where $\xi_t$ is a Student-t white noise with standard deviation $1$, and degree of freedom $\nu$. We generate two sample paths of size $2000$ from our synthetic model, one for which we choose $\nu$ so that the innovation term has infinite kurtosis ($\nu=4$), and one for which the innovation term is almost Gaussian ($\nu=100$). In each simulation, we estimate the entropy rate of the underlying process using the model-free, nonparametric and maximum-entropy approaches previously described on the first $T$ observations for $100 \leq T \leq 2000$, and we plot the relative error\footnote{Defined as estimated entropy rate minus true entropy rate divided by absolute value of true entropy rate, and expressed in percentage points.} as a function of $T$ in Figures (\ref{fig:convergence_df_4}) and (\ref{fig:convergence_df_100}). For the nonparametric estimation, our estimate for the spectral density is obtained using Welch's method \cite{welch1967use} with a Hanning window, a window size equals to $100$, and a $50$\% overlap. For the model-free approach, we set $m$ such that $2^{-m}$ is equal to $1/5$-th of the sample standard deviation. As for the ground truth, we recall that the differential entropy rate of any autoregressive process is the differential entropy rate $h(\{\xi_t\})$ of its innovation term,\footnote{Hint: $h(\{y_t\}) = h(y_t | y_{t-1}, \dots) = h(y_t | y_{t-1}, \dots, y_{t-p}) = \mathbb{E}\left(h(y_t | (y_{t-1}, \dots, y_{t-p}) = * \right)$} and by temporal independence of the innovation process, it is also equal to the differential entropy of any observation $h(\xi_t)$, which is available in closed-form for the Student-t distribution.

\begin{figure*}
  \centering
  \subfloat[$\nu=4$]{\label{fig:convergence_df_4}\includegraphics[width=0.45\textwidth]{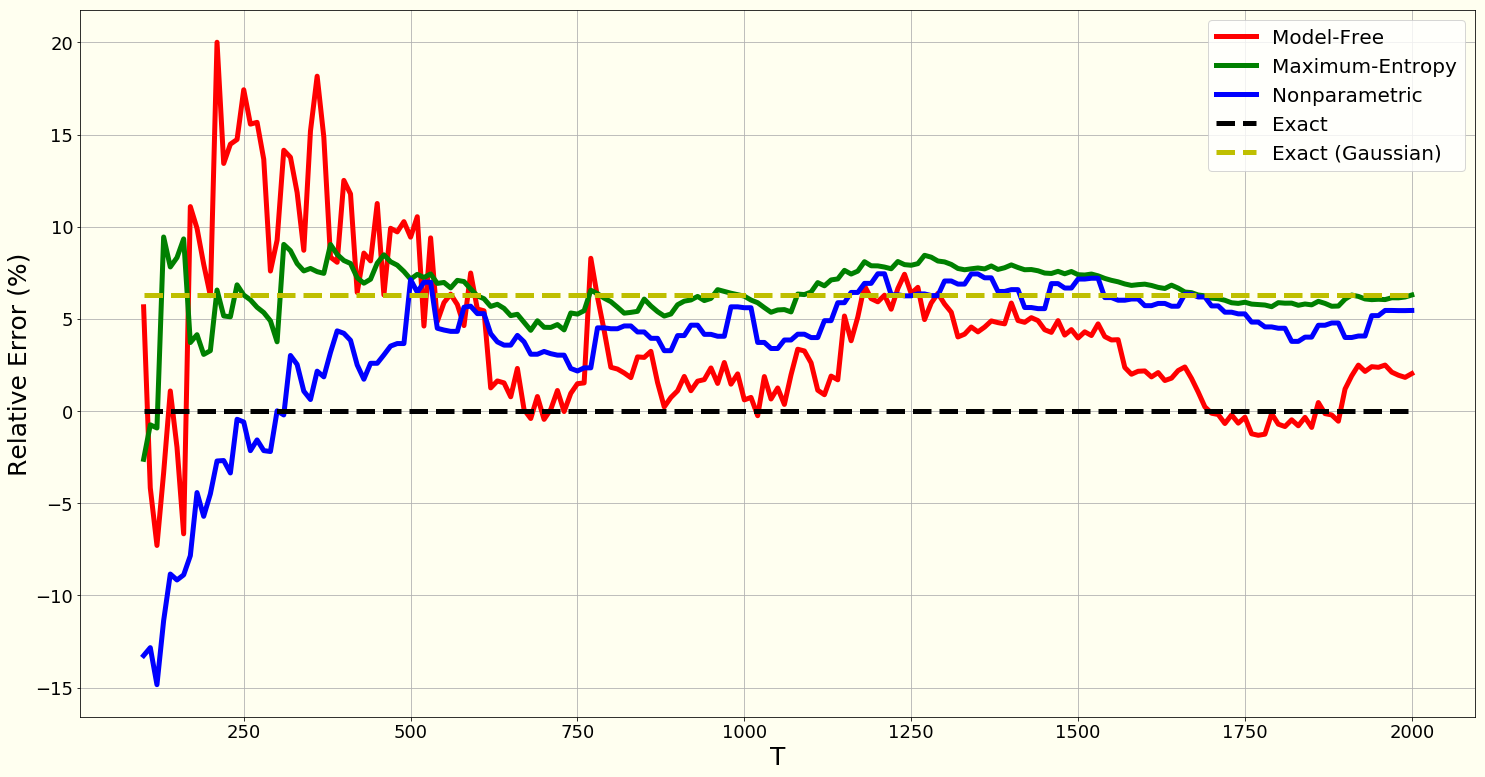}}
  \hfill
  \subfloat[$\nu=100$]{\label{fig:convergence_df_100}\includegraphics[width=0.45\textwidth]{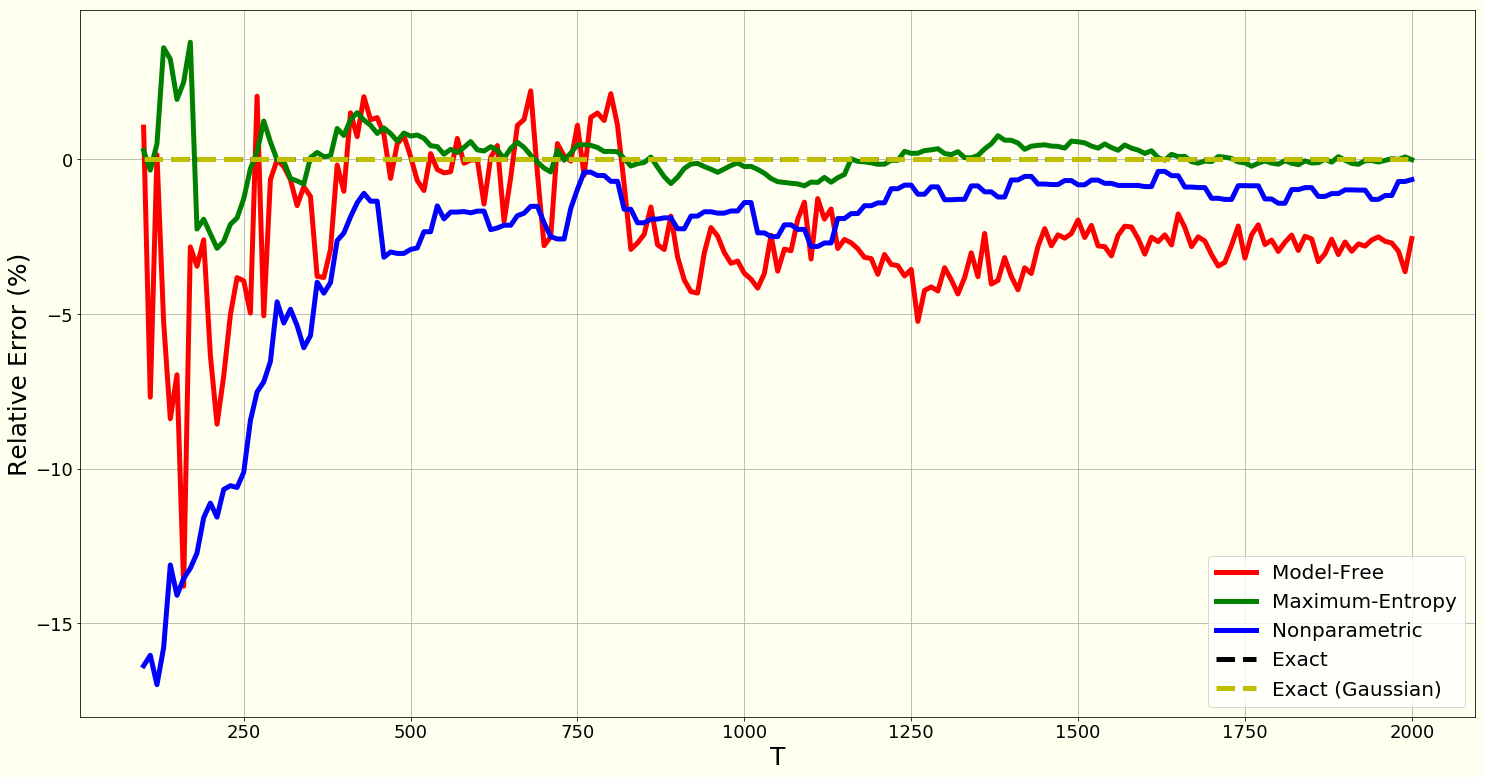}}
  \caption{Estimation of the entropy rate of an AR(1) process with Student-t noise with $\nu$ degrees of freedom, scale parameter chosen so that the innovation process is unit standard deviation, and for a sample size $T$. For an estimate $\hat{h}$, ground-truth $h$, the relative error is defined as $100*(\hat{h}-h)/|h|$. Exact (Gaussian) corresponds to using as estimate $\hat{h}$ the entropy-rate of the Gaussian AR(1) with identical coefficients and unit innovation standard deviation. For the nonparametric estimation, our estimate for the spectral density is obtained using Welch's method \cite{welch1967use} with a Hanning window, a window size equals to $T/20$, and a $50$\% overlap. For the model-free approach, we set $m$ such that $2^{-m}$ is equal to $1/5$-th of the sample standard deviation.}
\end{figure*}

Overall, it can be seen that all three approaches converge. As expected, both the maximum-entropy and the nonparametric approaches, which are the only ones assuming Gaussianity, converge to the entropy rate of the Gaussian AR(1) process that has the same mean and autocovariance function as our Student-t AR(1) process. Interestingly, even when the excess kurtosis of our synthetic model is infinite, the entropy rate of the closest Gaussian AR(1) model is only $6\%$ higher. In other words, the error we would make in maximum-entropy estimation by discarding fourth order moments does not exceed $6\%$ (in this family of examples), which is reached in the extreme case of infinite fourth moment. The two Gaussian approaches have near-identical performances, but the maximum-entropy approach has a considerably simpler implementation than the nonparametric approach, and is also faster, although both have linear time complexity.

The model-free approach on the other hand always converges to the ground truth, even when the excess kurtosis is infinite. In fact, for a fixed sample size $T$, the model-free approach converges faster when the excess kurtosis is large, which is understandable as this corresponds to lower entropy rates; in general, the higher the entropy rate, the more samples the model-free approach would require on average to achieve the same estimation accuracy. A good rule of thumb would be to avoid the model-free approach for univariate time series when $T<500$.

To summarize, as much as it appears more flexible than the maximum-entropy approach on the surface, in the univariate case, the nonparametric approach does not add much as far as estimating differential entropy rate is concerned. As for whether one should prefer the model-free or maximum-entropy approach in the univariate case, when data is scarce, one should always prefer the maximum-entropy approach, but when data abound, one should prefer the model-free approach. This viewpoint can in fact be generalized to much estimation problems ---when data abound in good quality, we should always be humble about postulating what is the `true' model of the world, and let data speak instead. When good quality data is scarce however, imposing a carefully crafted prior structure as part of inference is a must.\\

\noindent \textbf{Varying} $\pmb{n}$, \textbf{Fixed} $\pmb{T}$: Next, we consider empirically investigating how the accuracies of our three approaches to estimating differential entropy rates of an $\mathbb{R}^n$-valued discrete-time stationary ergodic processes scale with dimensionality $n$. To do so, we generate $T=2000$ samples of an $\mathbb{R}^n$-valued process $\{\pmb{z}_t\}$ whose coordinate processes are independent and each follows the Student-t AR(1) diffusion of Equation (\ref{eq:student_ar1}), for various $n$. For each $n$, we estimate the differential entropy rate of the corresponding sample using the model-free approach, the nonparametric approach, and the maximum-entropy approach. Results are illustrated in Figures (\ref{fig:id_eff_full}), (\ref{fig:id_eff_zoomed}), and (\ref{fig:id_eff_partial}). 

\begin{figure*}
  \centering
  \subfloat[All $3$ Approaches]{\label{fig:id_eff_full}\includegraphics[width=0.32\textwidth]{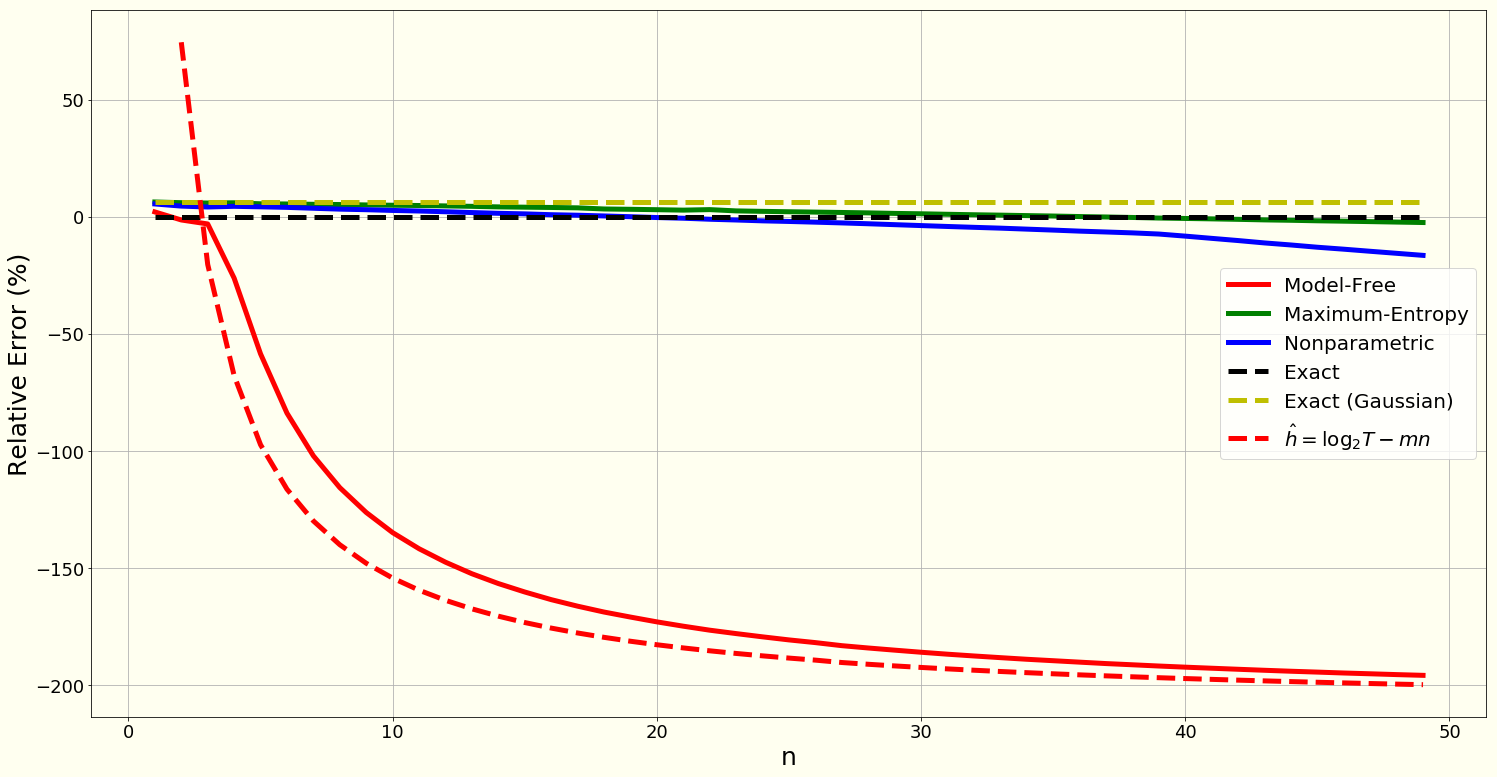}}
      \hfill
  \subfloat[Zoomed-In ($n \leq 5$)]{\label{fig:id_eff_zoomed}\includegraphics[width=0.32\textwidth]{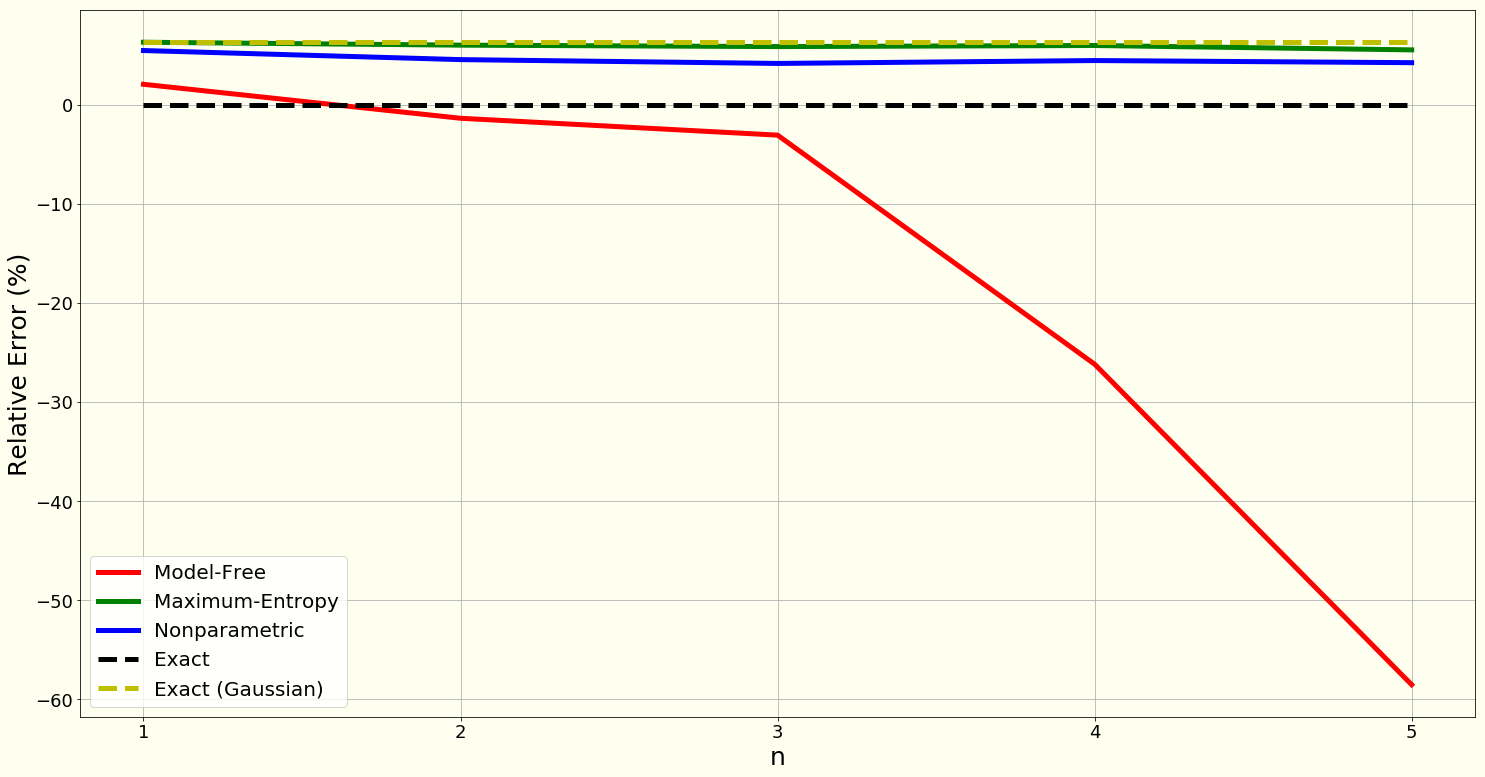}}
    \hfill
  \subfloat[Gaussian Approaches]{\label{fig:id_eff_partial}\includegraphics[width=0.32\textwidth]{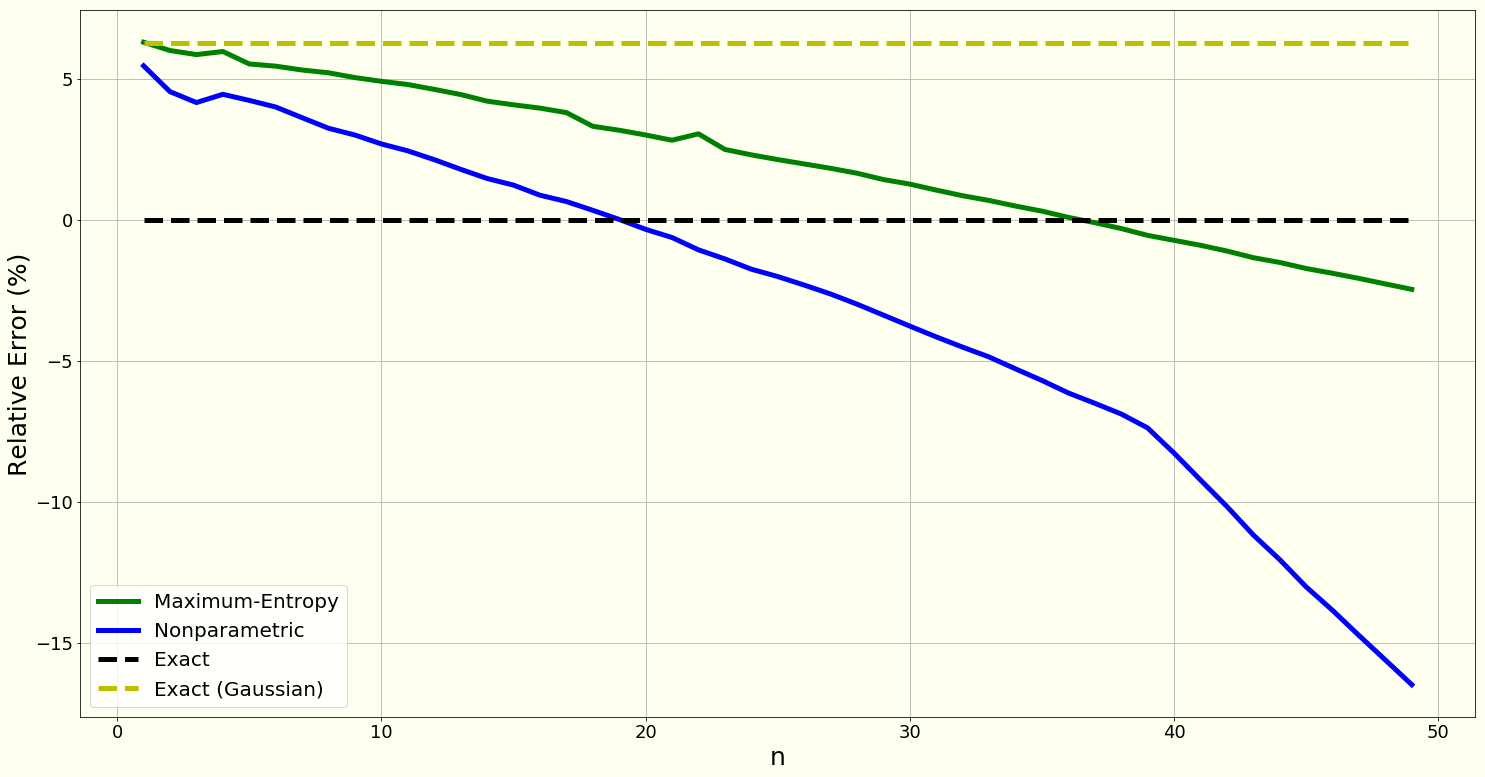}}
  
  \caption{Estimation of the entropy rate of an $\mathbb{R}^n$-valued stochastic process whose coordinate processes are independent, and each follows an AR(1) process with Student-t noise with $\nu=4$ degrees of freedom, and scale parameter chosen so that the innovation process is unit standard deviation. The sample size is $T=2000$. For an estimate $\hat{h}$, ground-truth $h$, the relative error is defined as $100*(\hat{h}-h)/|h|$. Exact (Gaussian) corresponds to using as estimate $\hat{h}$ the entropy-rate of the mean-zero Gaussian process with the same covariance function. For the nonparametric estimation, our estimate for the spectral density is obtained using Welch's method \cite{welch1967use} with a Hanning window, a window size equals to $100$, and a $50$\% overlap. For the model-free approach, we set $m$ such that $2^{-m}$ is equal to $1/5$-th of the smallest sample standard deviation across coordinate processes.}
\end{figure*} 

It can be seen from Figures (\ref{fig:id_eff_full}) and (\ref{fig:id_eff_zoomed}) that the model-free approach is grossly data-inefficient, and should certainly not be used beyond $n=3$ for a sample size $T\leq 2000$. This is in line with our previous back-of-the-envelop analysis that suggested that the sample size should increase exponentially with $n$ to maintain a fixed estimation accuracy in the model-free approach. To ease analysis of the performance of the model-free approach, we also plotted in Figure (\ref{fig:id_eff_full}) the performance of using as estimator $\log_2 T -mn$, which is what we would expect the model-free estimator (Equation (\ref{eq:approx_cond_ent})) to degenerate into when all characters in the discretized sample are distinct, which would occur when the sample size $T$ is not large enough for dimensionality $n$ and/or precision $m$. It can be seen in Figure (\ref{fig:id_eff_full}) that the behavior of the model-free approach is indeed dominated by the error of $\log_2 T -mn$, which further confirms that sample size $T=2000$, is insufficient beyond $n=3$, at least when $m$ is chosen so that discretization precision $2^{-m}$ is equal to $1/5$-th of the smallest sample standard deviation across coordinate processes. The answer is however not to decrease the discretization precision, as the model-free estimator is only a valid estimator for differential entropy rates when $2^{-m}$ is very small (see Corollary \ref{cor:approx_diff_ent_disc}, Equation (\ref{eq:approx_diff_ent_disc_multi})).

As for nonparametric and maximum-entropy approaches, their accuracies decrease (roughly) linearly with dimensionality $n$. However, unlike the one-dimensional case, in the multi-dimensional case the maximum-entropy approach is a lot more data-efficient than the nonparametric case, and the difference between the two grows with $n$. Considering that, additionally, the nonparametric approach is more complex to implement than the maximum-entropy approach, and is sensitive to choice of window and window size, the maximum-entropy should always be preferred over the nonparametric approach, and should also always be preferred over the model-free approach in the multivariate case. Whence, in following experiments, we always use the maximum-entropy approach to estimate differential entropy rates, unless stated otherwise.

Another important point worth noting about Figure (\ref{fig:id_eff_partial}) is that it gives us a sense of the order of magnitude of the relative error we would make by using the maximum-entropy approach as a function of $n$, and consequently, how large an $n$ we could afford for $T=2000$ ---which corresponds to about $8$ years of daily data--- while keeping the estimation error within reasonable bounds. This heuristic can in turn be used to guide the selection of the order $q$ of $\mathbb{ID}^q$. We find that, with $T=2000$, we can choose $q$ as large as $30$ while keeping relative estimation error below $5$\%.

\subsubsection{\textsc{Stylized Facts Consistency}}
\label{sct:id_sf}
We have previously shown that $\mathbb{ID}$ satisfies all 5 Stylized Facts. In this section, we aim to illustrate that the finite-sample maximum-entropy estimator also satisfies all 5 Stylized Facts. To do so, we generate a set of synthetic returns time series that exhibit both cross-sectional dependency and temporal dependency. First, we build the `innovation' component of returns as a factor model. More specifically, we generate a random orthogonal matrix $U$ of shape $N \times r$ with $N=50$ and $r=25$. We define $$X = UZ + \sigma_e E$$ where $Z$ (resp. $E$) is a standard Gaussian matrix with shape $(r, T)$ (resp. $(N, T)$) for $T=2000$. Columns of $X$ are thus i.i.d. Gaussian with mean $0$ and covariance matrix $$C=UU^T + \sigma_e^2 I.$$ We choose this structure to emulate a low-rank covariance matrix while avoiding numerical instabilities in the OLS estimation of tracking errors due to ill-conditioning. In this spirit, we choose $\sigma_e$ so that $C$ has determinant $10^{-10}$. We introduce temporal dependency from the innovation in an AR(1) fashion to obtain the synthetic time series of returns. Specifically, we define the $(N, T)$ matrix $Y$ such that its first column is the same as that of $X$, namely $Y[0] = X[0]$, and all other columns are defined as follows
\begin{align}
\label{eq:dv_exp}
Y[i] = \frac{1}{2}Y[i-1] + X[i].
\end{align}
Each row of $Y$ plays the role of a size $T$ path of a time series of returns of a different asset.\\

\noindent \textbf{Stylized Facts 1 \& 2}: We loop through rows sequentially, and for row $n$, we compute both the incremental diversification the corresponding asset adds to the reference pool defined by the first $n-1$ rows, and the correlation between returns of the asset and those of the best replicating portfolio of assets in the aforementioned reference pool. Results are illustrated in Figure (\ref{fig:id_sf12}), from which we note that incremental diversification tends to decrease with replication correlation, which is in line with Stylized Facts 1 and 2. \\
\begin{figure}
  \centering
  \includegraphics[width=0.45\textwidth]{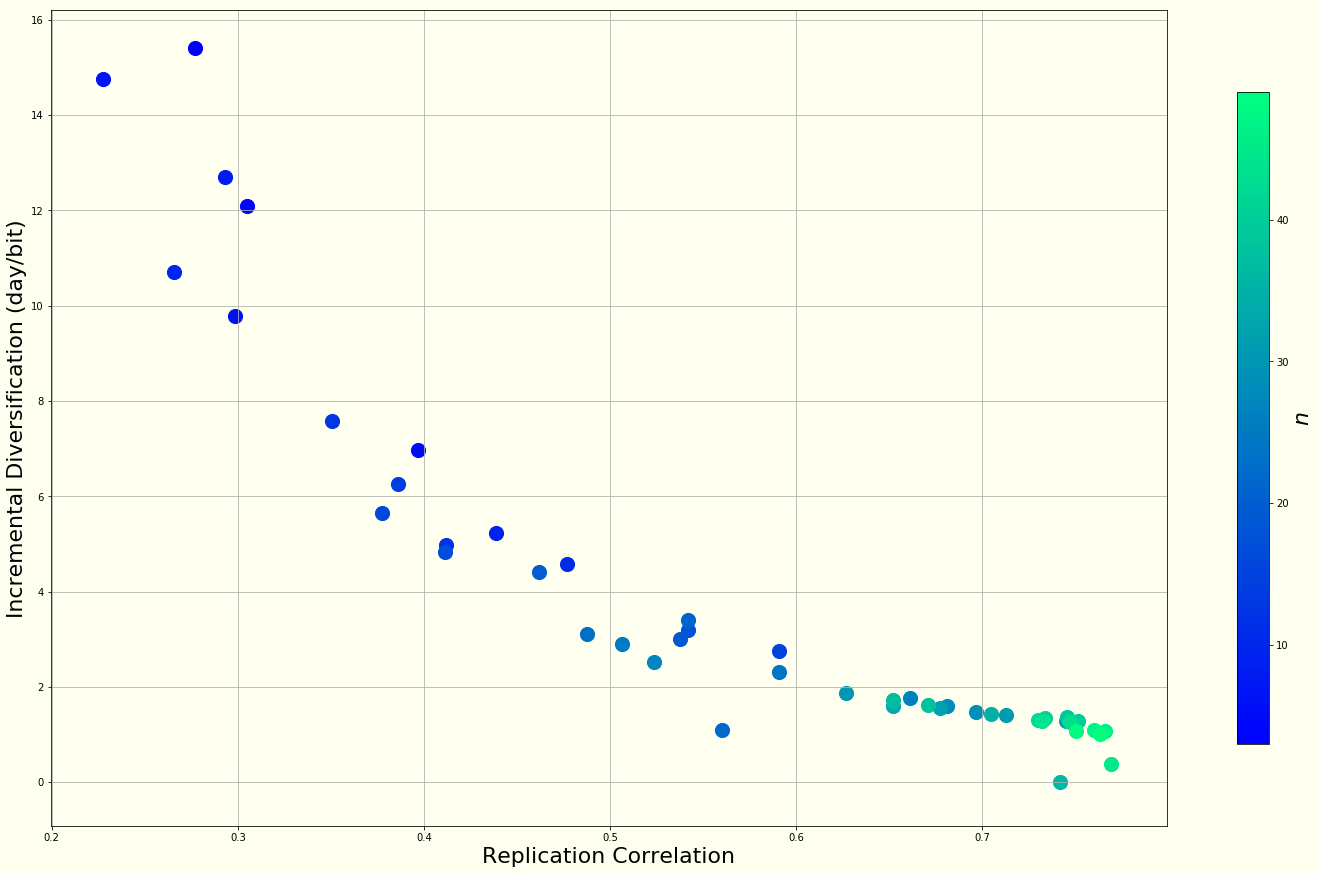}
  \caption{Illustration of the relationship between maximum-entropy estimation of incremental diversification and correlation with the best replicating portfolio in the toy experiment described in Section \ref{sct:id_sf}.}
  \label{fig:id_sf12}
\end{figure}

\noindent \textbf{Stylized Facts 3}: To assess consistency with Stylized Fact 3, we consider a simple momentum strategy on assets defined by rows of $Y$. For a window size $m$, the momentum strategy consists of investing proportionally to the returns of each asset in the pool over the past $m$ time periods. Returns of this momentum strategy can be written as a fixed function of current and past $m$ returns of assets defined by $Y$. We have previously shown that such assets add no incremental diversification to the reference pool, providing that $\mathbb{ID}$ can be computed exactly. When $\mathbb{ID}$ is estimated using finite samples however, incremental diversification is not necessarily $0$ but it is fairly small, as illustrated in Figure (\ref{fig:id_sf3}).\\
\begin{figure}
  \centering
  \includegraphics[width=0.45\textwidth]{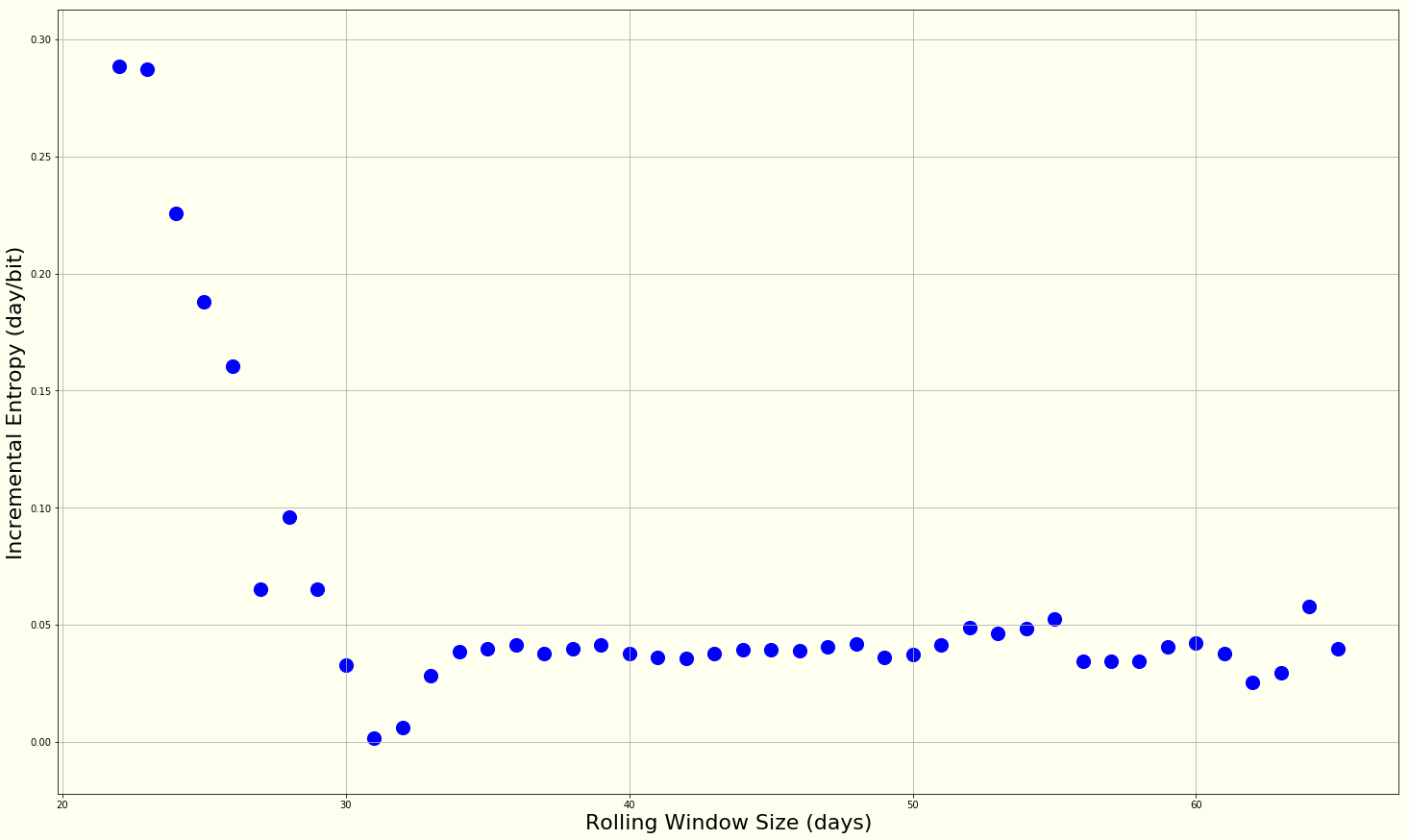}
  \caption{Maximum-entropy estimation of the incremental diversification a momentum strategy on a synthetic reference pool of assets adds to the reference pool. The generative model for returns of assets in the reference pool is described in Section \ref{sct:id_sf}. The momentum strategy consists of investing proportionally to asset performance over a certain window size in the past.}
  \label{fig:id_sf3}
\end{figure}

\noindent \textbf{Stylized Facts 4}: To illustrate that our approach allows for manager diversification, we consider the universe of assets whose returns are $Y$ (Equation (\ref{eq:dv_exp})), and we consider $40$ managers trading these assets long-only, and without leverage. Allocation processes are taken to be independent across managers. At each time period, each manager rebalances his portfolio according to an independent draw from a Dirichlet distribution, whose concentration parameter varies across managers but is the same across time. We generate concentration parameters randomly by drawing their coordinates i.i.d. from the uniform distribution on $[0, \alpha]$ for a configurable manager-specific parameter $\alpha$. $\alpha$ has a direct impact on the variance of the individual asset weights, and consequently managers' turnovers. If we have too high an $\alpha$ then weights won't vary much from one time period to the next, and we would expect the corresponding fund not to add diversification to the reference pool as its returns would be a linear combination of those of the reference pool (Stylized Fact 2), and more importantly we would also expect fund managers not to diversify each other much. Small $\alpha$ should be preferred to test consistency with Stylized Fact 4 (manager diversification).

We consider managers one at a time, and for each manager we estimate the incremental diversification his/her fund adds to the funds of all managers previously considered. First we use as $\alpha$ for the $i$-th manager $\frac{1}{1000} + \frac{5(i-1)}{1000}$. As illustrated in Figure (\ref{fig:id_s4_inc}) managers trading the same assets can provide incremental diversification to each other. Then we reverse the order in which we add managers, considering managers in decreasing order of $\alpha$ (or equivalently, increasing order of turnover), which we illustrate in Figure (\ref{fig:id_s4_dec}). Overall it can be seen that, although the number of managers previously added to the reference pool tends to reduce the incremental diversification a new manager adds, how active the new manager is (or equivalently his turnover) is a significantly bigger factor.\\
\begin{figure*}
\centering
\subfloat[Managers Added in Increasing Order of Turnover]{\label{fig:id_s4_inc}\includegraphics[width=0.45\textwidth]{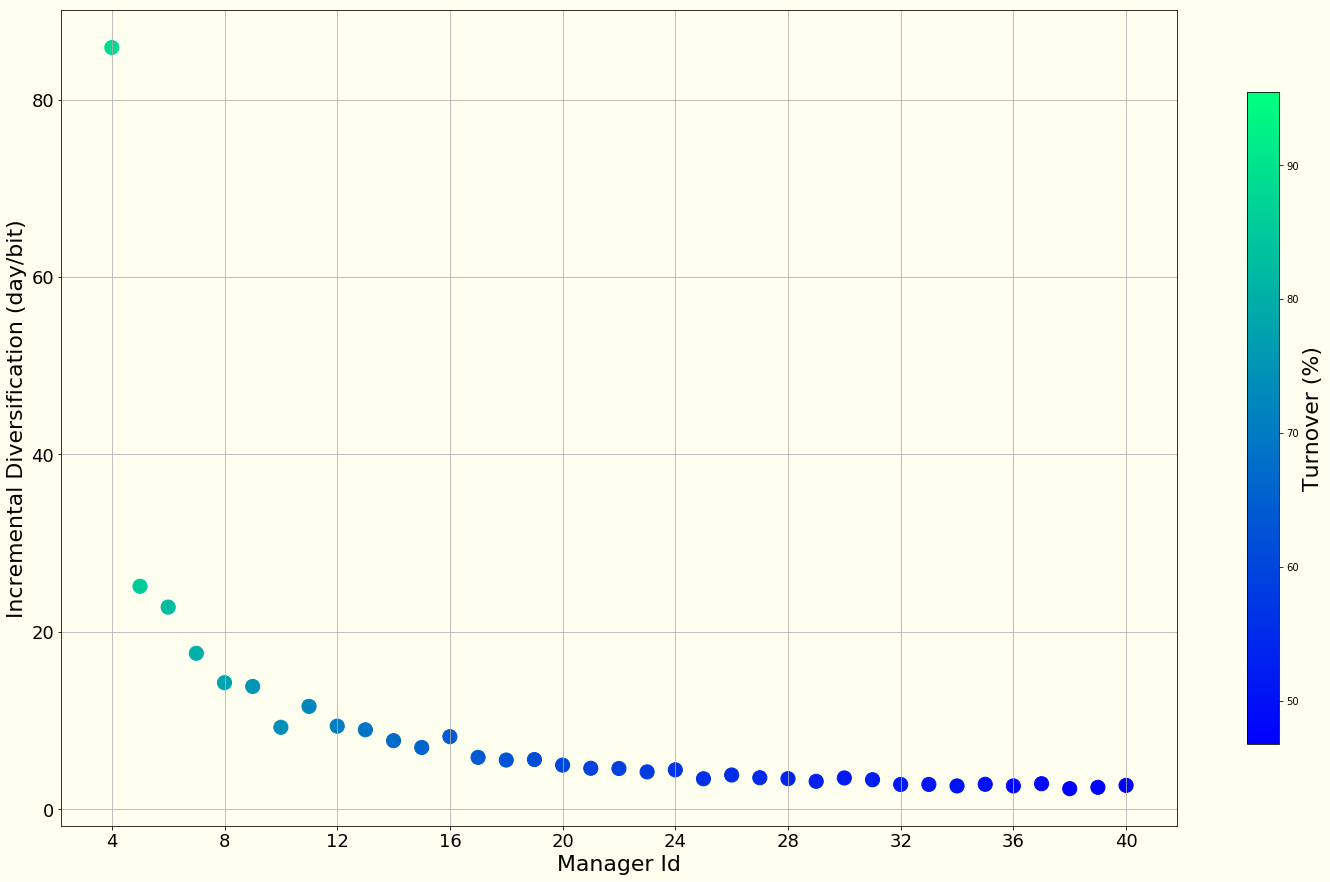}}
\hfill
\subfloat[Managers Added in Decreasing Order of Turnover]{\label{fig:id_s4_dec}\includegraphics[width=0.45\textwidth]{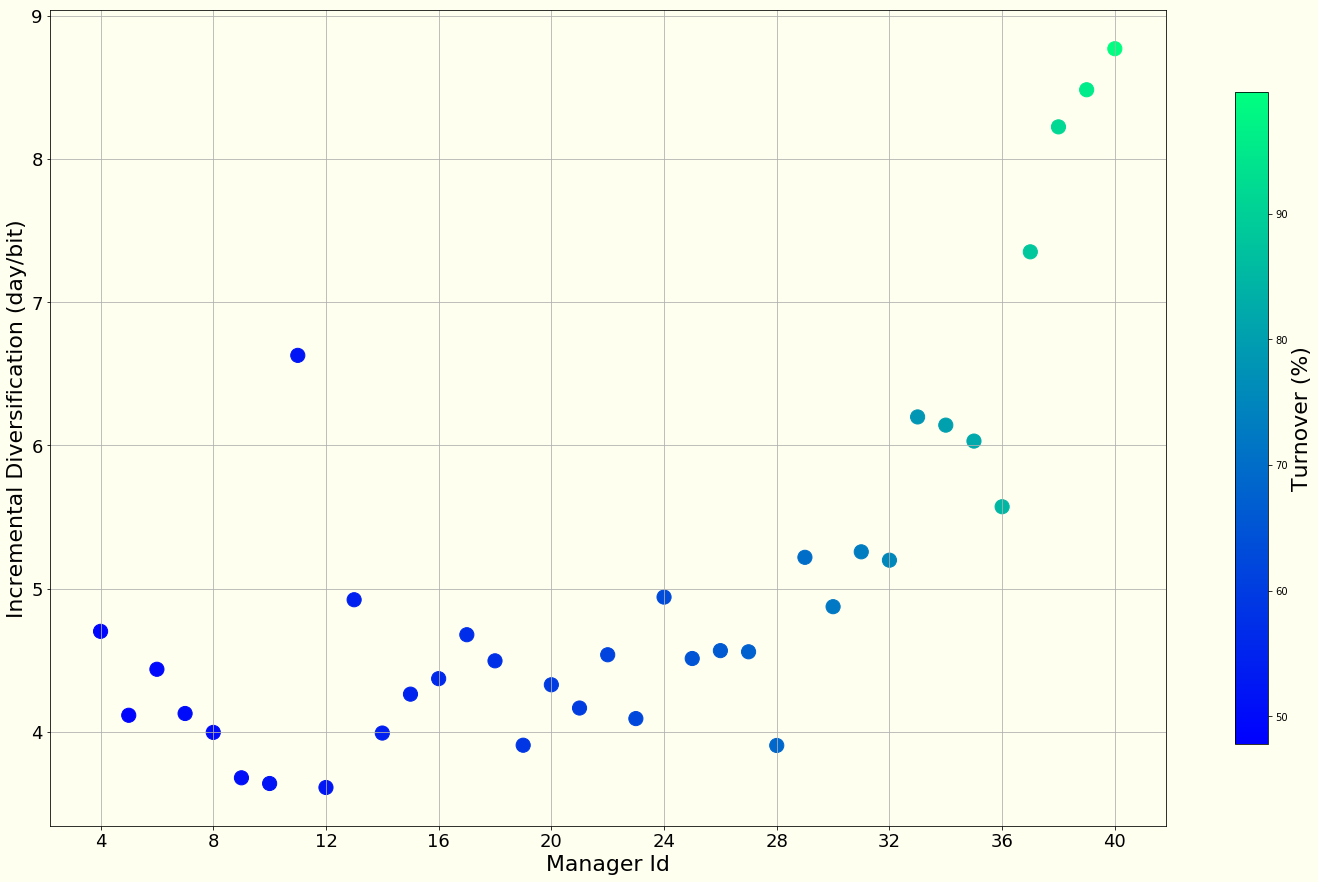}}
\caption{Maximum-entropy estimation of the incremental diversifications fund managers trading the same universe of assets add to each other in a synthetic experiment. The manager id represents the order in which the corresponding manager was added to the reference pool. Incremental diversification is computed relative to the pool of fund managers previously added. The generative model for returns of the universe of assets fund managers trade, as well as the generative models of managers' asset allocations are described in Section \ref{sct:id_sf}.}
\end{figure*}

\noindent \textbf{Stylized Facts 5}: Finally, in order to demonstrate that our finite-sample estimator of incremental diversification is invariant by rescaling, we construct an $N\times T$ matrix $Y^\prime$ such that each of its row is obtained by multiplying the corresponding row of $Y$ by a random scalar drawn from a standard normal. We loop through rows of $Y$ and $Y^\prime$ simultaneously, and we plot the incremental diversification row $n$ of $Y$ adds to the first $n-1$ rows of $Y$ against the incremental diversification row $n$ of $Y^\prime$ adds to the first $n-1$ rows of $Y^\prime$. This is illustrated in Figure (\ref{fig:id_sf5}), where it can be seen that our finite-sample estimator of incremental diversification is indeed invariant by both positive and negative rescaling. \\
\begin{figure}
  \centering
  \includegraphics[width=0.45\textwidth]{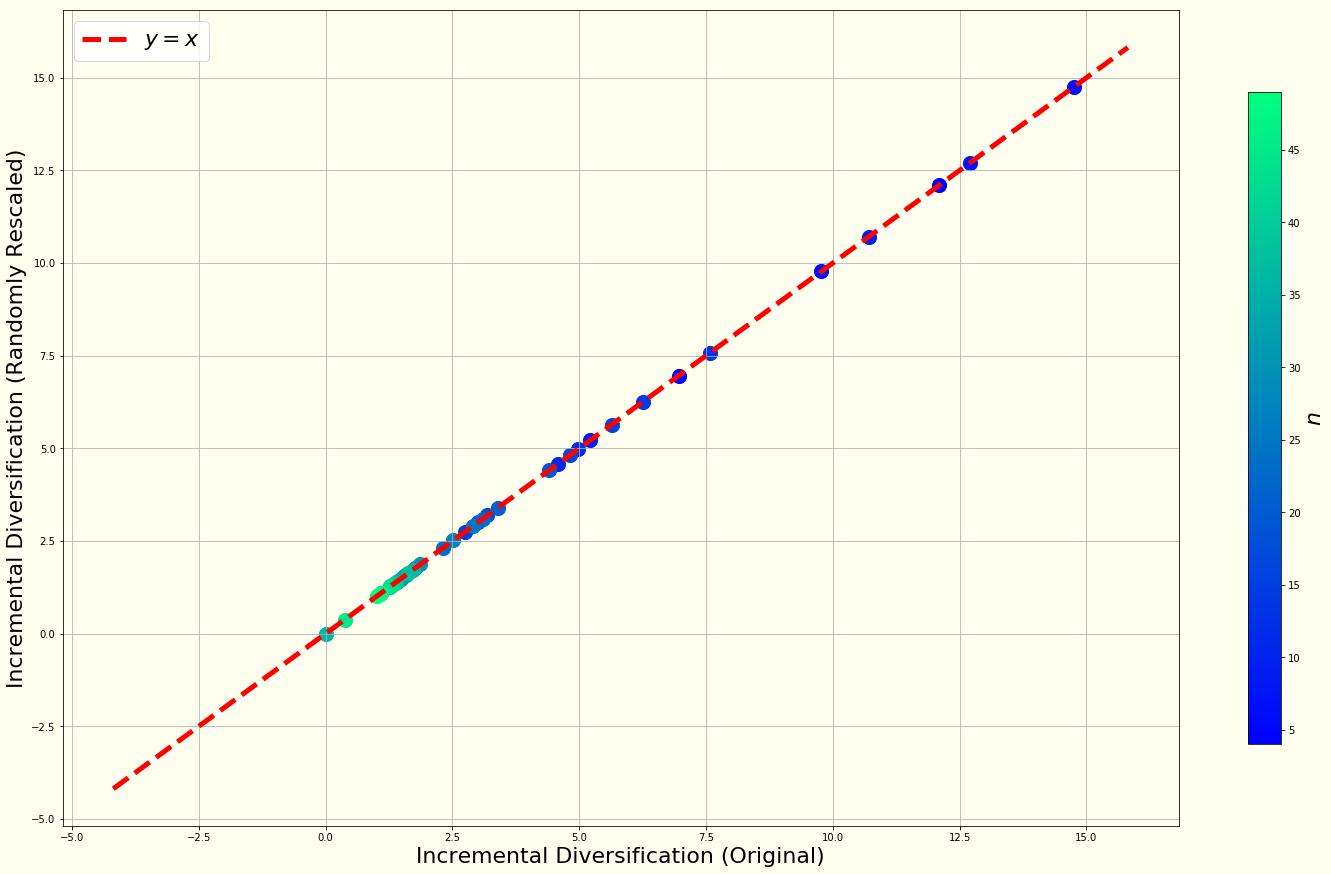}
  \caption{Comparison of the maximum-entropy estimations of the incremental diversifications an asset adds to a reference pool of $n$ assets, with and without randomly rescaling all returns time series.}
  \label{fig:id_sf5}
\end{figure}

\subsubsection{\textsc{Relation Between Pairwise Correlation and Pairwise Incremental Diversification}}
Considering the widespread, and somewhat excessive, use of correlation by both practitioners and academics as the canonical measure of dependency between assets, the alert reader must be wondering, in the case of two assets, how our measure of incremental diversification, namely pairwise mutual information timescale, relates to pairwise correlation. We aim to address this question empirically. For every pair of assets in the S\&P 100 index at the time of writing this paper,\footnote{See Table \ref{table:sp100_list} for the full list of companies.} we compute both the correlation between their daily returns and the incremental diversification one asset adds to the other. Results are illustrated in Figure (\ref{fig:id_pw}), in which it can be seen that, overall, incremental diversification tends to decrease with pairwise correlation, as expected.\footnote{To be more specific, pairwise mutual information timescale tends to decrease with the absolute value of pairwise correlation, but in this experiment most pairwise correlations are non-negative.} 

An interesting observation evidencing the limits of using correlation to quantify dependency between assets is that the width of the cloud of red points in Figure (\ref{fig:id_pw}) decreases with pairwise correlation. This makes intuitive sense. A strong pairwise correlation between two assets is a strong indication of (linear) dependency, and therefore also a strong indication that one asset can hardly diversify the other. A weak pairwise correlation, on the other hand, is a strong indication of lack of \emph{linear dependency} between returns corresponding to the \emph{same time period}, which \emph{does not imply} lack of dependency between returns of the two assets across time. When asset returns are both Gaussian and memoryless however, strong indication of lack of \emph{linear dependency} does imply lack of any kind of dependency between the two assets. In fact, if assets A and B have returns time series that are jointly stationary Gaussian and memoryless, it can be shown that there is a one-to-one map between incremental diversification and correlation, specifically $$\mathbb{ID}(A; B) = \mathbb{ID}(B; A) = \frac{-2}{\log_2 \left(1-\mathbb{C}\text{orr}(A, B)^2 \right)};$$ this is the blue curve illustrated in Figure (\ref{fig:id_pw}), which we will refer to as the \emph{correlation frontier}. Hence, the deviation of the cloud of red points in Figure (\ref{fig:id_pw}) away from the correlation frontier is strong evidence that the memoryless Gaussian assumption does not hold for S\&P 100 constituents or, equivalently, strong indication that returns of assets in the S\&P 100 exhibit nonlinear and/or temporal (e.g. lead-lag) relationship, and consequently pairwise correlation is inadequate to measure dependency between asset returns in practice as, unlike mutual information timescale, it cannot capture temporal or nonlinear dependencies. Noting that almost all points are below the correlation frontier, it follows that using pairwise correlation as a measure of dependency between assets tends to overestimate potential for diversification or, equivalently, underestimate similarities between assets.

Correlation can easily be adjusted to account for temporal and nonlinear dependencies, by first estimating pairwise incremental diversification, and then inferring the unadjusted correlation value that would be consistent with estimated incremental diversification, under the Gaussian memoryless assumption. We refer to the resulting quantity,
\begin{align*}
\mathbb{AC}\text{orr}(A, B) = \text{sign}\left(\mathbb{C}\text{orr}(A, B)\right) \sqrt{1-2^{\frac{-2}{\mathbb{ID}(A; B)}}},
\end{align*}
as the \emph{information-adjusted correlation coefficient}. 

It can be seen from Figure (\ref{fig:id_pw_2}) that, more often than not, accounting for temporal and nonlinear dependencies increases correlation. The lower the unadjusted correlation, the higher the difference between unadjusted and information-adjusted correlation coefficients; the difference can be as high as $0.2$.

Another interesting difference between pairwise incremental diversification and pairwise correlation is that the former is a lot more sensitive than the latter for smaller pairwise correlations, that is, when it matters, and is less sensitive for larger pairwise correlations, that is, when information redundancy is obvious.

\begin{figure*}
\centering
\subfloat[Relationship Between Pairwise Incremental Diversification and Pairwise Correlation.]{\label{fig:id_pw}\includegraphics[width=0.45\textwidth]{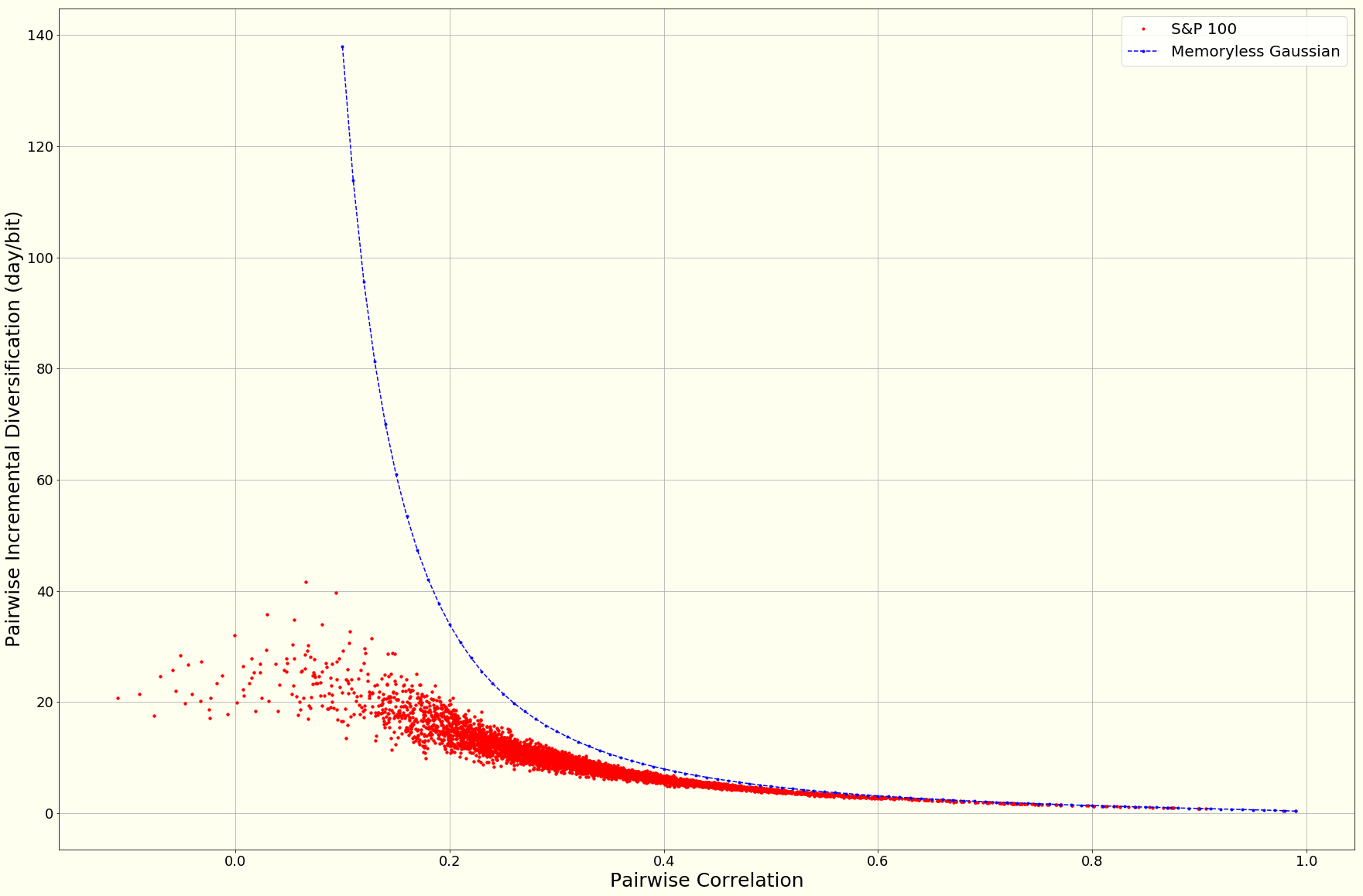}}
\hfill
\subfloat[Relationship Between Information-Adjusted Pairwise Correlation and Pairwise Correlation.]{\label{fig:id_pw_2}\includegraphics[width=0.45\textwidth]{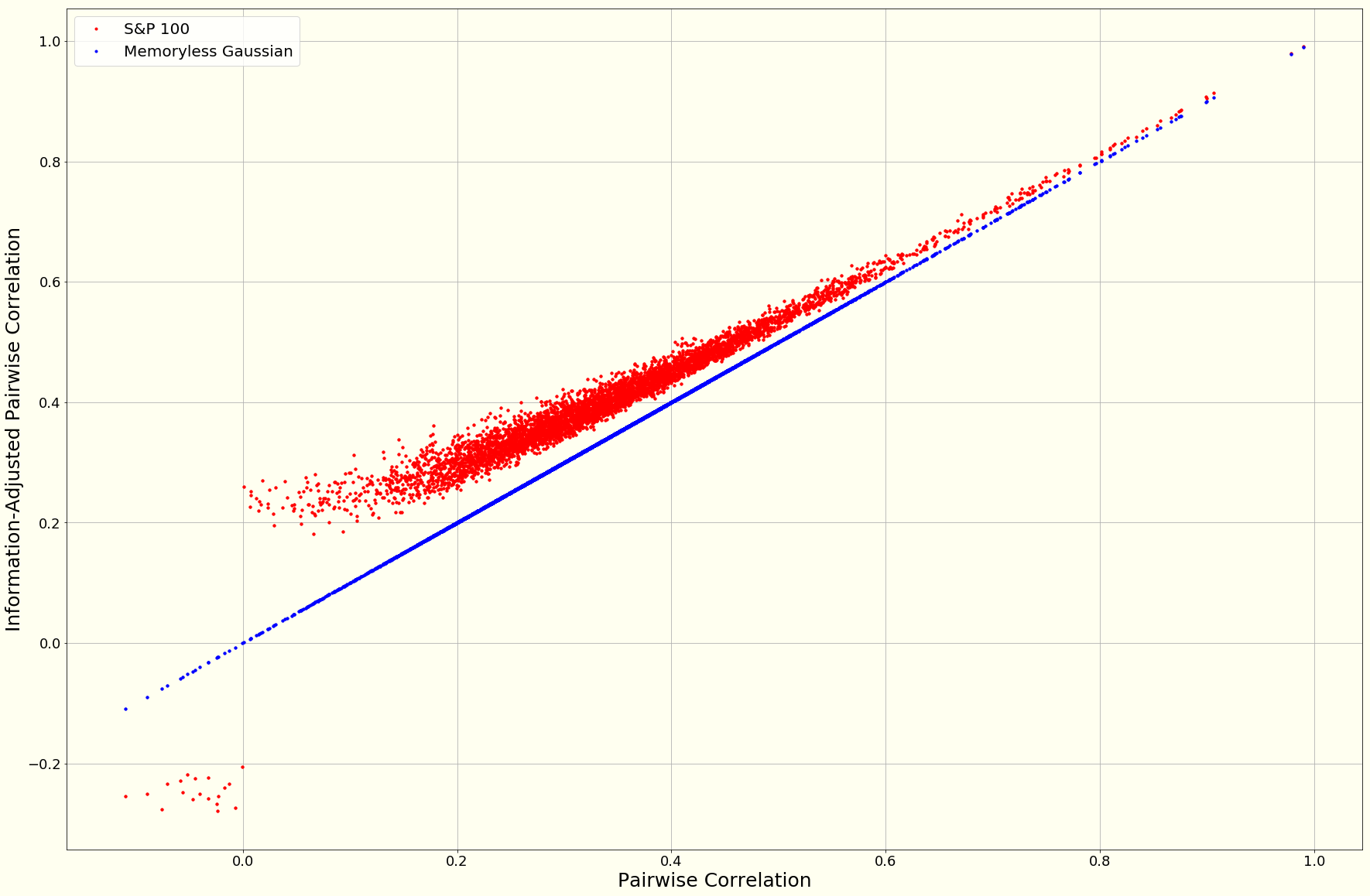}}
\caption{Evidence of temporal and nonlinear relationships between daily returns of constituents of the S\&P 100 index at the time of writing. The memoryless Gaussian case assumes the two assets have jointly-Gaussian and memoryless returns time series. In this case, denoting $\rho$ the correlation between the two assets, the incremental diversification $i$ one asset adds to the other is provably equal to $i=\frac{-2}{\log_2 \left(1-\rho^2\right)}$. The information-adjusted correlation is defined as the correlation value $\hat{\rho}$ that has the same sign as the unadjusted correlation $\rho$, and that, under the Gaussian memoryless assumption, would yield incremental diversification $\hat{i}$ identical to that estimated from the data: $\hat{\rho} = \text{sign}(\rho) \sqrt{1-2^{-(2/\hat{i})}}$.}
\end{figure*}

\subsubsection{\textsc{Asset Class Information Clustering}}
A common perception is that diversifying across asset classes provides considerably more benefits than within asset class diversification; the intuition being that assets within the same class share more economic drivers than assets across classes. In our last experiment on incremental diversification, we consider quantifying the benefits of asset class diversification. 

Specifically, using our measure of incremental diversification, we consider empirically evaluating how much more diversification can be obtained by choosing a new asset to add (to a reference pool made of assets belonging to the same asset class) from a different asset class, compared to choosing a new asset in the same asset class. We consider three pools of assets in three asset classes, namely constituents of the Dow Jones Industrial Average for Equities, $18$ of the most traded global currencies for FX, and $20$ of the most actively traded U.S. futures contracts for Futures. For each reference pool, we compute the average incremental diversification that an asset in the pool adds to the rest of the pool, as well as the average incremental diversification that an asset in another pool adds to the pool currently considered. Results are illustrated in Figure (\ref{fig:id_inf_clust}), from which it can be seen that cross asset class diversification usually provides at least as much benefits as within asset class diversification.

Figure (\ref{fig:id_inf_clust}) however paints a more granular story, one that can hardly be obtained with traditional tools. In effect, we note that U.S. futures on average add nearly as much incremental diversification to our reference pool of currencies as a currency in the pool adds to the rest of the pool on average. In other words, we find that using a U.S. futures as means of diversifying global currencies might not be much more impactful than considering trading one more currency. Interestingly, the reverse does not hold true. The investment manager considering diversifying U.S. futures, on average, would be much better off considering adding a foreign currency than another U.S. future. We find that the former would yield average incremental diversification $0.96$ days/bit, while the latter would result in $1.4$ days/bit incremental diversification. 

This could be used as empirical piece of evidence that global currencies already factor-in U.S. futures risk factors, but are also driven by additional risk factors that are unrelated to U.S. futures, and that offer cross asset class diversification opportunities for futures. We stress that we could not arrive to this conclusion by using average pairwise correlation as a measure of how much diversification one asset class can add to another, as practitioners routinely do; had we done so, the matrix of Figure (\ref{fig:id_inf_clust}) would have been symmetric, and we would have implicitly postulated that, irrespective of empirical evidence, currencies can only diversify U.S. futures as much as U.S. futures can diversify currencies. Our approach to quantifying incremental diversification on the other hand is flexible enough to reveal from the data that cross asset class diversification is not symmetric!

Another interesting observation we can make from Figure (\ref{fig:id_inf_clust}) is that foreign currencies provide the greatest cross asset class diversification benefits. The largest average incremental diversification ($1.4$ days/bit) is obtained by using global currencies to diversify U.S. futures, or by using U.S. blue chips to diversify global currencies.

\begin{figure}
  \centering
  \includegraphics[width=0.45\textwidth]{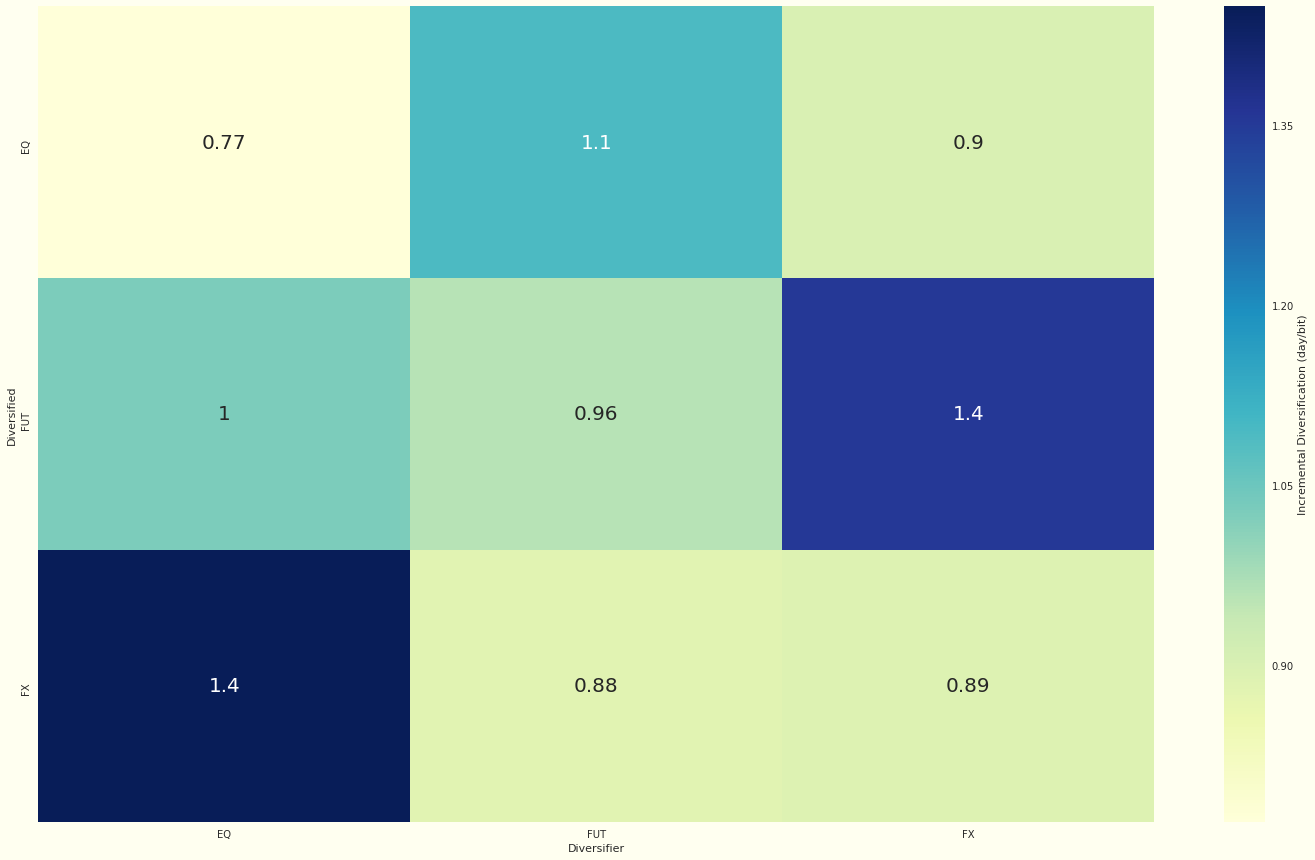}
  \caption{Comparison of within asset class diversification and cross asset class diversification. We consider three universes of assets, namely, constituents of the Dow Jones Industrial Average (EQ), $18$ of the most actively traded global currencies (FX), and $20$ of the most actively traded U.S. futures contracts. For each off-diagonal square in the matrix above, we compute the average incremental diversification an asset in the universe defined by the column (Diversifier) adds to the universe defined by the row (Diversified). For diagonal squares, we compute the average incremental diversification one asset in the universe defined by the column/row adds to the rest of the universe. The full list of assets considered as well as additional setup details are provided in Appendix \ref{sct:exp_setup}.}
  \label{fig:id_inf_clust}
\end{figure}

\section{\MakeUppercase{Quantifying Predictability of Returns Time Series}}
\label{section:predictability}

Intuitively, a time series of returns $\{y_t\}$ can be considered sufficiently predictable if there exists a stream of information available at time $t$ that reduces the uncertainty about future values at time $t+p, ~ p>0$. Determining whether future values of a returns time series can be predicted using \emph{any} piece of information that currently exists, whether we have access to it or not, would be impractical. Instead, we focus on quantifying whether a time series can be predicted using data that we do have access to, starting with all past values of the time series, and then generalizing to any stream of information we have access to.

\subsection{\textbf{\textsc{Auto-Predictability}}}
We say of a time series of returns that it is sufficiently auto-predictable when its past values sufficiently reduce the uncertainty about future values (i.e. when it has sufficient memory). 

When $\{y_t\}$ is (strongly) stationary,  the entropy rate $h\left(\{y_t\}\right)$ always exists and it can be shown\footnote{See Equation (12.38) in \cite{cover}.} that 
\begin{align}
h\left(\{y_t\}\right) = \underset{T \to +\infty}{\lim} h\left(y_T \vert y_{T-1}, \dots, y_1 \right).
\end{align}

Moreover, it can be shown that $h\left(y_T \vert y_{T-1}, \dots, y_1 \right)$ decreases with $T$.\footnote{Hint: $h\left(y_T \vert y_{T-1}, \dots, y_1 \right) \leq h\left(y_T \vert y_{T-1}, \dots, y_2 \right)$ and $h\left(y_T \vert y_{T-1}, \dots, y_2 \right) = h\left(y_{T-1} \vert y_{T-2}, \dots, y_1 \right)$ by strong stationarity.} Thus,
\begin{align} 
\label{eq:pred_metric}
\mathbb{PR}\left(\{y_t\}\right) := h\left(y_t\right) - h\left(\{y_t\}\right),
\end{align}
which we note does not depend on $t$ for stationary processes, can be regarded as the maximum reduction in the uncertainty of the return at any point in time $t$ that one can achieve by knowing all returns prior to $t$, which makes it suitable for quantifying predictability of returns. We name $\mathbb{PR}\left(\{y_t\}\right)$ the \emph{measure of auto-predictability} of time series $\{y_t\}$.
\begin{remark}\label{rmk:inv}The auto-predictability measure is invariant by affine transformations $$\forall \alpha, \beta \neq 0,  ~~ \mathbb{PR}\left(\{\alpha y_t + \beta \}\right) = \mathbb{PR}\left(\{y_t\}\right).$$
\end{remark}

Proposition \ref{prop:memory} confirms that memoryless time series are the least auto-predictable.
\begin{proposition}
\label{prop:memory}
Let $\{y_t\}$ be a stationary discrete-time stochastic process such that $\vert h\left(y_t\right) \vert < \infty$. Then, $$\mathbb{PR}\left(\{y_t\}\right) \geq 0.$$ Moreover, $$\mathbb{PR}\left(\{y_t\}\right) = 0$$ if and only if $(y_1, \dots, y_T)$ are jointly independent for any $T$.
\end{proposition}
\begin{proof}
$$\mathbb{PR}\left(\{y_t \}\right) =  \underset{T \to +\infty}{\lim} h(y_T) - h\left(y_T \vert y_{T-1}, \dots, y_1 \right),$$ $h(y_T) - h\left(y_T \vert y_{T-1}, \dots, y_1 \right) \geq 0$ for every $T$ and increases with $T$. Hence $\mathbb{PR}\left(\{ y_t \}\right)=0$ if and only if $h(y_T) - h\left(y_T \vert y_{T-1}, \dots, y_1 \right) = 0$ for every $T$, which holds if and only if  $(y_1, \dots, y_T)$ are jointly independent for every $T$.
\end{proof}
Another take on the measure of auto-predictability is obtained by noting that 
\begin{align} 
& \mathbb{PR}\left(\{y_t\}\right) \\ 
&= \underset{T \to +\infty}{\lim} ~ \frac{1}{T} D_{KL}\left[p(y_1, \dots, y_T) \vert \vert  p(y_1)\dots p(y_T)\right]. \nonumber
\end{align}

In other words, the measure of auto-predictability is the rate of KL-divergence between the returns time series and its memoryless equivalent. This parallel further confirms Proposition \ref{prop:memory} and Remark \ref{rmk:inv}. In fact Remark \ref{rmk:inv} can be generalized to $$\mathbb{PR}\left(\{ f(y_t) \}\right) = \mathbb{PR}\left(\{y_t\}\right)$$ for any smooth bijection $f$.

\subsection{\textbf{\textsc{Estimation}}}
Similarly to incremental diversification, the measure of auto-predictability can be estimated using a model-free approach, a nonparametric approach, or a maximum-entropy approach.\\

\noindent \textbf{Model-Free Estimation}: In the model-free case, we require the time series to be stationary and ergodic, but not necessarily Gaussian. It follows from Corollaries \ref{cor:approx_diff_ent_disc} and \ref{cor:lzcomp_sh} that
\begin{align}
\hat{H}(y_t^m) \left( 1-\frac{c(T)}{\frac{1}{k} \sum_{i=1}^k c_i(T)} \right),
\end{align}
where $\{y_t^m\}$ is the discretized version of $\{y_t\}$ with precision $m$, $\hat{H}(y_t^m)$ the na\"ive frequency estimator of discrete entropy $H(y_t^m)$, and $c(T)$ and $c_i(T)$ are as per Corollary \ref{cor:lzcomp_sh}, is a consistent (in $(m, T)$) estimator of the measure of auto-predictability.\\

\noindent \textbf{Nonparametric Estimation}: The spectral analysis approach requires assuming $\{y_t\}$ is a stationary and ergodic Gaussian process, in which case \begin{align}h(y_t) = \frac{1}{2} \log_2 \left(2\pi e \mathbb{V}\text{ar}(y_t) \right),\end{align} and the sample variance provides a consistent estimate of $\mathbb{V}\text{ar}(y_t)$ thanks to the ergodic assumption. Moreover, we recall that
\begin{align}
h\left( \{ y_t \} \right) = \frac{1}{4\pi} \int_0^{2\pi} \log_2 \left(4 \pi^2 e f(\omega)\right) d\omega,
\end{align}
where $f$ is the spectral density function of $\{ y_t \}$. Thus, $h\left( \{ y_t \} \right)$ can be estimated in the same manner as in the previous section, by first estimating the spectral density through a smoothed periodogram \cite{priestley1981spectral, book:512425, bachandjordan, welch1967use}, and then using quadrature techniques to estimate the integral. \\

\noindent \textbf{Maximum-Entropy Estimation}: The maximum-entropy estimation of the measure of auto-predictability is easily found from Section \ref{sct:maxent} to read $$ \mathbb{PR}\left( \{y_t \} \right) = \log_2 \hat{\gamma}(0) - \log_2 \left[ \frac{\text{det} \left(\hat{\Gamma}_p\right)}{\text{det} \left( \hat{\Gamma}_{p-1} \right)} \right],$$ where $p$, $\hat{\gamma}(0)$ and $\hat{\Gamma}_p$ are as per Section \ref{sct:maxent}.

\subsection{\textbf{\textsc{Exogenous Predictability}}}
The measure of auto-predictability can easily be extended to vector-valued time series as
\begin{align} 
\label{eq:vv_pred_metric}
\mathbb{PR}\left(\{\pmb{x}_t\}\right) := h\left(\pmb{x}_t\right) - h\left(\{\pmb{x}_t\}\right),
\end{align}
for which Proposition \ref{prop:memory} still holds. We further extend this measure of predictability to capture the reduction in uncertainty about future returns resulting from knowing not just current and past returns but, more generally, current and past values of any given set of factors or signals.
\begin{definition}
We denote measure of predictability of $\{y_t\}$ using $\{\pmb{x}_t\}$
\begin{align}
\mathbb{PR}\left(\{y_t\} \vert \{\pmb{x}_t\}  \right) := h\left(y_t\right) - h\left(\{y_t \} \vert \{\pmb{x}_t\}\right).
\end{align}
\end{definition}
$\mathbb{PR}\left(\{y_t\} \vert \{\pmb{x}_t\}  \right)$ represents the maximum amount of uncertainty reduction about a future return value one can achieve by observing factors or signals $\{\pmb{x}_t\}$ and past values of $\{y_t\}$ . Intuitively, we would expect $\mathbb{PR}\left(\{y_t\} \vert \{\pmb{x}_t\}  \right)$ to be the smallest when the two processes are independent and $\{y_t\}$ is memoryless. This is confirmed in the following Proposition.
 \begin{proposition}
\label{prop:memory2}
Let $\{y_t\}$ and $\{\pmb{x}_t\}$ be two stationary discrete-time stochastic processes such that $\vert h\left(y_t\right) \vert,  \vert h\left(\pmb{x}_t\right) \vert < \infty$. Then, $$\mathbb{PR}\left(\{y_t\} \vert \{\pmb{x}_t\}  \right) \geq \mathbb{PR}\left(\{y_t\}\right) \geq 0.$$ Moreover, $$\mathbb{PR}\left(\{y_t\} \vert \{\pmb{x}_t\} \right) = \mathbb{PR}\left(\{y_t\}\right)$$ if and only if $\{y_t\}$ and $\{\pmb{x}_t\}$ are independent, and $$\mathbb{PR}\left(\{y_t\} \vert \{\pmb{x}_t\} \right) = 0$$ if and only if $\{y_t\}$ and $\{\pmb{x}_t\}$ are independent and $(y_1, \dots, y_T)$ is jointly independent for every $T$.
 \end{proposition}
\begin{proof}
Similar to the proof of Proposition \ref{prop:memory}.
\end{proof}
As in the one-dimensional case, the measure of predictability of $\{y_t\}$ using $\{\pmb{x}_t\}$ can also be interpreted as the rate of KL-divergence between the joint process $\{y_t, \pmb{x}_t\}$ and the process whose distribution is identical to that of $\{y_t, \pmb{x}_t\}$, except that the first coordinate process (corresponding to  $\{y_t\}$) is independent from the other ones (corresponding to $\{\pmb{x}_t\}$) and memoryless.

\subsection{\textbf{\textsc{Illustration}}}
Let us consider the AR(1) time series $$y_t = \phi y_{t-1} + \epsilon_t.$$ Intuitively, the lower $\phi$, the closer this time series is to its innovation white noise, and therefore the less we would expect the time series to be auto-predictable. Moreover, when the time series is equal to its innovation white noise, that is when $\phi=0$, we would expect it not to be predictable. This is confirmed in Figures (\ref{fig:pred_ar2000}) and (\ref{fig:pred_ar10000}) where we simulated $\{y_t\}$ with a standard Student-t noise term with $\nu=4$ degrees of freedom. Once more, maximum-entropy and nonparametric estimations coincide almost perfectly. As for the model-free approach, it exhibits the right pattern, although it is less data-efficient than the other two approaches in that it has higher estimation variance for the same sample size $T$. When the measure of autopredictability is small, the model-free approach approach is not reliable and can even generate negative estimates; the nonparametric and maximum-entropy approaches should be preferred.
\begin{figure}
  \centering
  \includegraphics[width=0.45\textwidth]{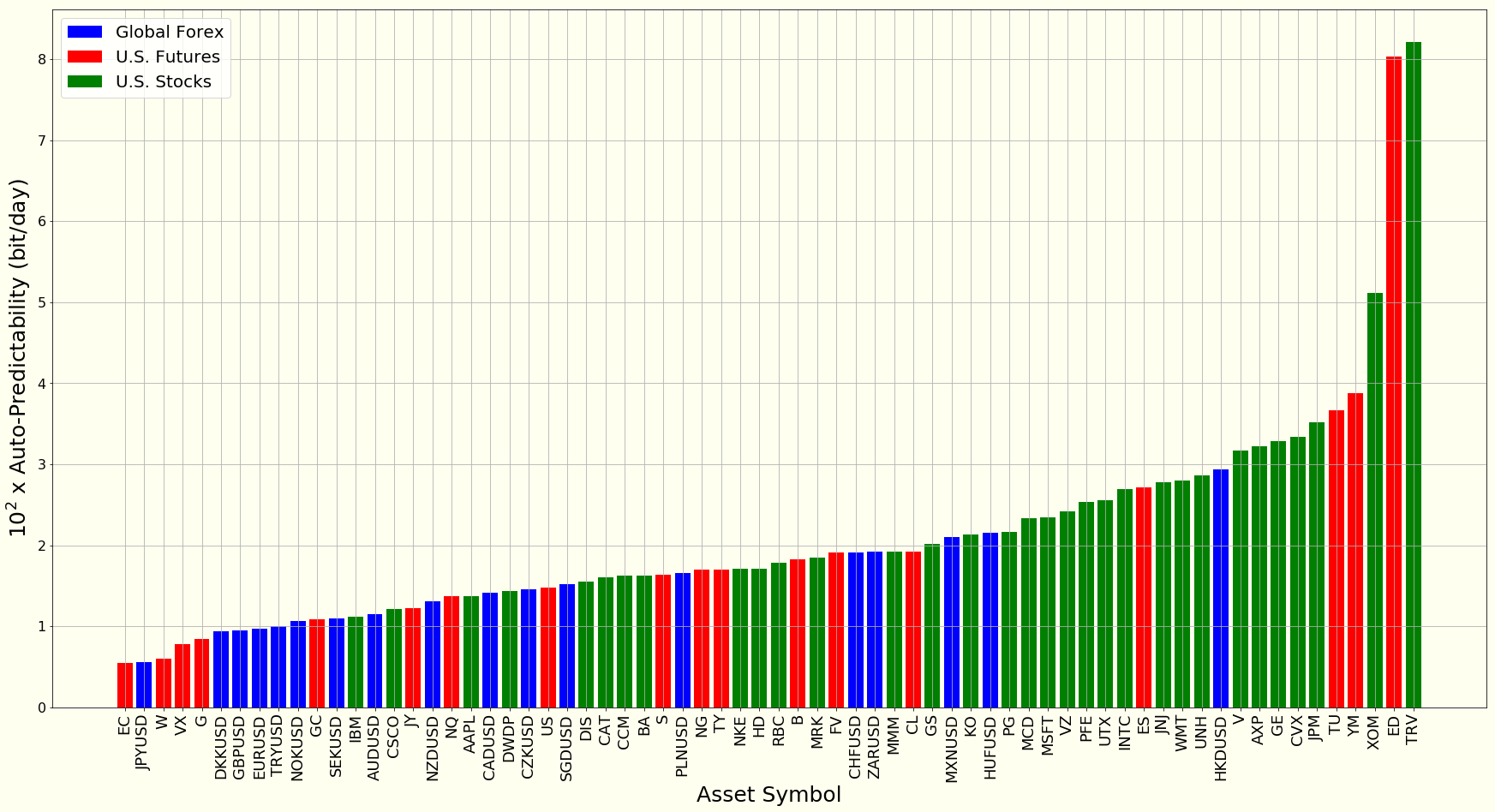}
  \caption{Maximum-entropy estimation of auto-predictability of the close-to-close daily returns of the most actively traded global currencies, U.S. stocks and U.S. futures. Asset symbols to names mappings as well as additional setup details are provided in Appendix \ref{sct:exp_setup}.}
  \label{fig:real_auto_pred}
\end{figure}
\begin{figure*}
  \centering
  \subfloat[$T=2000$]{\label{fig:pred_ar2000}\includegraphics[width=0.4\textwidth]{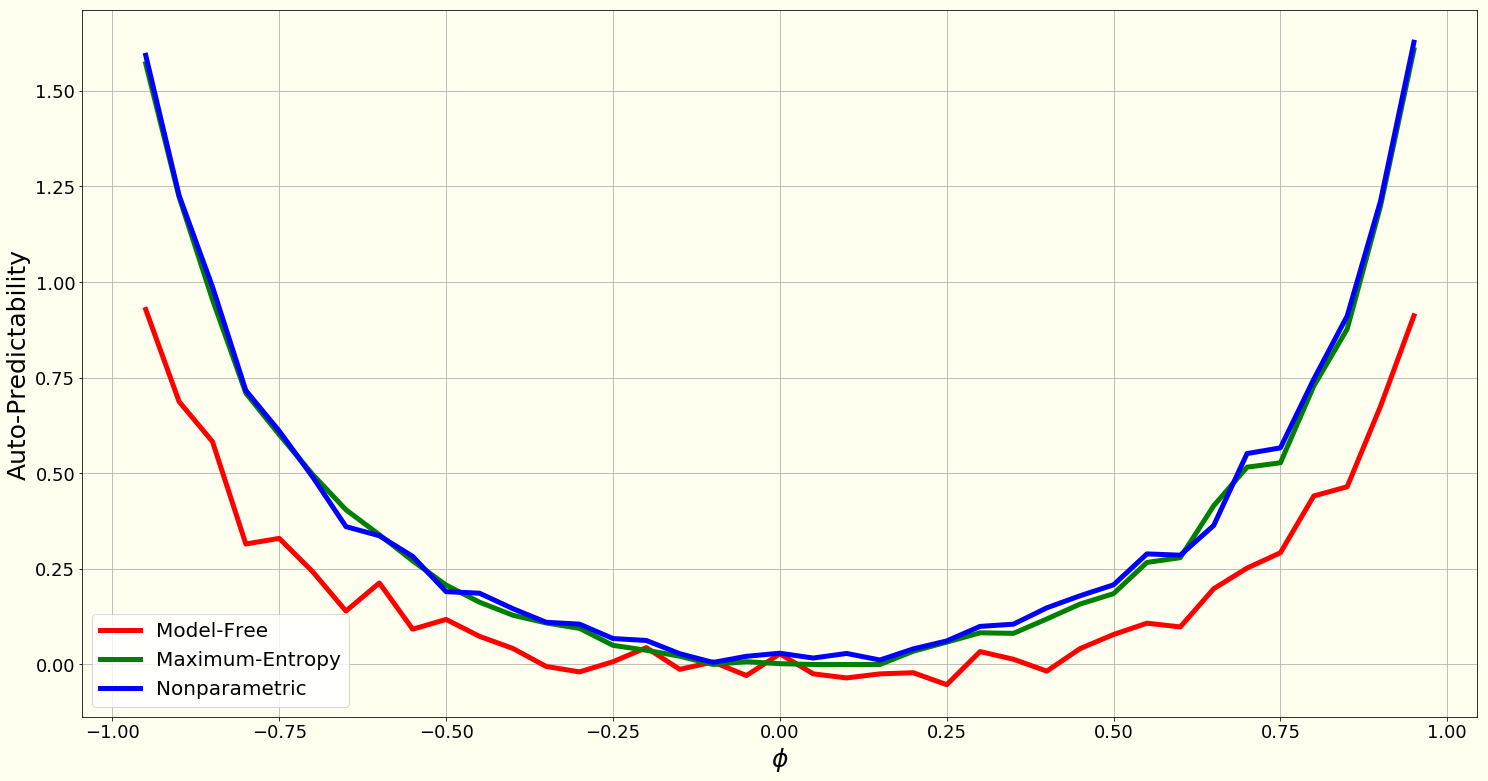}}
    \hfill
  \subfloat[$T=10000$]{\label{fig:pred_ar10000}\includegraphics[width=0.4\textwidth]{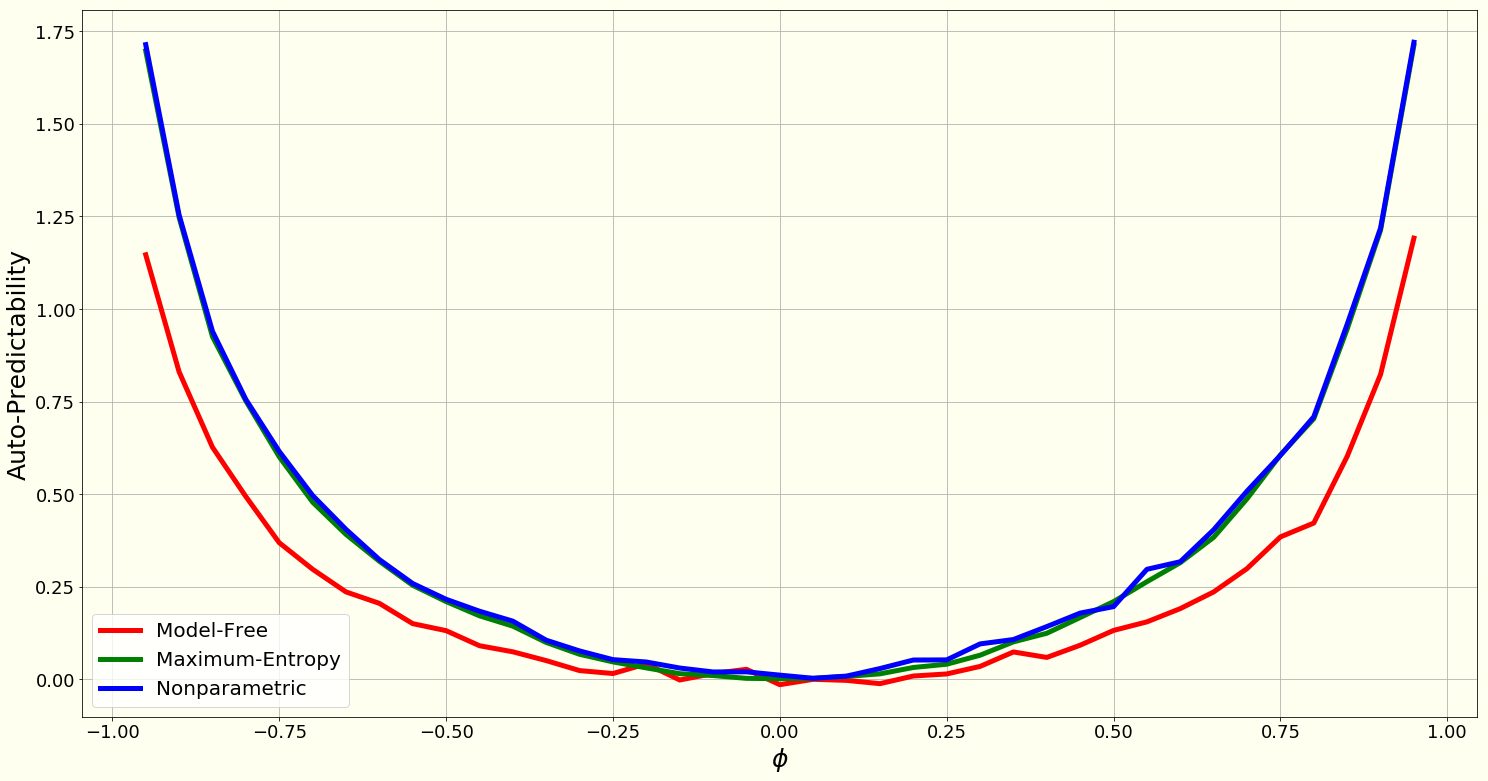}}
  \caption{Measure of auto-predictability of the  AR(1) $y_t = \phi y_{t-1} + \epsilon_t$ with standard Student-t noise with $4$ degrees of freedom. Estimations are based on a sample of size $T$. For the nonparametric estimation, our estimate for the spectral density is obtained using Welch's method \cite{welch1967use} with a Hanning window, a window size equals to $100$, and a $50$\% overlap. For the model-free approach, we set $m$ such that $2^{-m}$ is equal to $1/5$-th of the sample standard deviation.}
\end{figure*}

Next, we consider assessing how predictable the most actively traded global currencies, U.S. blue chip stocks and U.S. futures are. We use as proxy for the most active U.S. blue chip stocks constituents of the Dow Jones Industrial Average. For futures, we select the 20 most active futures by average daily volume over the year 2017 trading on ICE, CME Group and CBOE. Returns are daily close-to-close from January 1st 2008 to January 1st 2018. For currencies we use 5PM ET as cutoff, and for futures we use the exchange settlement price in lieu of close. Results are summarized in Figure (\ref{fig:real_auto_pred}). 

Overall, it can be seen that auto-predictabilities of financial assets are fairly low as expected, much lower than the values obtained in our synthetic experiment (Figures (\ref{fig:pred_ar2000}) and (\ref{fig:pred_ar10000})). However, returns time series are not white noises, they do exhibit memory, some more than others. As it turns out, stocks are more predictable than currencies and futures. The auto-predictability of futures varies a lot from one contract to the other. For instance Wheat futures (W) are one of the least predictable across all three asset classes, whereas Eurodollar futures (ED) are the second most predictable assets out of the $68$ considered. Predictabilities of currencies vary in a much tighter range. 

\section{\MakeUppercase{Quantifying Impact on Distribution Tails}}
\label{section:tails}

Our measure of the impact an asset's returns $\{y_t\}$ could have on the tails of those in a reference pool $\{\pmb{x}_t\}$ should implicitly or explicitly address two concerns: how big are tail events which the new asset can undergo, and how do tail events of the new asset compare to those of the reference pool in magnitude? To facilitate statistical estimation, we once more assume that $\{y_t, \pmb{x}_t\}$ is a jointly stationary and ergodic discrete-time process. 

Traditionally, in the one-dimensional case, whether a distribution has heavy tails is often associated with whether its even moments higher than $2$ are greater than those of the Gaussian distribution with the same mean and variance. Examples such distributions are the so-called \emph{leptokurtic} distributions, defined as distributions whose fourth central moment, also referred to as  \emph{kurtosis}, defined as
\begin{align}
\label{eq:uni_kurt}
\text{Kurt}(y_t) = \mathbb{E} \left[ \left( \frac{y_t - \mathbb{E}(y_t)}{\sqrt{\mathbb{V}\text{ar}(y_t)}}\right)^4 \right]
\end{align}
is higher than that of a Gaussian, which we recall is $3$. This measure was extended to the multidimensional case in \cite{mardia} as 
\begin{align}
\label{eq:mult_kurt}
& \text{Kurt}(\pmb{x}_t) =  \mathbb{E} \left( \psi(\pmb{x}_t)^2  \right)
\end{align}
where $$\psi(\pmb{x}_t) := (\pmb{x}_t - \mathbb{E}(\pmb{x}_t))^T \mathbb{C}\text{ov}\left(\pmb{x}_t, \pmb{x}_t \right)^{-1} \left(\pmb{x}_t - \mathbb{E}(\pmb{x}_t) \right),$$
which we recall is equal to $n(n+2)$ for multivariate normal vectors of length $n$. The kurtosis ratio $$\mathbb{KR}(\pmb{x}_t) := \frac{ \mathbb{E} \left( \psi(\pmb{x}_t)^2  \right)}{n(n+2)}$$ can therefore be regarded as a measure of the tails of returns of assets in the reference pool.

Noting that, when $\pmb{x}_t$ is a multivariate Gaussian, $\psi(\pmb{x}_t)$ follows a $\chi^2$ distribution with $n$ degrees of freedom, and recalling that the $p$-th moment of a $\chi^2$ distribution with $n$ degrees of freedom is $\prod_{i=0}^{p-1} (n + 2i),$ it follows that the kurtosis ratio can be generalized to higher moments into the \emph{tail ratio}, which we define as
\begin{align}
\label{eq:tail_ratio}
\mathbb{TR}(\pmb{x}_t; p) := \frac{\mathbb{E} \left( \psi(\pmb{x}_t)^p  \right)}{\prod_{i=0}^{p-1} (n+ 2i)}, ~~ p \geq 2.
\end{align}
The kurtosis ratio corresponds to the special case $p=2$. In general however, the larger $p$, the more sensitive the tail ratio is to extreme events.

A natural approach for quantifying the impact of a new asset on the tails of a reference pool is to compare the tail ratio of the reference pool with and without the new asset, for instance through the difference $$\mathbb{TR}(\pmb{x}_t; p) - \mathbb{TR}(y_t, \pmb{x}_t; p).$$

The main limitation with this idea is that, when the number of assets $n$ in the reference pool is very large, inverting the covariance matrix $\mathbb{C}\text{ov}\left(\pmb{x}_t, \pmb{x}_t \right)$ would be numerically intractable or unstable (i.e. prone to ill-conditioning). To circumvent this limitation, we swap $\pmb{x}_t$ for best replicating portfolio returns $y_t^*:= \pmb{x}_t^T \omega^* +  \left( 1- 1^T\omega^* \right)r_f $,  we swap $y_t$ for the innovation or tracking error $\epsilon_t := y_t - y_t^*$, and we measure impact on tails by comparing the tail ratio of the best replicating portfolio $y_t^*$, with and without the innovation term $\epsilon_t$
\begin{align}
\label{eq:def_it}
\mathbb{IT}(y_t, \pmb{x}_t; p) = \mathbb{TR}(y_t^*; p) - \mathbb{TR}(\epsilon_t, y_t^*; p).
\end{align}

To confirm that $\mathbb{IT}$ indeed measures the impact of the new asset on the existing reference pool, we review a few scenarios. When $(y_t, \pmb{x}_t)$ are jointly Gaussian, so are $(\epsilon_t, y_t^*)$ and $\mathbb{IT}(y_t, \pmb{x}_t; p) = 0$ for all $p \geq 2$. When $\pmb{x}_t$ is Gaussian, so is $y_t^*$, and $\mathbb{TR}(y_t^*; p)=1$ so that $\mathbb{IT}(y_t, \pmb{x}_t; p)$ only reflects the non-Gaussianity of tails of the innovation term $\epsilon_t$.  Reciprocally, when $\mathbb{TR}(y_t^*; p) \neq 1$, this can only be because $\pmb{x}_t$ is non-Gaussian, more precisely because $\pmb{x}_t$ has heavy (resp. light) tails if $\mathbb{TR}(y_t^*; p) > 1$ (resp. $\mathbb{TR}(y_t^*; p) < 1$). More generally, $\mathbb{IT}(y_t, \pmb{x}_t; p) < 0$ holds when the innovation of the new asset (relative to the reference pool) makes its tails heavier and $\mathbb{IT}(y_t, \pmb{x}_t; p) \geq 0$ otherwise. 

Our working assumption is that, when $\{\epsilon_t\} \neq 0$, the extent to which the tails of $y_t$ impact those of $\pmb{x}_t$ is fully reflected in the best replicating portfolio and the tracking error, and we use $\mathbb{IT}(y_t, \pmb{x}_t; p)$ to quantify the impact on portfolio tails. When $\{\epsilon_t\} = 0$, the new asset can be perfectly replicated using existing assets and consequently it cannot impact tails of the reference pool. In such a case, by convention we set $$\mathbb{IT}(\beta^T\pmb{x}_t, \pmb{x}_t; p)=0$$ for every $\beta$ and $p$.

As for estimation, $\mathbb{IT}(y_t, \pmb{x}_t; p)$ can be estimated in a consistent, fast, and robust manner, by noting that, thanks to our stationary ergodic assumption, the expectation in Equation (\ref{eq:tail_ratio}) can be replaced by sample average, and that $\epsilon_t$ and $y_t^*$ are decorrelated, so that only variances need to be estimated, which can be done consistently using sample variances.

\subsection{\textbf{\textsc{Illustration}}}
We compare the tails of stocks and currencies. As proxy for stocks we use constituents of the Dow Jones Industrial Average (DJIA), and as proxy for currencies, we use $18$ of the most liquid electronically traded foreign currencies against the U.S. dollar. For each asset we compute daily close-to-close\footnote{We use 5PM ET as daily cutoff for currencies.} returns between January 1st 2008 and January 1st 2018. Ranked sample kurtoses are illustrated in Figure (\ref{fig:it_kurt}). For each currency pair, we compute its impact on the tails of constituents of the DJIA; this is illustrated in Figure (\ref{fig:it_fx_on_dj}). For each constituent of the DJIA, we compute its impact on tails of our universe of currencies; this is illustrated in Figure (\ref{fig:it_dj_on_fx}). 

The excessive kurtosis of CHFUSD can be partly attributed to the $-9\%$ daily move in September 2011 due to the Swiss Franc starting to peg the Euro, and the $17\%$ daily move in January 2015 due the Swiss Franc unpegging the Euro. Considering how unusual such moves and the corresponding sample kurtosis (250) are, we would expect CHFUSD to have an adverse impact on the tails of DJIA constituents; this is indeed captured by our measure of tail impact as can be seen in Figure (\ref{fig:it_fx_on_dj}). Similarly, Cisco Systems, Inc. (CSCO), which has the second highest sample kurtosis, is found by our measure of tail impact to have the worst impact on the tails of our basket of currencies  as illustrated in Figure (\ref{fig:it_dj_on_fx}). In general however, comparing sample kurtoses is not a granular or accurate enough approach to assessing impact on tails. For instance, when the sample kurtosis of the new asset is approximately equal to the median sample kurtosis of assets in the reference pool, it is not clear from sample kurtoses alone whether the new asset will positively or negatively impact tails. This is the case for JPYUSD and PLNUSD, whose sample kurtoses ($7.21$ and $7.72$ respectively) are the closest to the median sample kurtosis of DJIA constituents, namely $7.62$. Our approach on the other hand is granular enough to find that, although their kurtoses are the closest to the median kurtoses of DJIA constituents, JPYUSD has the most beneficial impact on DJIA tails, while PLNUSD has one of the worst impacts on DJIA tails. This is because tail events of JPYUSD are mostly reflected in those of DJIA constituents, while tails of PLNUSD are more idiosyncratic. Interestingly, there are $12$ currency pairs whose sample kurtoses are lower than that of JPYUSD, but that have a worse impact on DJIA tails than JPYUSD.

\begin{figure*}
  \centering
  \subfloat[Ranked sample kurtoses of various U.S. blue chip stocks and currency pairs.]{\label{fig:it_kurt}\includegraphics[width=0.32\textwidth]{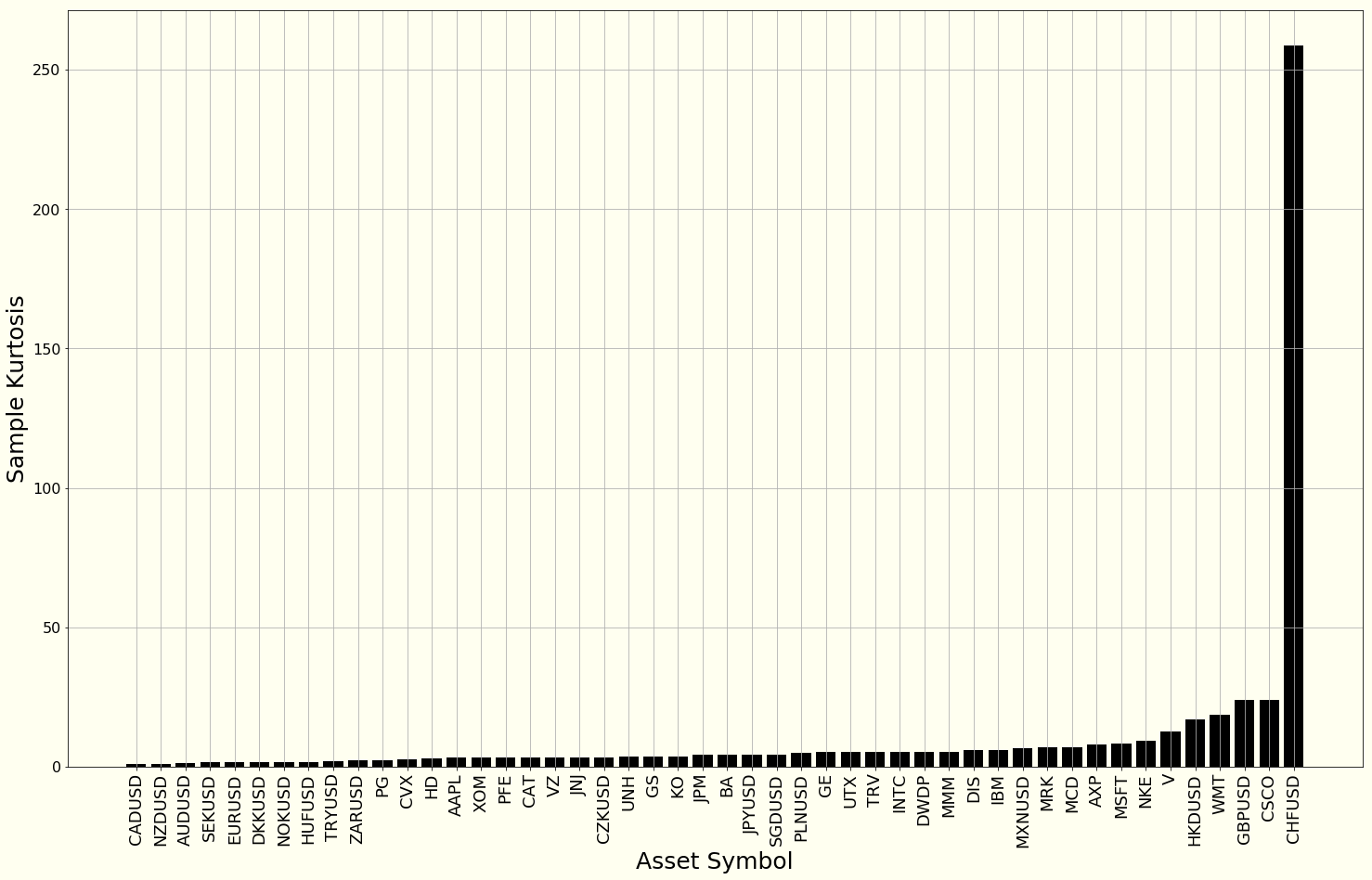}}
      \hfill
  \subfloat[Impact of $18$ of the most liquid electronically traded foreign currencies on the tails of constituents of the DJIA.]{\label{fig:it_fx_on_dj}\includegraphics[width=0.32\textwidth]{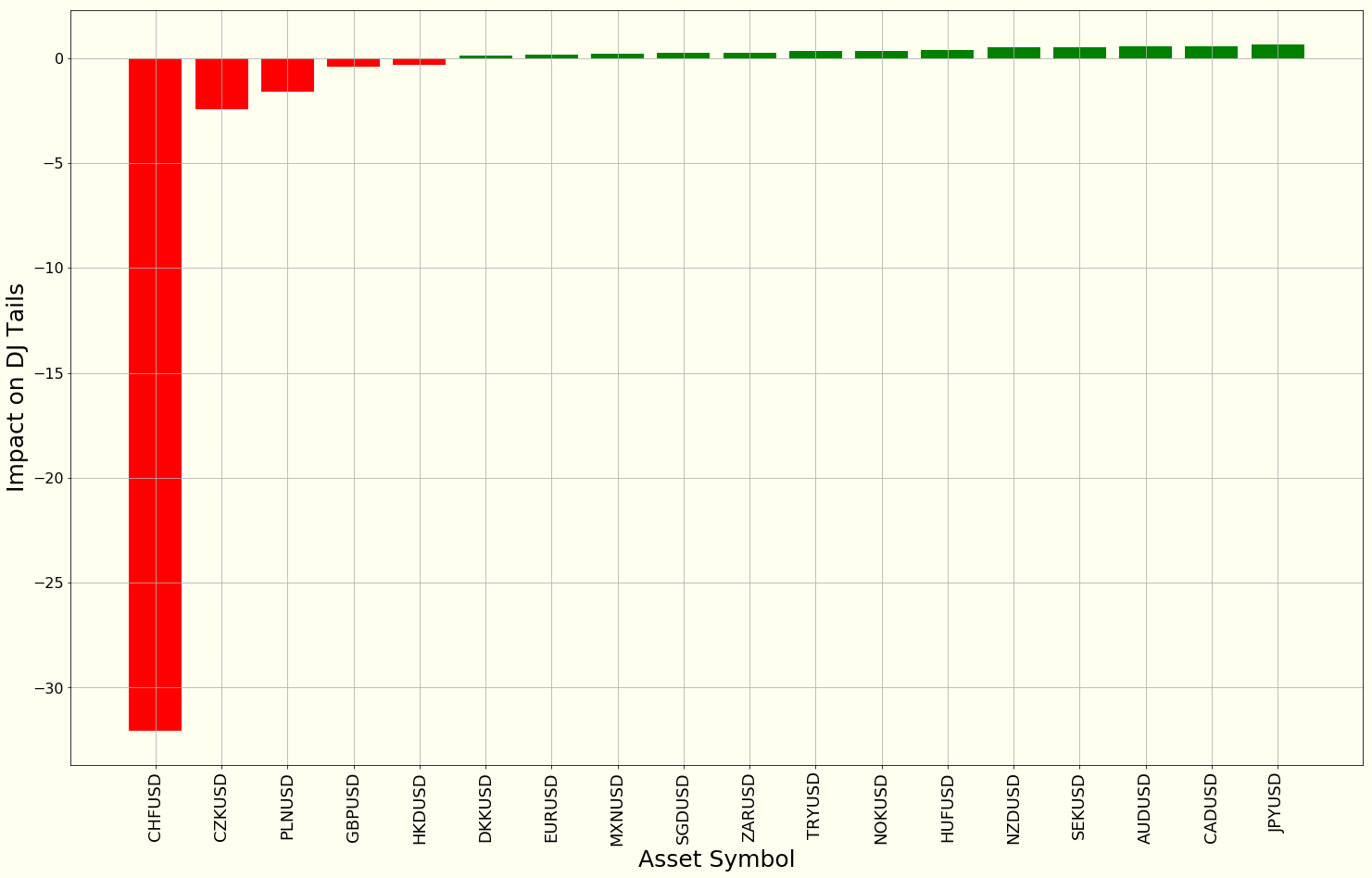}}
    \hfill
  \subfloat[Impact of each DJIA constituent on the tails of a basket of $18$ of the most liquid electronically traded foreign currencies against the U.S. dollar.]{\label{fig:it_dj_on_fx}\includegraphics[width=0.32\textwidth]{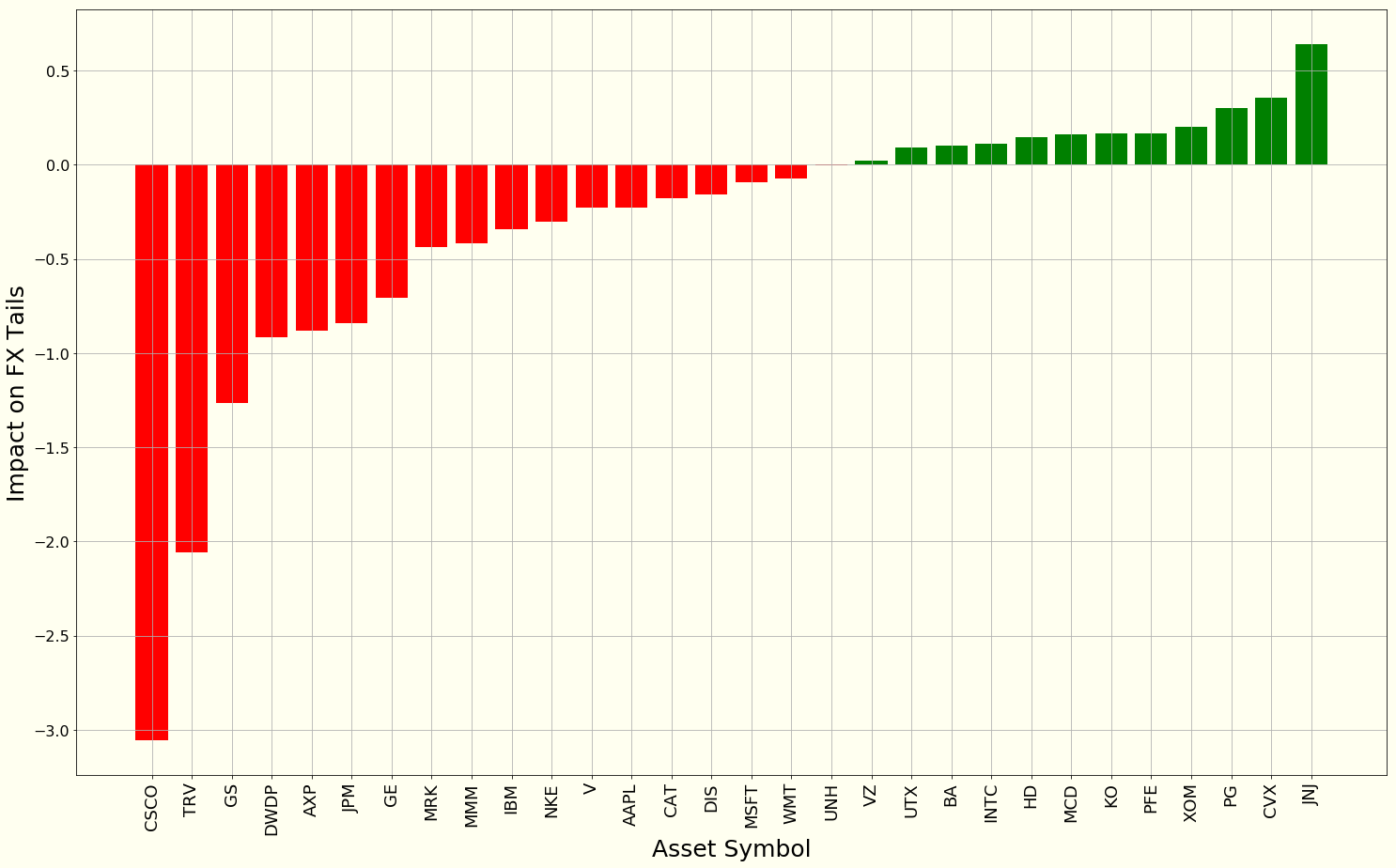}}
  
  \caption{Analysis of tails of stocks and currencies. All assets are characterized by their daily close-to-close returns. Impacts on tails are evaluated using Equation (\ref{eq:def_it}) with $p=2$. Asset symbols to names mappings as well as additional setup details are provided in Appendix \ref{sct:exp_setup}.}
\end{figure*}

\section{\MakeUppercase{Quantifying Suitability for Passive Investment}}
\label{section:passive}

Whether a new asset is suitable for passive investment or not boils down to whether one can achieve a decent level of risk-adjusted returns without changing the investment decision too often, an extreme example of which being buying and holding the asset or short-selling the asset and holding onto the short position for a long time. As a usefulness criteria, suitability for passive investment primarily caters to very large investment managers such as pension funds, social security funds and the likes who, because of their size, cannot afford to be active in the market too often, at the risk of paying excessive transaction costs and eroding their returns as a result. What qualifies as `decent' risk-adjusted returns can therefore be motivated by risk-adjusted returns of assets pension funds and similarly large investment managers are fond of, such as blue chip stocks, ETFs, index funds and other passive funds.

In the interest of providing a unified treatment of both long and short passive strategies, we introduce the \emph{bidirectional Sharpe ratio} of an asset A with stationary returns series $\{y_t\}$, which we define as
\begin{align}
\label{eq:bidirectional_sharpe}
\text{BSR}(A) := \frac{\vert \mathbb{E}(y_t) \vert -r_c- r_f}{\sqrt{\mathbb{V}\text{ar}(y_t)}},
\end{align}
where $\mathbb{E}(y_t)$ is the expected gross return of the asset (i.e. prior to any cost such as transaction cost, exchange and brokerage fees, short-selling borrowing cost etc.), $r_c$ is the total operating cost incurred per time period and per unit of wealth that can be attributed to holding the asset long when $\mathbb{E}(y_t) > 0$, and short when $ \mathbb{E}(y_t) < 0$, and $r_f$ is the risk-free rate. When $\mathbb{E}(y_t) < 0$, buying and holding the asset is a losing strategy but, because $y_t$ is the gross/frictionless return (all operating costs are captured in $r_c$), selling the asset would actually be a winning strategy gross of operating costs,\footnote{Which in this case could include borrowing cost, bid-ask spread, brokerage cost etc.} the expected return net of operating costs would read $$-\mathbb{E}(y_t) - r_c,$$ and the expected net excess return over the risk-free rate would read $$-\mathbb{E}(y_t) - r_c-r_f.$$When $\mathbb{E}(y_t) > 0$, buying and holding the asset is a winning strategy gross of operating costs,\footnote{Which in this case could include cost of carry, bid-ask spread, brokerage cost etc.} and the expected net excess return over the risk-free rate reads $$\mathbb{E}(y_t) - r_c-r_f.$$ Overall, the \emph{bidirectional Sharpe ratio} therefore represents a measure of the best expected net excess return per unit of risk that can be obtained by passively investing in an asset, long or short.

Our working assumption is that the extent to which an asset is suitable for passive investment is the extent to which its bidirectional Sharpe ratio is typical of those of a reference pool of assets known to be sought after by large investment managers. Quantifying suitability for passive investment therefore requires accurately estimating a bidirectional Sharpe ratio, and quantifying its `similarity' to those of assets that appeal to large asset managers. 

A critical issue needs addressing as part of our approach to quantifying suitability for passive investment. In the interest of clarity, let us consider a statistical test of suitability for passive investment that has false positive probability $p$, that is, the probability that the test concludes that an asset is suitable for passive investment when in fact it is not. When the test is run independently for $m$ assets, the probability to misclassify at least one asset as suitable for passive investment when it is not is $\left(1- \left(1-p\right)^m\right) \approx mp$ where the approximation holds for small $p$. When the $m$-th test is attempted \emph{because} the first $m-1$ failed, the false positive probability on the $m$-th test is no longer $p$, but increases to $\left(1- \left(1-p\right)^m\right) \approx mp$. This situation routinely occurs in the investment process, when fund managers and portfolio managers explore a wide range of strategies, most of which are discarded on the basis of their weak risk-adjusted performances, \emph{until} they find one that appears well suited to their needs. Thus, a good measure of suitability for passive investment should therefore adjust for the number of failed attempts, if any, that led to the selection of the asset. It is not so much that our \emph{estimation} of the bidirectional Sharpe ratio can be corrupted by multiple trials, but rather that our \emph{use} of it to infer suitability for passive investment should account for the number of failed trials previously attempted. This issue is commonly referred to as \emph{backtest overfitting}, and is addressed in Section \ref{sct:testing_spi}.\\

\subsection{\textbf{\textsc{Estimating Bidirectional Sharpe Ratios}}}
\label{sct:stat_meaningfulness}
We estimate  a bidirectional Sharpe ratio by replacing $\mathbb{E}(y_t)$ and $\mathbb{V}\text{ar}(y_t)$ in Equation (\ref{eq:bidirectional_sharpe}) by their sample estimates. It follows from our ergodic assumption that the resulting estimator is consistent. In order to make the bidirectional Sharpe ratio independent of the sample frequency, we prefer estimating the annualized bidirectional Sharpe ratio, which we obtain by computing monthly returns, which we assume are independent across time, using Equation (\ref{eq:bidirectional_sharpe}) to estimate the monthly bidirectional Sharpe ratio, and then multiplying the monthly bidirectional Sharpe ratio by $\sqrt{12}$. We prefer working with monthly returns and normalizing by $\sqrt{12}$ over working with daily returns and normalizing by $\sqrt{252}$ because the latter requires assuming independence of daily returns which, as can be seen in Figure (\ref{fig:real_auto_pred}), rarely holds in practice. When returns exhibit positive auto-correlation, which the success of momentum strategies \cite{chan1996momentum, hong2000bad, jegadeesh2001profitability} would imply occurs often, the annualized bidirectional Sharpe ratio estimator using daily returns will overshoot. We find that monthly returns of stocks, currencies and futures exhibit little to no auto-predictability.

\subsection{\textbf{\textsc{Characterizing the Bidirectional Sharpe Ratio of Passive Investments}}}
Once the bidirectional Sharpe ratio of the new asset has been estimated, we need to determine how similar it is to those of assets we know to be suitable for passive investment. To do so, we assume that all assets suitable for passive investment are alike in the sense that their bidirectional Sharpe ratios are independently drawn from the same latent distribution, which we aim to empirically estimate from data. To this end, we adopt a Bayesian approach. Every asset suitable for passive investment is assumed to have bidirectional Sharpe ratio $r$ that is independently drawn from the same distribution. We place as prior on $r$
\begin{align}
\label{eq:r_prior}
r \vert \mu, \tau \sim \mathcal{N}\left(\mu, \tau^{-1}\right),
\end{align}
which we complement with the conjugate Normal-Gamma prior on $(\mu, \tau)$, which we recall means that
\begin{align}
\label{eq:bmu_prior}
\mu \vert \tau  \sim \mathcal{N}\left(\mu_0, \left(\tau \nu_0\right)^{-1}\right),
\end{align}
and 
\begin{align}
\label{eq:btau_prior}
\tau \sim \text{Gamma}\left(\alpha_0, \beta_0\right).
\end{align}
Upon estimating the bidirectional Sharpe ratios $\hat{r}_1, \dots, \hat{r}_n$ of $n$ assets that are known to be suitable for passive investment (e.g. passive funds, stocks and bonds ETFs, blue chip stocks, etc.), the predictive distribution $$r \vert \hat{r}_1, \dots, \hat{r}_n,$$ forms our best guess, in light of observed data, about the characteristic distribution of bidirectional Sharpe ratios of assets that are suitable for passive investment. 

We recall that the predictive distribution is available in closed-form and reads
\begin{align}
\label{eq:posterior_bsr_2}
r \vert \hat{r}_1, \dots, \hat{r}_n \sim t_{2\alpha_n}\left( \mu_n, s_n \right),
\end{align}
where 
\begin{align}
s_n =& ~ \frac{\beta_n (\nu_n + 1)}{\alpha_n \nu_n} \\
\mu_n =& ~\frac{\nu_0 \mu_0 +  n \bar{r}_n}{\nu_0 + n}\nonumber \\
\nu_n =& ~\nu_0 + n \\
\alpha_n =& ~\alpha_0 + \frac{n}{2} \\
\beta_n =& ~\beta_0 + \frac{1}{2} \sum_{i=1}^n \left(\hat{r}_i - \bar{r}_n \right)^2 \nonumber \\
&+ \frac{n \nu_0}{\nu_0 + n} \frac{\left(\bar{r}_n - \mu_0 \right)^2}{2},
\end{align}
with $$\bar{r}_n = \frac{1}{n}\sum_{i=1}^n \hat{r}_i,$$ and where $t_u(v, w)$ is the Student-t distribution with $u$ degrees of freedom, location parameter $v$, and scale parameter $w$. The associated probability density function therefore reads
\begin{align}
\label{eq:pred_distro}
p(r \vert \hat{r}_1, \dots, \hat{r}_n) =& \frac{1}{\sqrt{2\pi \alpha_n}} \frac{\Gamma \left( \alpha_n  + \frac{1}{2}\right)}{\Gamma\left(\alpha_n \right)} \\
\times &\left( 1 + \frac{1}{2\alpha_n} \left( \frac{r-\mu_n}{s_n}\right)^2\right)^{-\frac{2\alpha_n +1 }{2}}, \nonumber
\end{align}
where $\Gamma$ is the gamma function.

Prior parameters $(\nu_0, \alpha_0, \beta_0)$ can be set to express uninformativeness. Moreover, we recommend setting $\mu_0 = 0$ so as to avoid expressing (a priori) that assets suitable for passive investment should be expected to return more or less than the risk-free rate, net of operating costs, and rather rely on the data to determine (a posteriori) whether assets suitable for passive investment typically outperform the risk-free rate.

\subsection{\textbf{\textsc{Testing for Suitability for Passive Investment}}}
\label{sct:testing_spi}
Once the distribution of bidirectional Sharpe ratios of assets known to be suitable for passive investment has been estimated as the posterior predictive distribution $$r \big\vert \hat{r}_1, \dots, \hat{r}_n $$ of Equation (\ref{eq:posterior_bsr_2}), we are ready to test whether a new asset A is suitable for passive investment. \\

\noindent \textbf{One Trial Allowed}: We begin by assuming that A is the only asset that we will put to our suitability for passive investment test. 

In general, the log-predictive posterior $$\log_e p\left( r \big\vert \hat{r}_1, \dots, \hat{r}_n \right)$$ reflects the `likelihood' that a bidirectional Sharpe ratio $r$ is consistent with observations $\hat{r}_1, \dots, \hat{r}_n$, and therefore the extent to which the asset whose bidirectional Sharpe ratio is $r$ is suitable for passive investment. Denoting $\hat{\text{BSR}}(A)$ our point estimate of A's bidirectional Sharpe ratio, we define \emph{measure of suitability for passive investment} the log-predictive posterior evaluated at $r=\hat{\text{BSR}}(A)$:
\begin{align}
\label{eq:spi_def}
\mathbb{SPI}(A) := \log_e p\left(\hat{\text{BSR}}(A) \big\vert \hat{r}_1, \dots, \hat{r}_n \right).
\end{align}
As for a statistical hypothesis test of suitability for passive investment, we note that
\begin{align}
\label{eq:stat_test}
\mathbb{P} \left( r \leq \hat{\text{BSR}}(A) \big\vert \hat{r}_1, \dots, \hat{r}_n \right),
\end{align}
reflects the probability that an asset suitable to passive investment, as per observations $\hat{r}_1, \dots, \hat{r}_n$, presents less passive investment opportunities than A, or equivalently the probability that A presents more passive investment opportunities than what would be expected of an asset suitable for passive investment, as per $\hat{r}_1, \dots, \hat{r}_n$. Whence, the fact that the probability in Equation (\ref{eq:stat_test}) is very small can be regarded as an indication that asset A is not suitable for passive investment. Formally, we define the following statistical hypothesis test:
\begin{align}
\label{eq:suit_test}
\begin{cases}
&H_0: \text{Asset A is suitable for passive investment}. \\
&p: p\text{-value}. \\
&\text{Decision: Reject } H_0 \text{ when } \\ 
&\mathbb{P} \left( r \leq \hat{\text{BSR}}(A) \big\vert \hat{r}_1, \dots, \hat{r}_n \right) < p. 
\end{cases}
\end{align}
The $p$-value $p$ is the probability that our test makes a false positive (or type I) error,\footnote{That is, in this case, the probability of rejecting the null hypothesis of suitability for passive investment when it is in fact true.} and should therefore be set to a small value, for instance $5\%$.\\

\noindent \textbf{Multiple Trials Allowed}: When running multiple tests of suitability for passive investment, care should be taken while assessing the false positive or type I error rate. 

If $m$ tests of suitability for passive investment are independently run on $m$ assets, then the overall false positive rate, defined as the expected number of false positive errors divided by the number of tests, remains the test's $p$-value. In practice however, this might not be the best metric to rely on to measure the efficacy of our testing procedure, as we will typically ignore assets that are not deemed suitable for passive investment, and only act on the rest. Thus, a more reliable measure of accuracy should focus on errors we make on test results we do act on.

If a test to act on is chosen uniformly at random among all $m$ tests, then the expected false positive rate is also $p$. In general, nevertheless, when the test result to act on is selected among all available $m$ tests using a different strategy, one cannot conclude. To see why, let us denote $t_1, \dots, t_m$ the independent Bernoulli random variables such that $t_i=1$ if test $i$ does not reject $H_0$, and $t_i=0$ otherwise, which we denote passive investment indicator. If a selection strategy observes some or all test results in order to choose which test to rely on, then its passive investment indicator is a Bernoulli random variable that takes form 
$$t_* = \mathcal{S}\left( t_1, \dots, t_m \right),$$
for some $\mathcal{S}: \{0, 1\}^m \to \{0, 1\}$. Clearly, the probability that $t_* = 1$, and consequently the false positive rate, depends on $\mathcal{S}$, despite the fact that $t_i$ are i.i.d. Whence, our statistical test needs to be adapted to have a known and configurable false positive rate for any number of trials $m$.

In order to adapt our statistical test to multiple trials, let us consider a selection strategy typical of an investment manager looking for a new investment opportunity. We assume the investment manager keeps testing assets until he/she finds one that is suitable for passive investment. Let us assume $m$ assets have been independently tested by the investment manager. A necessary (but not sufficient) condition for the $m$-th asset $A_m$ to pass the test of suitability for passive investment, providing the previous ones $A_1, \dots, A_{m-1}$ failed, is that its bidirectional Sharpe ratio be the largest
\begin{align}
\hat{\text{BSR}}(A_m) = \underset{i \leq m}{\max} ~ \hat{\text{BSR}}(A_i).
\end{align}
Thus, $$ \mathbb{P} \left( r \leq \underset{i \leq m}{\max} ~ \hat{\text{BSR}}(A_i) \big\vert \hat{r}_1, \dots, \hat{r}_n \right)$$ is ill-suited to measure the likelihood that the $m$-th asset is suitable for passive investment as it compares the best of $m$ independent attempts at finding an asset suitable for passive investment to a single attempt $r$ at generating a bidirectional Sharpe ratio similar to the observed $\hat{r}_1, \dots, \hat{r}_n$. A fairer comparison would be between the best of $m$ independently drawn bidirectional Sharpe ratios $r_i$ similar to the observed $\hat{r}_1, \dots, \hat{r}_n$, and the best of $m$ independent attempts at finding an asset suitable for passive investment:
\begin{align}
& \mathbb{P} \left( \underset{i \leq m}{\max} ~ r_i \leq \underset{i \leq m}{\max} ~ \hat{\text{BSR}}(A_i) \big\vert \hat{r}_1, \dots, \hat{r}_n \right) \nonumber \\
& =  \mathbb{P} \left( \bigcap\limits_{i=1}^m  \left( r_i \leq  \hat{\text{BSR}}(A_m) \right) \big\vert \hat{r}_1, \dots, \hat{r}_n \right) \nonumber \\
& = \prod_{i=1}^m  \mathbb{P} \left( r_i \leq  \hat{\text{BSR}}(A_m)  \big\vert \hat{r}_1, \dots, \hat{r}_n \right) \nonumber \\
& = \mathbb{P} \left( r \leq  \hat{\text{BSR}}(A_m)  \big\vert \hat{r}_1, \dots, \hat{r}_n \right)^m .
\end{align}

Hence, the investment manager willing to test whether the best of $m$ assets chosen at random is suitable for passive investment should use the following statistical test:
\begin{align}
\label{eq:suit_test_2}
\begin{cases}
&H_0: \text{The best of } m \text{ assets } A_1, \dots, A_m,\\ 
&\text{ independently chosen at random, is}  \\
&\text{ suitable for passive investment.}  \\
&p: p\text{-value}. \\
&\text{Decision: Reject } H_0 \text{ when }  \\ 
& \mathbb{P} \left( r \leq  \underset{i \leq m}{\max} ~ \hat{\text{BSR}}(A_i) \big\vert \hat{r}_1, \dots, \hat{r}_n \right) < p^{\frac{1}{m}}. 
\end{cases}
\end{align}

\begin{remark}
We stress that the false positive or type I error rate of the test above is always $p$, irrespective of $m$. When the asset with the highest bidirectional Sharpe ratio is the $m$-th asset $A_m$, accounting for previously failed attempts (Test (\ref{eq:suit_test_2})) is essentially the same as assuming a single trial (Test (\ref{eq:suit_test})), except that suitability for passive investment is rejected at the considerably higher threshold $p^{\frac{1}{m}}$. When $$\mathbb{P} \left( r \leq  \hat{\text{BSR}}(A_m) \big\vert \hat{r}_1, \dots, \hat{r}_n \right) < p,$$ or  $$\mathbb{P} \left( r \leq  \hat{\text{BSR}}(A_m) \big\vert \hat{r}_1, \dots, \hat{r}_n \right) \geq p^{\frac{1}{m}},$$ both tests agree. However, when $$\mathbb{P} \left( r \leq  \hat{\text{BSR}}(A_m) \big\vert \hat{r}_1, \dots, \hat{r}_n \right) \in [p, p^{\frac{1}{m}}[,$$ both tests disagree, and the one not accounting for previously failed attempts is wrong---this is the manifestation of so-called \emph{backtest overfitting}. To see how pervasive this issue is, we note that when $p=0.05$ and $m=10$, $[p, p^{\frac{1}{m}}[ = [0.05, 0.74[$! When $m=50$, the overfitting range $[p, p^{\frac{1}{m}}[$ widens to $[0.05, 0.94[$! 
\end{remark}

The measure of suitability for passive investment $\mathbb{SPI}$ previously introduced can also be extended to the multiple-trials case as the log-predictive posterior $$\log_e p\left( r^* \big\vert \hat{r}_1, \dots, \hat{r}_n \right)$$ where $r^* := \underset{i \leq m}{\max} ~ r_i$ and $r_i$ are i.i.d. drawn from the posterior distribution $r \big\vert \hat{r}_1, \dots, \hat{r}_n$. Specifically, denoting $F\left(r \big\vert \hat{r}_1, \dots, \hat{r}_n \right)$ the cumulative density function associated to predictive density $p\left(r \big\vert \hat{r}_1, \dots, \hat{r}_n\right)$ (Equation (\ref{eq:pred_distro})), we define measure of suitability for passive investment of the best of $m$ assets $A_1, \dots, A_m,$ the quantity
\begin{align}
&\mathbb{SPI}(A_1, \dots, A_m) \\
:&= \log_e \frac{\partial F\left(r \big\vert \hat{r}_1, \dots, \hat{r}_n \right)^m}{\partial r} \Bigg\vert_{r= \underset{i \leq m}{\max} ~ \hat{\text{BSR}}(A_i)} \nonumber \\
&= \log_e p\left( \underset{i \leq m}{\max} ~ \hat{\text{BSR}}(A_i) \big\vert \hat{r}_1, \dots, \hat{r}_n \right) + \log_e m  \nonumber \\
& + (m-1) \log_e F\left( \underset{i \leq m}{\max} ~ \hat{\text{BSR}}(A_i) \big\vert \hat{r}_1, \dots, \hat{r}_n \right). \nonumber
\end{align}

The previous analysis was based on a selection strategy that keeps testing assets for suitability for passive investment until one such asset is found. When the investment manager does not stop at the first asset suitable for passive investment, but instead continuously tests assets/strategies, selecting the ones suitable for passive investment and discarding the others, the same analysis can be applied. However, instead of $A_1, \dots, A_m$ representing all assets tested thus far, they should be all assets tested since the last asset suitable for passive investment was found.

\begin{remark}
We consciously make the conservative/overly penalizing assumption that bidirectional Sharpe ratios of tested assets are independent. In practice, we understand that assets tested by the investment manager might be positively correlated, for instance when optimizing a parametric family of trading strategies. Dealing with the general case accounting for dependencies between tests is beyond the scope of this paper. We note however that, when the setup of the investment manager allows for the estimation of the effective sample size $m_\text{eff}$ (i.e. the number of independent assets corresponding to $A_1, \dots, A_m$), our approach can be used as is, with $m_\text{eff}$ instead of $m$. That being said, we are of the opinion that statistical hypothesis testing should be used parsimoniously, and in particular, care should be taken to only test a new asset if it is sufficiently unrelated to previously tested assets.
\end{remark}

\subsection{\textbf{\textsc{Illustration}}}
To illustrate our approach, we use as reference set of assets suitable for passive investment U.S. blue chip stocks, specifically, constituents of the Dow Jones Industrial Average at the time of writing of this paper. Figure (\ref{fig:spi_dj_post}) illustrates the posterior distributions $p(r \vert \hat{r}_1, \dots, \hat{r}_n)$ for various cost and risk-free assumptions.
\begin{figure}
\centering
\includegraphics[width=0.45\textwidth]{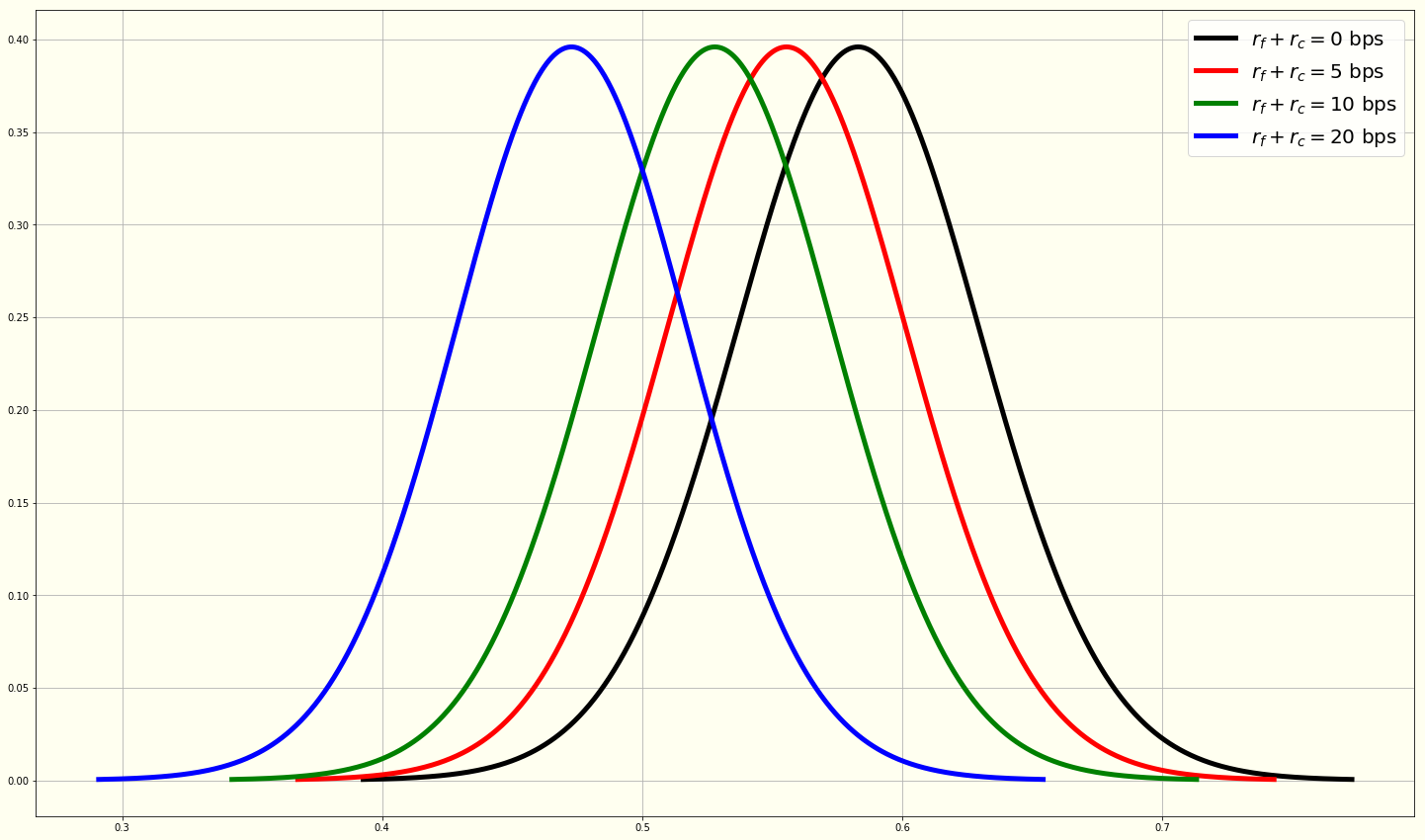}
\caption{Posterior distributions of the annualized bidirectional Sharpe ratio of an asset suitable for passive investments, using as reference set constituents of the Dow Jones Industrial Average, and under various monthly cost and risk-free assumptions. The full list of assets considered as well as additional setup details are provided in Appendix \ref{sct:exp_setup}.}
\label{fig:spi_dj_post}
\end{figure}
Using the foregoing reference set, we consider testing whether the most actively traded global currencies and U.S. futures are suitable for passive investment at a $p$-value of $5\%$, and for an aggregate risk-free rate and operating cost of $10$ basis points per month, and we compute their suitability for passive investment (Equation (\ref{eq:spi_def})). Futures returns are obtained by continuously adjusting the front contract using the proportional back adjustment method, and rolling on the first day of the delivery month.

Interestingly, no currency or currency future\footnote{The only two currency futures here are CME Yen future (JY) and CME Euro future (EC).} is found to be suitable for passive investment. This makes intuitive sense. Indeed, had a foreign currency been found to be suitable for passive investment, this would have suggested that it would have had tendency to either appreciate relative to the U.S. dollar in the long run or tendency to depreciate against the U.S. dollar in the long run. Either way, any significant exchange rate trend would have profound economic implications for the corresponding monetary zone, which would force its central bank to intervene.

Out of the $20$ futures considered, only $6$ were found to be suitable for passive investment, namely the CBOT $10$-year U.S. Treasury Note (TY), the NYMEX Natural Gas (NG), the CBOT Soybeans (S), the CBOT $5$-year U.S. Treasury Note (FV), the CME E-mini Dow Jones, and the CBOE VIX.

\begin{figure}
\centering
\includegraphics[width=0.45\textwidth]{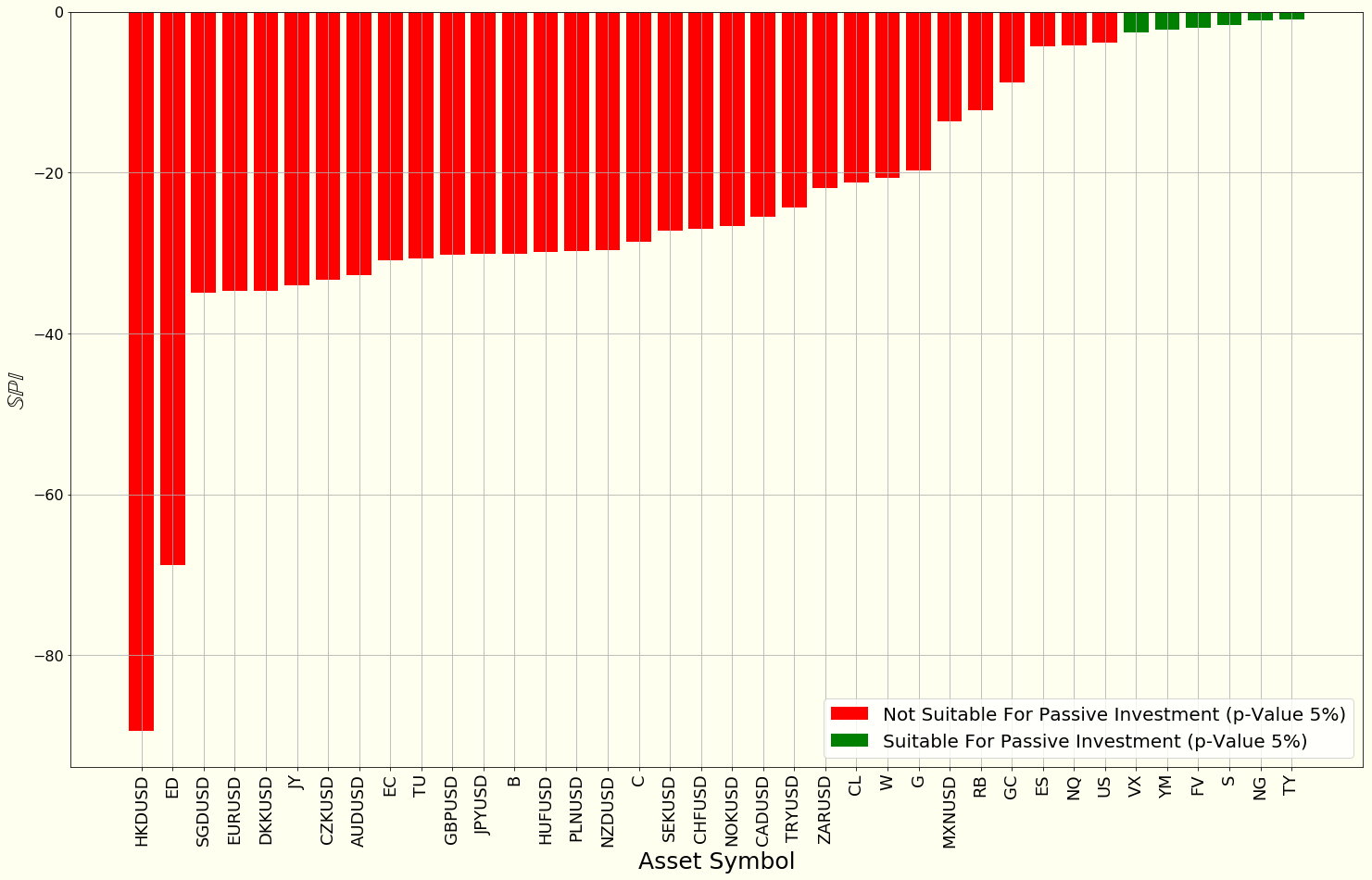}
\caption{Suitability for passive investment (Equation (\ref{eq:spi_def})) of the most actively traded global currencies and U.S. futures, sorted in increasing order. The sum of the monthly risk-free rate and other costs is assumed to be $10$bps. We use as reference assets suitable for passive investment constituents of the Dow Jones Industrial Average. Assets in red (resp. green) are the ones that fail (resp. pass) our one-trial test of suitability for passive investment at $p$-value $5$\%. Symbols to names mappings and additional setup details are provided in Appendix \ref{sct:exp_setup}.}
\label{fig:suit_test}
\end{figure}

\section{\MakeUppercase{Conclusion}}
\label{section:conclusion}

In this paper, we provide a quantitative framework for answering a basic, yet fundamental question: what makes an asset useful to an investment manager? The notion of asset in the aforementioned question includes all investments resulting in a periodic stream of returns, realized or marked-to-market. This allows us to develop a unified framework within which stocks, currencies, fixed income, commodities, ETFs, real-estate, futures, derivatives, LP interest in hedge funds or fund of funds, and any static or dynamic allocation to any combination of these, to name but a few, can be evaluated consistently, provided that the investment manager has a long enough history of returns for the corresponding product. Our framework assumes that an asset is fully characterized by its stream of returns, that two assets that have identical time series of returns are identical for all practical investment purposes, and consequently we solely rely on an asset's time series of returns to answer the foregoing question.\\

\noindent \textbf{\textsc{Summary and Contributions:}} We argue that the usefulness of a new asset to an investment manager is relative to the pool of assets he/she already has access to and factors he/she would like to avoid exposure to. Indeed, if the new asset can easily be replicated using existing assets and factors, intuitively it is of little use as the investment manager can do without.

We identify four key criteria a new asset should exhibit to be considered useful to an investment manager, two primary and two secondary, each corresponding to a motivation an investment manager might have for broadening the universe of assets he/she trades, and all four are independent from the investment manager's asset allocation strategy. 

As primary criteria, we propose that, to be useful, a new asset should sufficiently diversify the pool of assets and factors the investment manager already has access to, and the new asset's returns time series should be sufficiently predictable. These two criteria are primary criteria in that an investment manager, active or passive, would hardly find an asset valuable if it is either redundant (i.e. doesn't provide sufficient incremental diversification) or its returns time series is pure white noise. 

Additionally, we propose as secondary criteria that, to be useful, a new asset should not have an excessive adverse impact on the tails of assets the investment manager currently trades, and it should be suitable for passive investment. The first secondary criteria caters to investment managers interested in broadening the universe of assets they trade so as to mitigate risk concentration, thereby reducing their exposure to idiosyncratic moves. The second secondary criteria caters to large investment managers that, because of their size, might not have the luxury of changing their investment decisions often. We consider the last two criteria secondary in that they are required by some, but not all, investment managers, to consider a new asset useful. We propose a mathematical framework for quantifying all four criteria, and provide scalable algorithmic solutions.

We introduce the \emph{mutual information timescale} as measure of how much incremental diversification a new asset adds to a reference pool of assets and factors. Simply put, the mutual information timescale quantifies the amount of time required to see a bit of mutual/shared information between the new asset and the reference pool; the higher the mutual information timescale, the longer it would take to see a bit of mutual/shared information between the new asset and the reference pool, and consequently the more unrelated the new asset is to the reference pool. Additionally, as a measure of incremental diversification, mutual information timescale satisfies key features of practical importance. Specifically, i) the easier it is to replicate the new asset with the reference pool, the lower the mutual information timescale, ii) mutual information timescale is consistent with the idea, at the core of fund-of-funds, that fund managers trading the same universe of assets can diversify each other, and iii) the amount of incremental diversification a new asset adds to a reference pool is invariant by rescaling any asset (new or old) up or down through leverage and by change of direction (long/short). Incidentally, mutual information timescale can also be used to generalize Pearson's correlation coefficient to capture both nonlinear and temporal dependencies in asset returns; we call the new coefficient \emph{information-adjusted correlation}.

We use as measure of predictability of returns the maximum reduction in uncertainty about future returns that can be achieved by knowing past returns and possibly other set of signals. Crucially, our approach does not make any assumption on how one would go about predicting future values of returns of the new asset. It addresses \emph{how predictable} a returns time series is, independently from \emph{how to best predict} a returns times series.

We measure impact on tails by comparing the tails of returns of the portfolio of assets in the reference pool that best replicates the new asset, to the tails of returns of the replication error/innovation. In simpler terms, we measure whether the `beta' component of the new asset's returns time series with respect to the reference pool has heavier tails than its `alpha'/idiosyncratic component.

As for quantifying suitability for passive investment, we propose proceeding in two steps. First, we quantify how much risk adjusted net return above the risk-free rate one can get by investing passively in the new asset, long or short. We call the corresponding metric bidirectional Sharpe ratio. We build a statistical test to quantify whether the estimated bidirectional Sharpe ratio is `good enough', or equivalently, whether it is typical of assets we know to be suitable for passive investment, for instance U.S. blue chip stocks. Our statistical hypothesis test of suitability for passive investment is robust to backtest-overfitting in that it properly accounts for the number of failed trials previously attempted.

The pertinence of proposed approaches is demonstrated on a wide ranges of experiments on synthetic and real data.\\


\noindent \textbf{\textsc{Empirical Findings:}} By applying the proposed techniques to real data we are able to recover well known stylized facts, as well as new findings.

By comparing our measure of incremental diversification to correlation (in the two-assets case), we find evidence that daily returns of constituents of the S\&P 100 exhibit a nonlinear relationship and/or a relationship that is both cross-sectional and temporal in nature (i.e. exhibits lead-lag or mutual excitation across time), that pairwise correlation cannot capture, but mutual information timescale does.

We find that cross asset class diversification works better than within asset class diversification, as expected. However, how much more incremental diversification one can get through cross asset class diversification, as opposed to within asset class diversification, varies drastically as a function of the asset class(es) to diversify and the diversifying asset class. For instance, we find that using U.S. futures to diversify a basket of actively traded global currencies adds as much value as trading more currencies, but using currencies to diversify U.S. futures adds far more diversification than trading more U.S. futures.

We also find that, as well-known to practitioners, time series of asset returns are not white noises; they have memory and, consequently, are predictable, some more than others. We find that currencies are less predictable than stocks and futures overall, stocks are more predictable than futures overall, and the predictabilities of futures vary significantly from one underlying to the next.

In relation to impact on tails, we find that currencies have the heaviest tails in isolation. However, adding currencies to a basket of U.S. blue chip stocks, more often than not, has a positive impact on tails, while adding blue chips to currencies on average has a negative effect on tails (surprisingly).

As for suitability for passive investment, we find that foreign currencies are (unsurprisingly) not suitable for passive investment, which suggests that one can only make money trading currencies through active management.\\


\noindent \textbf{\textsc{Coming Up:}} Exchange-traded assets, typically represented through ticker symbols, are by far the most commonly used representation of financial markets through which investment managers seek to find investment opportunities. As much as it is the only representation that is consistent with the institutional segmentation of the economy, it is certainly not the only possible tradeable representation of financial markets, and most likely not the tradeable representation of financial markets that is the most useful to investment managers, or said differently, that is the most conduicive to finding investment opportunities.

Indeed, any set of time series of investment decisions $(\{\omega_t^1\}, \dots, \{\omega_t^n\})$, where $\omega_t^i$ denotes a portfolio weight vector,\footnote{That is, a vector representing how much one should invest in buying or selling exchange-traded assets per unit of wealth.} forms a valid tradeable representation of financial markets. Each time series of investment decisions $\{\omega_t^i\}$, when executed, will result in a time series of returns, and consequently can be regarded as an asset. Buying $C>0$ dollars of the asset characterized by $\{\omega_t^i\}$ at time $t$ is equivalent to investing $C$ dollars at time $t$ according to portfolio weights $\omega_t^i$ and, every time the target portfolio weights change, rebalancing accordingly.  Selling $C>0$ dollars of the asset characterized by $\{\omega_t^i\}$ at time $t$ is equivalent to buying $C>0$ dollars of the asset characterized by the portfolio weights time series $\{-\omega_t^i\}$. 

Clearly, there are infinitely many such tradeable representations of financial markets, and it is unlikely that the one defined by ticker symbols is the most useful to investment managers, or the most conducive to finding investment opportunities; companies don't IPO so that hedge funds can generate alpha, government and municipalities don't issue bonds so that hedge funds can generate alpha.

Pit.AI Technologies will use the framework developed in this paper to lauch a marketplace for incrementally constructing an alternative tradeable representation of financial markets, engineered from the ground up to be useful to investment managers, that quantitative investment managers can use as building blocks in their investment process, and where machine learning researchers can earn money for doing provably useful work.  To stay up to date, watch our GitHub repo (https://github.com/devisechain/Devise) and follow us on medium (https://medium.com/pit-ai-technologies).  \\

\bibliographystyle{unsrt}
\bibliography{bibliography}

\documentclass[../main.tex]{subfiles}
\newpage
\newpage
\onecolumn
\begin{appendices}
\section{Algorithms}
\begin{algorithm} 
\caption{Direct Model-Free Estimation of Differential Entropy Rate.} 
\label{alg:direct_comp} 
\begin{algorithmic} 
\INPUTS \\
$m$: discretization precision. \\
$\pmb{Z} = (\hat{\pmb{z}}_1, \dots, \hat{\pmb{z}}_T)$: sample path of $\mathbb{R}^n$-valued process  $\{\pmb{z}_t\}$. 
\OUTPUTS \\
An estimate of $h\left(\{\pmb{z}_t\} \right)$.
\ASSUMPTIONS \\
\textbf{A1:} $\{\pmb{z}_t\}$ is stationary and ergodic.\\
\PROCEDURE
\Statex \textbf{Step 1.} Discretize  $\left[\hat{\pmb{z}}_1, \dots,  \hat{\pmb{z}}_T \right]$ using the scheme of Theorem (\ref{theo:discrete_cont_entropy}) with precision $2^{-m}$, to obtain the sequence of $T$ discrete tuples (or characters) $(\hat{\alpha}_1, \dots, \hat{\alpha}_T)$.
\Statex \textbf{Step 2.}  Compute the Lempel-Ziv complexity $\hat{c}_\alpha(T)$ of $(\hat{\alpha}_1, \dots, \hat{\alpha}_T)$ using Listing 1.
\Statex \textbf{Step 3.} Draw $k\geq1$ sequences $(\hat{\alpha}_1^i, \dots, \hat{\alpha}_T^i)$ where each character is sampled independently and uniformly at random with replacement from $(\hat{\alpha}_1, \dots, \hat{\alpha}_T)$.
\Statex \textbf{Step 4.} Compute the Lempel-Ziv complexity $\hat{c}_{\alpha^i}(T)$ of $(\hat{\alpha}_1^i, \dots, \hat{\alpha}_T^i)$ using Listing 1.
\Statex \textbf{Step 6.} Compute the frequency of occurrence of each character in $(\hat{\alpha}_1, \dots, \hat{\alpha}_T)$ and compute the corresponding estimate of discrete entropy $\hat{H}(\hat{\alpha}_t)$.
\Statex \textbf{Step 5.} 
\begin{align}
\label{eq:approx_cond_ent}
h\left(\{\pmb{z}_t\} \right) \approx &  \frac{\hat{c}_\alpha(T)}{\frac{1}{k} \sum_{i=1}^k \hat{c}_{\alpha^i}(T)} \hat{H}(\hat{\alpha}_t) -mn
\end{align}
\end{algorithmic}
\end{algorithm}

\begin{algorithm} 
\caption{Nonparametric Estimation of Differential Entropy Rate.} 
\label{alg:direct_spec} 
\begin{algorithmic} 
\INPUTS \\
$\pmb{Z} = (\hat{\pmb{z}}_1, \dots, \hat{\pmb{z}}_T)$: sample path of $\mathbb{R}^n$-valued process  $\{\pmb{z}_t\}$.
\OUTPUTS \\
An estimate of $h\left(\{\pmb{z}_t\} \right)$
\ASSUMPTIONS \\
\textbf{A1:} $\{\pmb{z}_t\}$ is stationary and ergodic.\\
\textbf{A2:} $\{\pmb{z}_t\}$ is a Gaussian process.\\
\PROCEDURE
\Statex \textbf{Step 1.} Compute an estimate $\hat{g}$ for the matrix-valued spectral densities function  of $\{\pmb{z}_t\}$ as a smoothed periodogram, for instance using Welch's method \cite{welch1967use}.
\Statex \textbf{Step 2.} Approximate Equation (\ref{eq:fourier_entropy_rate}) using previously estimated spectral density functions and Bayesian Quadrature.
\end{algorithmic}
\end{algorithm}

\begin{algorithm} 
\caption{Maximum-Entropy Estimation of Differential Entropy Rate.} 
\label{alg:maxent} 
\begin{algorithmic} 
\INPUTS \\
$\pmb{Z} = (\hat{\pmb{z}}_1, \dots, \hat{\pmb{z}}_T)$: sample path of $\mathbb{R}^n$-valued process  $\{\pmb{z}_t\}$.
\OUTPUTS \\
An estimate of $h\left(\{\pmb{z}_t\} \right)$.
\ASSUMPTIONS \\
\textbf{A3:} All maximum-entropy constraints are of the autocovariance type.\\

\PROCEDURE
\Statex \textbf{Step 1.} Define $p= \left\lfloor 12 \left(\frac{T}{100}\right)^{\frac{1}{4}} \right\rfloor$, and for $0 \leq h \leq p$ compute sample cross-covariance terms $\hat{C}(h)$.
\Statex \textbf{Step 2.} Denote $\hat{\Sigma}_p$ the corresponding sample autocovariance matrix (Equation (\ref{eq:sample_autocov})).
\Statex \textbf{Step 3.} $$ h\left(\{\pmb{z}_t \} \right) \approx \frac{1}{2} \log_2 \left(2 \pi e\right) + \frac{1}{2} \log_2 \left[ \frac{\text{det} \left(\hat{\Sigma}_p\right)}{\text{det} \left( \hat{\Sigma}_{p-1} \right)} \right].$$
\end{algorithmic}
\end{algorithm}

\begin{algorithm} 
\caption{Estimation of Order-$q$ Incremental Entropy.} 
\label{alg:order_q} 
\begin{algorithmic} 
\INPUTS \\
$Y = (\hat{y}_1, \dots, \hat{y}_T)$: sample path of returns of new asset A. \\
$\pmb{X} = (\hat{\pmb{x}}_1, \dots, \hat{\pmb{x}}_T)$: sample path of returns of the $n$ assets and factors in the reference pool P. \\
$q$: Sparsity parameter.
\OUTPUTS \\
An estimate of $h_q\left(\{y_t\} \vert \{\pmb{x}_t\} \right)$.
\ASSUMPTIONS \\
\textbf{A1:} $\{y_t, \pmb{x}_t\}$ is stationary and ergodic.\\
\PROCEDURE
\Statex \textbf{Step 0.} Normalize $Y$ and $\pmb{X}$ so that each column has sample variance $2/(\pi e)$.
\Statex \textbf{Step 1.} Sample $k$ random partitions of $\{1, \dots, n\}$ into subsets of size $q$.
\Statex \textbf{Step 2.} For each subset $i$ in random partition $j$, $1 \leq j \leq k$, define $\pmb{X}_{ij}$ by selecting the column of $\pmb{X}$ whose indexes are in subset $i$.
\Statex \textbf{Step 3.} For each $i$, $j$, use either one of Algorithms \ref{alg:direct_comp}, \ref{alg:direct_spec}, or \ref{alg:maxent} to estimate entropy rates, first with $\pmb{Z}=Y$ then with $\pmb{Z}=\pmb{X}_{ij}$, and finally with $\pmb{Z}=[Y, \pmb{X}_{ij}]$, and denote $I_{ij}$ the difference between the sum of the first two estimated entropy rates and the last.
\Statex \textbf{Step 4.} $$\mathbb{ID}^q\left(A; P\right)  \approx \underset{i, j}{\min} ~ 1/I_{ij}.$$
\end{algorithmic}
\end{algorithm}

\begin{table*}
\begin{lstlisting}[linewidth=\textwidth,language=Python, caption=Sample Python code computing the Lempel-Ziv complexity of a sequence of characters.]
def lz76_complexity(S):
    """
    Compute the Lempel-Ziv complexity of a sequence of characters as per [1]. 
    
    :param S: List of characters forming the sequence whose complexity we 
        would like to evaluate. Characters can be any Python object with a
        string representation.

    Reference: 
        [1] On the complexity of finite sequences, A Lempel, J Ziv
            IEEE Transactions on information theory, (1976)

    Examples:
        >>> print(lz76_complexity('0001101001000101')[0])
            6

        >>> print(lz76_complexity('0001101001000101')[1])
            ['0', '001', '10', '100', '1000', '101']
    """
    n = len(S)
    exhaustive_history = [S[0]]
    complexity = 1
    hi, i, u, v, vmax = 1, 0, 1, 1, 1
    
    while u+v <= n:
        if S[i+v-1] == S[u+v-1]:
            v += 1
        else:
            vmax = max(v, vmax)
            i += 1
            if i == u:
                complexity += 1
                u += vmax
                v, i = 1, 0
                exhaustive_history += ["".join([str(_) for _ in S[hi:hi+vmax]])]
                hi += vmax
                vmax = v
            else:
                v = 1  
          
    if v != 1:
        exhaustive_history += ["".join([str(_) for _ in S[hi:]])]
        complexity += 1
    
    return complexity, exhaustive_history
\end{lstlisting}
\end{table*}
\newpage 
\section{Experimental Setup}
\label{sct:exp_setup}
Throughout this paper, unless stated otherwise, asset returns are daily close-to-close returns. For foreign exchange rates, we use 5PM Eastern Time as daily cutoff for the close. For futures, we use exchange settlement prices instead of the close. All futures contracts are continuously adjusted front month contracts. The adjustment method used is the backward ratio method, and rolling occurs on the first day of the delivery month. All financial data used range from January 1st 2008 to January 1st 2018. All futures considered in this paper are listed in Table \ref{table:futures_list}, all stocks considered in this paper are listed in Table \ref{table:dj_list}, and all currency pairs considered in this paper are listed in Table \ref{table:fx_list}.

\begin{table}[h!]
\centering
\begin{tabular}{ |l|l| }
  \hline
Symbol & Description \\
  \hline
B   & ICE Brent Crude Oil \\
C   & CBOT Corn \\
CL & NYMEX WTI Crude Oil \\
EC & CME Euro FX \\
ED & CME Eurodollar \\
ES & CME S\&P 500 Index E-Mini \\
FV & CBOT 5-year US Treasury Note \\
G    & ICE Gasoil	\\
GC & NYMEX Gold \\
JY  & CME Japanese Yen JPY \\
NG & NYMEX Natural Gas \\
NQ & CME NASDAQ 100 Index Mini \\
RB & NYMEX Gasoline \\
S    & CBOT Soybeans \\
TU & CBOT 2-year US Treasury Note	\\
TY & CBOT 10-year US Treasury Note \\
US & CBOT 30-year US Treasury Bond \\
VX & CBOE VIX Futures \\
W   & CBOT Wheat \\
YM & CME E-mini Dow Jones \\
  \hline
\end{tabular}
\caption{List of the $20$ U.S. future contracts used in experiments throughout the paper.}
\label{table:futures_list}
\end{table}

\begin{table}[h!]
\centering
\begin{tabular}{ |l|l|l|l| }
  \hline
Company									& Exchange	& 	Symbol	& Industry \\
  \hline
Apple Inc.									& NASDAQ	& 	AAPL	&	Consumer Electronics \\
The American Express Company				& NYSE		& 	AXP		&	Consumer Finance \\	
Boeing Company							& NYSE		&	BA		&	Aerospace, Defense \\
Caterpillar, Inc.								& NYSE		&	CAT		&	Construction and Mining Equipment \\	
Cisco Systems, Inc.							& NASDAQ	& 	CSCO	&	Computer Networking \\	
Chevron Corporation							& NYSE		&	CVX		&	Oil \& Gas \\
The Walt Disney Company					& NYSE		&	DIS		&	Broadcasting, Entertainment \\	
DowDuPont Inc.							& NYSE		&	DWDP	&	Chemical Industry \\
General Electric	 Company						& NYSE		&	GE		&	Conglomerate \\
The Goldman Sachs Group, Inc.	 				& NYSE		&	GS		&	Banking, Financial Services \\
The Home Depot, Inc.						& NYSE		&	HD		&	Home Improvement Retailer \\	
International Business Machines Corporation (IBM)	& NYSE		&	IBM		&	Computers, Technology \\
Intel  Corporation							& NASDAQ	&	INTC		&	Semiconductors \\
Johnson \& Johnson	Inc.    					& NYSE		&	JNJ		&	Pharmaceuticals \\
J.P. Morgan Chase  \& Co						& NYSE		&	JPM		&	Banking, Financial Services \\	
The Coca-Cola Company						& NYSE		&	KO		&	Beverages \\
McDonald's  Corporation						& NYSE		&	MCD		&	Fast Food \\
3M Company								& NYSE		&	MMM	&	Conglomerate \\
Merck \& Company Inc.						& NYSE		&	MRK		&	Pharmaceuticals \\	
Microsoft Corporation						& NASDAQ	& 	MSFT	&	Software \\
Nike, Inc.									& NYSE		&	NKE		&	Apparel \\
Pfizer, Inc.								& NYSE		&	PFE		&	Pharmaceuticals \\
The Procter \& Gamble Company				& NYSE		&	PG		&	Consumer Goods \\
The Travelers Companies, Inc.					& NYSE		&	TRV		&	Insurance \\
UnitedHealth Group Incorporated				& NYSE		&	UNH		&	Managed Health Care \\
United Technologies Corporation				& NYSE		&	UTX		&	Conglomerate \\
Visa Inc.									& NYSE		&	V		&	Consumer Banking \\
Verizon Communications Inc.					& NYSE		&	VZ		&	Telecommunication \\
Walmart Inc.								& NYSE		&	WMT	&	Retail \\
Exxon Mobil Corporation						& NYSE		&	XOM		&	Oil \& Gas \\
  \hline
\end{tabular}
\caption{Constituents of the Dow Jones Industrial Average at the time of writing this paper.}
\label{table:dj_list}
\end{table}

\begin{table}[h!]
\centering
\scalebox{.4}{
\begin{tabular}{ |l|l| }
  \hline
Symbol & Description \\
  \hline
AAPL 	& Apple Inc. \\
ABBV 	& AbbVie Inc.  \\
ABT  	& Abbott Laboratories  \\
ACN 		& Accenture plc \\
AGN  	& Allergan plc \\
AIG  		& American International Group Inc. \\
ALL	 	& Allstate Corp. \\
AMGN 	& Amgen Inc. \\
AMZN 	& Amazon.com \\
AXP 		& The American Express Company\\
BA 		& Boeing Company \\
BAC		& Bank of America Corp \\
BIIB		& Biogen \\
BK		& The Bank of New York Mellon \\
BKNG	& Booking Holdings \\
BLK		& BlackRock Inc \\
BMY		& Bristol-Myers Squibb \\
BRK.B	& Berkshire Hathaway \\
C		& Citigroup Inc \\
CAT		& Caterpillar Inc \\
CELG	& Celgene Corp \\
CHTR	& Charter Communications \\
CL		& Colgate-Palmolive Co. \\
CMCSA	& Comcast Corporation \\
COF		& Capital One Financial Corp. \\
COP		& ConocoPhillips \\
COST	& Costco \\
CSCO	& Cisco Systems, Inc. \\
CVS		& CVS Health \\
CVX		& Chevron Corporation \\
DHR		& Danaher Corporation \\
DIS		& The Walt Disney Company \\
DUK		& Duke Energy \\
DWDP	& DowDuPont Inc. \\
EMR		& Emerson Electric Co. \\
EXC		& Exelon \\
F		& Ford Motor \\
FB		& Facebook \\
FDX		& FedEx \\
FOX		& 21st Century Fox \\
FOXA	& 21st Century Fox \\
GD		& General Dynamics \\
GE		& General Electric Company \\
GILD		& Gilead Sciences \\
GM		& General Motors \\
GOOG	& Alphabet Inc \\
GOOGL	& Alphabet Inc \\
GS		& The Goldman Sachs Group, Inc. \\
HAL		& Halliburton \\
HD		& The Home Depot, Inc. \\
HON		& Honeywell \\
IBM		& International Business Machines Corporation (IBM)\\
INTC		& Intel Corporation \\
JNJ		& Johnson \& Johnson Inc \\
JPM		& J.P. Morgan Chase \& Co \\
KHC		& Kraft Heinz \\
KMI		& Kinder Morgan \\
KO		& The Coca-Cola Company \\
LLY		& Eli Lilly and Company \\
LMT		& Lockheed Martin \\
LOW		& Lowe's \\
MA		& MasterCard Inc \\
MCD		& McDonald's  Corporation \\
MDLZ	& Mondelēz International \\
MDT		& Medtronic Inc. \\
MET		& Metlife Inc. \\
MMM	& 3M Company \\
MO		& Altria Group \\
MON		& Monsanto \\
MRK		& Merck \& Company \\
MS		& Morgan Stanley \\
MSFT	& Microsoft \\
NEE		& NextEra Energy \\
NKE		& Nike, Inc. \\
ORCL	& Oracle Corporation \\
OXY		& Occidental Petroleum Corp. \\
PEP		& Pepsico Inc. \\
PFE		& Pfizer, Inc. \\
PG		& The Procter \& Gamble Company \\
PM		& Phillip Morris International \\
PYPL	& PayPal Holdings \\
QCOM	& Qualcomm Inc. \\
RTN		& Raytheon Company \\
SBUX	& Starbucks Corporation \\
SLB		& Schlumberger \\
SO		& Southern Company \\
SPG		& Simon Property Group, Inc. \\
T		& AT\&T Inc \\
TGT		& Target Corp. \\
TWX		& Time Warner Inc. \\
TXN		& Texas Instruments \\
UNH		& UnitedHealth Group Incorporated \\
UNP		& Union Pacific Corporation \\
UPS		& United Parcel Service \\
USB		& US Bancorp \\
UTX		& United Technologies Corporation \\
V		& Visa Inc. \\
VZ		& Verizon Communications Inc. \\
WBA		& Walgreens Boots Alliance \\
WFC		& Wells Fargo \\
WMT	& Wal-Mart \\
XOM		& Exxon Mobil Corporation \\
  \hline
\end{tabular}}
\caption{Constituents of the S\&P 100 index  at the time of writing this paper.}
\label{table:sp100_list}
\end{table}

\begin{table}[h!]
\centering
\begin{tabular}{ |l|l| }
  \hline
Symbol & Description \\
  \hline
AUDUSD  	& Price of $1$ Australian Dollar in U.S. Dollars \\
CADUSD   & Price of $1$ Canadian Dollar in U.S. Dollars \\
CHFUSD 	& Price of $1$ Swiss Franc in U.S. Dollars \\
CZKUSD 	& Price of $1$ Czech Koruna in U.S. Dollars \\
DKKUSD 	& Price of $1$ Danish Krone in U.S. Dollars \\
EURUSD 	& Price of $1$ Euro in U.S. Dollars \\
GBPUSD 	& Price of $1$ British Bound in U.S. Dollars \\
HKDUSD   & Price of $1$ Hong-Kong Dollar in U.S. Dollars	\\
HUFUSD 	& Price of $1$ Australian Dollar in U.S. Dollars \\
JPYUSD  	& Price of $1$ Japanese Yen in U.S. Dollars \\
MXNUSD 	& Price of $1$ Mexican Peso in U.S. Dollars \\
NOKUSD 	& Price of $1$ Norwegian Krone in U.S. Dollars \\
NZDUSD 	& Price of $1$ New Zealand Dollar in U.S. Dollars \\
PLNUSD   & Price of $1$ Poland Zloty in U.S. Dollars \\
SEKUSD 	& Price of $1$ Swedish Krona in U.S. Dollars	\\
SGDUSD 	& Price of $1$ Singapore Dollar in U.S. Dollars \\
TRYUSD 	& Price of $1$ Turkish Lira in U.S. Dollars \\
ZARUSD 	& Price of $1$ South African Rand in U.S. Dollars \\
  \hline
\end{tabular}
\caption{List of currency pairs used in experiments throughout the paper.}
\label{table:fx_list}
\end{table}
\end{appendices}

\end{document}